\documentclass[11pt]{article}
\usepackage{amssymb,amsmath}
\usepackage{bm} 
\usepackage{bbold}
\usepackage{booktabs} 
\usepackage{array}
\usepackage{latexsym}
\usepackage{graphicx}
\usepackage{color}
\usepackage{xcolor}
\usepackage{slashed}
\usepackage{datetime}
\usepackage[nosort]{cite}
\usepackage{verbatim}
\usepackage{enumerate}

\usepackage{chngpage} 
\allowdisplaybreaks
\usepackage{psfrag}

\usepackage{mciteplus}

\usepackage[colorlinks=true,      linkcolor=blue,      urlcolor=blue,
            filecolor=blue,      citecolor=blue,       pdfstartview=FitH,
						pdfpagemode=UseNone,      bookmarksopen=true]{hyperref}  
\usepackage[all]{hypcap}     

\numberwithin{equation}{section} 
\allowdisplaybreaks  

\newcommand\field[1]{{\ensuremath{\mathbb{{#1}}}}}
\newcommand{\RR}{\field{R}}

\def\psib{{\overline{\psi}}}

\def\cD{{\cal D}}

\def\cM{{\cal M}}

\def\cO{{\cal O}}

\def\cR{{\cal R}}
\def\cS{{\cal S}}

\def\psib{{\overline{\psi}}}

\def\uuv{{\underline{v}}}
\def\uur{{\underline{r}}}
\def\uux{{\underline{x}}}
\def\uui{{\underline{i}}}

\def\uua{{\underline{a}}}

\def\gvr{{\Gamma^{\uuv \uur}}}
\def\grr{{\Gamma^{\uur}}}
\def\gvv{{\Gamma^{\uuv}}}
\def\gii{{\Gamma^{\uui}}}

\def\gxx{{\Gamma^{\uux}}}
\def\gvi{{\Gamma^{\underline{vi}}}}
\def\gvx{{\Gamma^{\underline{vx}}}}
\def\pd{{\partial}}
\def\kh{{\hat{k}_i}}
\def\ief{{ingoing Eddington-Finkelstein coordinates}}
\newcommand\om{\omega}

\newcommand\ov{\over }
\def\cmt{{\widetilde{\cM}}}
\def\chib{{\overline{\chi}}}
\def\le{\left}
\def\ri{\right}
\newcommand{\ppp}[1]{ {\left(\psi_+^{(+)}\right)^{(#1)}}}
\newcommand{\pmm}[1]{ {\left(\psi_-^{(-)}\right)^{(#1)}}}

\newcommand{\ppm}[1]{ {\left(\psi_+^{(-)}\right)^{(#1)}}}
\newcommand{\pmp}[1]{ {\left(\psi_-^{(+)}\right)^{(#1)}}}

\def\tg{\widetilde{\Gamma}}

\newcommand{\be}{\begin{equation}}
\newcommand{\ee}{\end{equation}}
\newcommand{\bea}{\begin{eqnarray}}
\newcommand{\eea}{\end{eqnarray}}
\newcommand{\bwt}{\begin{widetext}}
\newcommand{\ewt}{\end{widetext}}

\newcommand{\bi}{\begin{itemize}}
\newcommand{\ei}{\end{itemize}}
\newcommand{\ben}{\begin{enumerate}}
\newcommand{\een}{\end{enumerate}}
\newcommand{\bca}{\begin{cases}}
\newcommand{\eca}{\end{cases}}
\newcommand{\bln}{\begin{align}}
\newcommand{\eln}{\end{align}}
\newcommand{\bst}{\begin{split}}
\newcommand{\est}{\end{split}}

\begin{document}


\begin{flushright}
QMUL-PH-19-22\\
\end{flushright}

\vspace{3mm}

\begin{center}

{\huge {\bf
Fermionic pole-skipping in holography
}}
\vspace{14mm}

{\large
\textsc{Nejc \v{C}eplak,~ Kushala Ramdial,~ David Vegh}}

\vspace{12mm}
 Centre for Research in String Theory, School of Physics and Astronomy,\\
Queen Mary University of London, 327 Mile End Road, London, E1 4NS, United Kingdom\\

\vspace{4mm}

\texttt{\href{mailto:n.ceplak@qmul.ac.uk}{n.ceplak@qmul.ac.uk}\,,  \href{mailto:k.ramdial@se15.qmul.ac.uk}{k.ramdial@se15.qmul.ac.uk}\,, \href{mailto:d.vegh@qmul.ac.uk}{d.vegh@qmul.ac.uk}}
\vspace{13mm}

\textsc{Abstract}

\end{center}

\begin{adjustwidth}{13.5mm}{13.5mm} 

\vspace{1mm}
\noindent
We examine thermal Green's functions of fermionic operators in quantum field theories with gravity duals. The calculations are performed on the gravity side using ingoing Eddington-Finkelstein coordinates.
We find that at negative imaginary Matsubara frequencies and special values of the wavenumber, there are multiple solutions to the bulk equations of motion that are ingoing at the horizon and thus the boundary Green's function is not uniquely defined.
At these points in Fourier space a line of poles and a line of zeros of the correlator intersect.
We analyze these `pole-skipping' points in three-dimensional  asymptotically anti-de Sitter spacetimes where exact Green's functions are known.  We then generalize the procedure to higher-dimensional spacetimes.
We also discuss the special case of a fermion with half-integer mass in the BTZ background.
We discuss the implications and possible generalizations of the results.

\end{adjustwidth}

\thispagestyle{empty}
\newpage



\baselineskip=14.5pt
\parskip=3pt

\tableofcontents

\baselineskip=15pt
\parskip=3pt

\section{Introduction}
\label{sec:Introduction}

Despite immense progress in our understanding of quantum field theories, a complete description of strongly interacting theories is still lacking.
The gauge/gravity correspondence \cite{Maldacena:1997re, Gubser:1998bc, Witten:1998qj} opened up a new path towards studying certain strongly coupled large-N quantum field theories by investigating their dual, weakly coupled, gravitational theories on curved backgrounds.

A basic quantity of interest in finite temperature quantum field theories is the retarded two-point function of an operator. It measures how the system in equilibrium responds to perturbations.
The prescription of how to compute the correlators in holographic theories in real-time was formulated in \cite{Son:2002sd}  (see also \cite{Horowitz:1999jd,Herzog:2002pc,Skenderis:2008dh,Skenderis:2008dg, Son:2009vu, Iqbal:2009fd, vanRees:2009rw, Glorioso:2018mmw, Liu:2018crr, deBoer:2018qqm}). A thermal state in the field theory corresponds to a black hole in an asymptotically anti-de Sitter (AdS) spacetime on the gravity side.
Boundary operators are dual to fields in the bulk (i.e. the curved background). The AdS/CFT dictionary relates the Green's function of a boundary operator $\cO$,
to solving the equations of motion for the corresponding bulk field $\phi$. Near the black hole event horizon the second-order equation of motion has an ingoing and an outgoing solution. In order to calculate the retarded (advanced) Green's function, one should pick the ingoing (outgoing) solution \cite{Son:2002sd}. This wavefunction is then evolved in the radial direction outwards to the spatial boundary of AdS where the Green's function can be read off.
Using this prescription, the retarded Green's function is uniquely defined in terms of the bulk solution satisfying the prescribed boundary conditions. One of the important conditions for this is the uniqueness of the ingoing solution in the interior.

In principle the prescription for calculating the retarded Green's function $G_R(\omega,k)$ is straightforward. However, evolving the ingoing solution to the boundary turns out to be computationally challenging. While it can be done explicitly in the simplest cases (e.g. the BTZ black hole \cite{ Son:2002sd, Iqbal:2009fd,Banados:1992gq, Banados:1992wn, Birmingham:2001hc, Cardoso:2001hn, Birmingham:2001pj, Blake:2019otz}), typically one has to use numerical methods to obtain the solutions. Generically, the retarded Green's function depends in a complicated way on the details of the state in the quantum field theory.
Simplifications occur in the low-frequency and low-wavenumber limit of the correlator. In this case,  the form of the retarded Green's function is dictated by near-horizon physics in the bulk and its qualitative features are independent of the rest of the geometry (see e.g. the results on shear viscosity \cite{Kovtun:2004de}).

Recently it has been observed that certain properties of the correlators {\it away from the $\om~=~0$, $k~=~0$ point in Fourier space} can  already be seen in the near horizon behavior of the solutions \cite{ Grozdanov:2017ajz,Blake:2017ris, Blake:2018leo, Grozdanov:2018kkt}.
Initially it was observed that at special complex values of the frequency and the momentum, the retarded Green's function contained information about the chaotic behavior of the theories (see \cite{Shenker:2013pqa, Shenker:2013yza, Shenker:2014cwa, Roberts:2014isa, Maldacena:2015waa}).
Such behavior was dubbed "pole-skipping" as it occurs where a line of poles intersects a line of zeros in the Green's function of the dual boundary operator (see also \cite{Balm:2019dxk} for related phenomena in the case of Fermi surfaces).

As it was shown in \cite{Blake:2019otz, Grozdanov:2019uhi, Natsuume:2019sfp, Natsuume:2019xcy} pole-skipping is not limited  only to the components of the energy-momentum tensor but can also be observed in other fields in the theory.
Its gravitational origin stems from the fact that at these points there is no unique ingoing solution at the interior of the bulk spacetime. With this, the holographic retarded Green's function ceases to be uniquely defined and becomes multivalued.
The interesting aspect of this phenomenon is that the bulk computation is limited to the horizon and has no knowledge of the boundary. In this sense a local calculation in the bulk constrains the structure of boundary Green's functions.

In this work we build on the findings of \cite{Blake:2019otz} and describe the pole-skipping for minimally coupled spinor fields on asymptotically AdS backgrounds.
Looking at the exact Green's function for fermions in the BTZ black hole background found in \cite{Iqbal:2009fd}, one can observe that there are special points at which the lines of poles intersect the lines of zeros. They occur precisely at the fermionic Matsubara frequencies\footnote{Euclidean Green's functions are defined at Matsubara frequencies which are real numbers.
With a slight abuse of notation, we will refer to the purely imaginary $\omega_n^F$ frequencies as fermionic Matsubara frequencies. }
\begin{align}
\label{eq:fermmats}
\omega = \omega_n^F := - 2 \pi i T \left(n + \frac12\right)\,, \qquad n = 0, 1, 2, \ldots\,.
\end{align}
This nicely complements the fact that scalar and energy-momentum pole-skipping points occur at bosonic Matsubara frequencies $\omega= \omega_n^B = - 2 \pi i T n$, with $n = 1, 2, 3, \ldots$.
In light of this, it has been conjectured in \cite{Blake:2019otz} that
this behavior has a bulk interpretation in terms of non-unique ingoing solutions. Here we will explicitly show that this is indeed the case.

We emphasize that only the energy-momentum near-horizon behavior is clearly related to chaos in holographic theories. There, one can observe pole-skipping at the first {\it positive} bosonic Matsubara frequency $\om = +2 \pi i T$ where the right-hand side is precisely the Lyapunov exponent that characterizes out-of-time order higher-point functions.
In other examples, such as the case of the scalar field, pole-skipping occurs on the lower-half frequency plane. Since we will encounter pole-skipping at fermionic Matsubara frequencies, it is even less likely that this phenomenon can be related to quantum chaos in a straightforward manner.
However, these features might be important in holographic theories in general.

The paper is organized in the following way.
In section~\ref{sec:Review} we review the pole-skipping phenomenon in the case of a minimally coupled scalar field.
In section~\ref{sec:fermion} we define a minimally coupled fermion field on an anti-de Sitter background and discuss spinors in holography.
Then in section~\ref{sec:AdS3fermionic} we  look at  pole-skipping in 3-dimensional bulk spacetimes.
The generalization to higher dimensions is given in section~\ref{sec:generalpoleskipping}, while in section~\ref{sec:Examples} we discuss some examples. Most notably we use the results to calculate the fermionic pole-skipping points for the BTZ black hole and compare our results with the known retarded Green's function. We examine the special cases of boundary operators with half-integer conformal dimensions and relate them to anomalous pole-skipping points.
We conclude with a discussion in section~\ref{sec:Discussion}.
In appendix \ref{app:gammas} we  present explicit representations of gamma matrices that can be useful in practical applications.
In appendix~\ref{app:form} we examine the form of the Green's function near a generic pole-skipping point and discuss the appearance of anomalous points.
Some of the more detailed calculations omitted in the main text are collected in  appendix \ref{app:details}.
 In appendix \ref{app:ExactBTZ} we  review the calculation of the exact Green's function for the BTZ black hole. We also  consider the equality of the retarded and advanced Green's function at the pole-skipping points. Finally we calculate the form of the retarded Green's function in the special cases where the mass of the fermion takes a half-integer value.

\section{Review of pole-skipping}
\label{sec:Review}

In this section we present the general form of the background metric in \ief \, and review the systematic procedure to extract the locations of the pole-skipping points in the case of a minimally coupled scalar field, which was developed in \cite{Blake:2019otz}.

We start by assuming that the action for the background fields is given by
\begin{align}\label{eq:backgroundaction}
S = \int d^{d+2}x \sqrt{- g} \left( R - 2 \Lambda \right)+ S_{matter}\,,
\end{align}
where $\Lambda = - d(d+1)/2L^2$ is the cosmological constant and $L$ is the AdS radius, which we henceforth set to $L = 1$. The term $S_{matter}$ allows for additional matter content which can also contribute to the curvature of the background.

We further assume that the equations of motion for this action admit a planar black hole solution given by the metric
\begin{align}
ds^2 = - r^2 f(r) dt^2 + \frac{dr^2}{r^2 f(r)} + h(r) d\vec{x}^2\,,
\end{align}
where $r$ is the radial direction. The boundary of spacetime is located at $r\rightarrow \infty$. Furthermore, $t$ denotes time and $x_i$ with ${i=1\ldots d}$ are the (flat) coordinates of the $d$ spatial dimensions. The combination $(t,\vec{x}) \in \RR^{1,d}$  also denotes the  Minkowski coordinates of the corresponding boundary theory.
The exact form of  the two functions $f(r)$ and $h(r)$ in general depends on the matter content of the theory.
Since we want our spacetime to be asymptotically anti-de Sitter, they have to approach $f(r) \rightarrow 1$ and $h(r) \rightarrow r^2$ as $r \rightarrow \infty$.

We assume that the background has a horizon at $r= r_0$, i.e. the emblackening factor vanishes at this radius: $f(r_0) = 0$. We also assume that the Taylor series of the functions $f$ and $h$ have finite radii of convergence near the horizon.
 The Hawking temperature of the black hole is given by
\begin{align}
\label{eq:hawk}
4 \pi T = r_0^2\, f'(r_0)\,.
\end{align}
In order to extract the pole-skipping points, it is convenient to introduce the \ief , defined by
\begin{align}
v = t + r_*\,, \qquad \frac{dr_*}{dr} = \frac{1}{r^2 f(r)}\,,
\end{align}
in which the background metric takes the form
\begin{align}\label{eq:backgroundmetric}
ds^2 = - r^2 f(r) dv^2 + 2dv\,dr +  h(r) d\vec{x}^2\,.
\end{align}
The vacuum solutions ($S_{matter} = 0$) with such properties are characterized by
\begin{align}
f(r) = 1 - \left(\frac{r_0}{r}\right)^{d+1}\,,\qquad h(r) = r^2\,,
\end{align}
which are the BTZ black hole \cite{Banados:1992gq, Banados:1992wn} if $d=1$ and the planar  AdS-Schwarzschild black hole solution if $d \geq 2$.

\subsection{Minimally coupled scalar field in the bulk}
The simplest instance for which one can observe pole-skipping is a minimally coupled scalar field in an asymptotically anti-de Sitter spacetime. To that end, we add to the action of the background \eqref{eq:backgroundaction} the action of a massive scalar in a curved background, which is given by
\begin{equation}
S_\varphi =  - \dfrac{1}{2} \int d^{d+2}x \sqrt{-g}\left(g^{\mu\nu}\partial_\mu\varphi\partial_\nu\varphi + m^{2}\varphi^{2}\right).
\end{equation}
The Green's function can be extracted by finding solutions to the equation of motion
\begin{equation}
 \label{EOMScalar}
\partial_{\mu}(\sqrt{-g}g^{\mu\nu}\partial_\nu\varphi) - m^{2}\sqrt{-g}\varphi = 0\,.
\end{equation}
Note that this is a second order differential equation for a single scalar field. As such it has two free parameters that we need to fix with boundary conditions. \

\noindent
The scaling dimension $\Delta$ and the mass $m$ of the scalar  field are related via
\begin{align}
\Delta\left(\Delta -d -1 \right) = m^2,\,
\end{align}
where we take the larger of the two roots to be the scaling dimension in the standard quantization. \

\noindent
If we wish to calculate the retarded Green's function, we need to choose the ingoing solution at the horizon \cite{Son:2002sd}. To do so, we consider the ansatz $\varphi = \phi(r)e^{-i\omega v + i\vec{k}\cdot\vec{x}}$ and perform a  series expansion of $\phi(r)$ around the horizon. We find that this boundary condition gives a unique ingoing solution to (\ref{EOMScalar}) for generic values of $\omega$ and $k$ up to an overall normalization.\

\noindent
The next step is to expand this solution near the boundary as
 \begin{equation}
 \label{eq:scalarasy}
 \phi = \phi_{A}(\omega,k)r^{\Delta -d - 1} + \phi_{B}(\omega,k)r^{-\Delta} + \dots
 \end{equation}
  to obtain the boundary retarded Green's function up to the possible existence of contact terms by
 \begin{equation} \label{Eq:GFscalar}
  G_{\mathcal{OO}}^{R}(\omega,k) = (2\Delta - d - 1)\frac{\phi_{B}(\omega,k)}{\phi_{A}(\omega,k)}.
  \end{equation}

\subsection{Pole-skipping points}

Here we briefly explain why imposing boundary conditions at special values of frequency $\omega$ and momentum $k$ is not sufficient to uniquely  (up to an overall factor) specify a solution $\varphi$ to the equation (\ref{EOMScalar}) and give the locations of the pole-skipping points for a minimally coupled scalar field. We closely follow \cite{Blake:2019otz}, where these calculations were initially performed. See that work and the references therein for more details.

After performing the Fourier transform and switching to the Eddington-Finkelstein coordinate system, (\ref{EOMScalar}) becomes
\begin{align} \label{Eq:EOMEF}
{d \ov dr} \left[h^{d \ov 2} \left(r^2 f \partial_{r} \phi - i\omega \phi \right)\right] - i \omega h^{d \ov 2} \partial_{r} \phi - h^{{d \ov 2} - 1} \left(k^2 + m^2 h \right) \phi = 0.
\end{align}
We look for solutions that are regular at the horizon. Such solutions can be written as a Taylor series expansion $\phi(r) =\phi_{0} + \phi_{1} (r - r_{0}) + \dots$ around $r = r_0$.
Near the horizon, there exist two power law solutions $\phi = (r - r_{0})^\alpha$ with
\begin{align}
\alpha_1 = 0,  \qquad \alpha_2 = {i\omega \ov 2\pi T},\,
\end{align}
which do not depend on $k$ and $m$. For generic values of $\omega$, only the solution with exponent $\alpha_1$ is regular and is therefore taken to be the ingoing solution. (Note that  $\alpha_1$ had to be zero, because the horizon is not a distinguished location in infalling coordinates.)
However, at the special values of frequency $\omega_n = -2i \pi T n$ with $n \in \{1, 2, \dots\}$, the second exponent becomes $\alpha_2 = n$ and naively both solutions seem to be regular at the horizon.  A more detailed calculation shows that one of the solutions contains logarithmic divergences which spoil the regularity, so there is still a unique regular ingoing solution.
One then finds that all such logarithmic divergences vanish for some particular values of the momentum $k$. This means that for finely tuned values of $\omega$ and $k$,  there is  no unique ingoing solution to (\ref{Eq:EOMEF}), which renders $G_{\mathcal{OO}}^{R}(\omega, k)$ ill-defined.

To see this explicitly we expand  \eqref{Eq:EOMEF} in a series around the horizon and solve the resulting equation order by order. At the zeroth order, one obtains a relation between the lowest two field coefficients $\phi_0$ and $\phi_1$ by
\begin{align} \label{eomathorizon}
- \left( k^2 + m^2 h(r_{0}) + {i\omega dh'(r_{0}) \ov 2} \right) \phi_{0} + \left(4\pi T - 2i \omega \right) h(r_{0}) \, \phi_{1} = 0,
\end{align}
which, for generic values of $k$ and $\omega$, fixes $\phi_{1}$ in terms of $\phi_{0}$. Higher order terms of the series expansion of the equation of motion allow us to relate all the field coefficients $\phi_n$ in terms of only $\phi_0$. Thus we explicitly construct a  unique regular solution with an undetermined overall normalization in the form of the factor $\phi_0$.

If the frequency takes the value of the first bosonic Matsubara frequency  $\omega = \omega_1 = -2i \pi T $, this method fails as (\ref{eomathorizon}) reduces to
\begin{align}
\left(k^2 + m^2 h(r_0) + \pi d T h'(r_0) \right)\, \phi_0 = 0.
\end{align}
For generic values of $k$, the above equation sets $\phi_0 = 0$. All higher order coefficients $\phi_n$ are then related to $\phi_1$, which can be taken as the undetermined normalization of the unique solution.
However, by finely tuning both the frequency $\omega$ and the momentum $k$ to take the values
\begin{align} \label{Eq:SpecialPoints}
\omega_1 = -2i \pi T, \qquad k_{1}^{2} = -m^2 h(r_0) - \pi dTh'(r_0),
\end{align}
the equation (\ref{eomathorizon}) becomes  trivially satisfied.  In this case, both $\phi_0$ and $\phi_1$ remain undetermined and all higher coefficients of the series expansion of the scalar field  $\phi_n$ are determined in terms of both $\phi_0$ and $\phi_1$. The regular solution then has two independent parameters and is thus not unique. Consequently, at \eqref{Eq:SpecialPoints}, the boundary Green's function is not uniquely defined.

One finds that there are pole-skipping points at  higher Matsubara frequencies as well. At $\omega = \omega_n= -2\pi i T n$, there are   $2n$ wavenumbers $k_{n}$ at which we observe pole-skipping.
In order to locate these points, one needs look at higher orders in the expansion of \eqref{Eq:EOMEF} around the horizon. Setting all the coefficients of the expansion to zero results in a coupled set of algebraic equations that can be written as
\begin{align}
\label{Eq:scalarmatrix}
M(\omega, k^2) \cdot \phi \equiv \begin{pmatrix}
M_{11} & 2\pi T - i\omega & 0 & 0 &\ldots  \\
M_{21} & M_{22}  & 4\pi T - i\omega & 0 &\ldots  \\
M_{31} & M_{32}& M_{33} & 6\pi T - i\omega &\ldots  \\
\ldots & \ldots & \ldots & \ldots & \ldots
\end{pmatrix} \begin{pmatrix}
\phi_{0} \\
\phi_{1} \\
\phi_{2} \\
\vdots
\end{pmatrix} =0\,,
\end{align}
where the coefficients are generically of the form $M_{ij}(\om, k^2) = i\om \,a_{ij} + k^2 b_{ij} + c_{ij}$, with $a_{ij}$, $b_{ij}$, and $c_{ij}$ determined by the background metric.

At generic values of frequency, (\ref{Eq:scalarmatrix}) is easily solved in an iterative manner. In fact, these are the equations that allow us to express all $\phi_n$ as functions of $\phi_0$.
However, at $\omega = \omega_n$, it is not possible to construct an ingoing solution in this way as the coefficient of $\phi_n$ vanishes in the $n^{th}$ row of (\ref{Eq:scalarmatrix}). We then obtain a
closed set of equations for the  coefficients $\tilde{\phi} = (\phi_0, \ldots, \phi_{n-1})$, which is of the form
\begin{align}
\mathcal{M}^{(n)} (\om_n, k^2)\,\cdot\,\widetilde{\phi} = 0\,,
\end{align}
where $\mathcal{M}^{(n)} (\om_n, k^2)$ is the submatrix of $M(\omega, k^2)$ consisting of the first $n$ rows and first $n$ columns.
For generic values of $k$, the matrix $\mathcal{M}^{(n)} (\om_n, k^2)$ is invertible, setting $\tilde{\phi} = 0$.
With that, $\phi_n$ takes the role of the free parameter and the remaining equations in \eqref{Eq:scalarmatrix} can be used to relate $\phi_m$, with $m >n$, to $\phi_n$, thus obtaining a unique ingoing solution up to an overall normalization, which is now $\phi_n$.

If, on the other hand, the value of $k$ is such that the matrix $\mathcal{M}^{(n)} (\om_n, k^2)$ is not invertible, then we  get an additional non-trivial ingoing solution which is parametrized by a free parameter that we can choose to be $\phi_0$. The regular solution has two free parameters ($\phi_0$ and $\phi_n$) and the boundary Green's function is again not unique. The values of $k$ for which  $\cM^{(n)}$  is not invertible are the same as the ones at which the determinant of the matrix vanishes. Pole-skipping at higher Matsubara frequencies can therefore be observed at the special locations
\begin{align}\label{eq:spp}
\omega_n = -2\pi i Tn, \qquad k^2 = k^2_{n}, \qquad \text{det} \, \mathcal{M}^{(n)} (\omega_n, k^2_{n}) = 0.
\end{align}

In summary, at special points in Fourier space \eqref{eq:spp}, imposing the ingoing boundary condition at the horizon is not enough to select a unique solution to the wave equation  and consequently, $G_{\mathcal{OO}}^{R}(\omega, k)$ is infinitely multivalued. As we  show in appendix \ref{app:form}, the  Green's function has a line of poles and a line of zeros that pass through these special points. This is why these locations have  been dubbed 'pole-skipping' points because the poles do not appear as they collide with the zeros \cite{Grozdanov:2017ajz, Blake:2017ris, Blake:2018leo, Grozdanov:2018kkt}. There also exists an interesting phenomenon where we naively observe pole-skipping, but the points are \emph{anomalous}, meaning that in the boundary correlator there are no intersecting lines of zeros and poles. We discuss these in more detail in appendix~\ref{app:form}.

\section{Minimally coupled fermion in the bulk}
\label{sec:fermion}
The aim of this paper is to locate the pole-skipping points for a general fermionic field in an asymptotically anti-de Sitter background. To do so, we must add to the  background  the action of a minimally coupled fermion field given by \cite{Henningson:1998cd, Mueck:1998iz}
\begin{align}
S_f = \int  d^{d+2}x \sqrt{- g}\, i\overline{\psi} \left( \Gamma^M D_M  - m  \right)\psi + S_{bdy}\,,
\end{align}
where $S_{bdy}$ is a boundary term that does not alter the equations of motion, the fermion conjugate is defined as $\psib = \psi^\dagger \Gamma^0$, and the covariant derivative acting on fermions is defined by
\begin{align}
\label{eq:covder}
D_M = \partial_M + \frac14 \left(\omega_{ab}\right)_M\Gamma^{ab}\,.
\end{align}
In what follows we will denote the curved indices  by upper-case Latin letters while flat space indices are denoted by lower-case Latin letters%
\footnote{A further comment on notation. A general flat space tensor  has lower-case Latin letter indices, but  particular values for the indices are underlined, for example $\uuv, \uur$, or $\uux$. This is to distinguish them from curved space indices where a generic tensor has upper-case Latin letters, but a particular value is lower-case letter that is not underlined, for example $u, v$, or $x$. 
}.
The resulting equation of motion for the spinor $\psi$ is then the Dirac equation
\begin{align}
\label{eq:Diraceq}
\left( \Gamma^M D_M - m\right) \psi = 0\,.
\end{align}
Recall that for a theory in $d+2$ spacetime dimensions, the number of components of a spinor is given by
\begin{align}
\label{eq:numofdof}
N = 2^{\lfloor \frac{d+2}{2}\rfloor}\,,
\end{align}
where $\lfloor q \rfloor$ denotes the highest integer that is less than or equal to $q$.
This makes the Dirac equation \eqref{eq:Diraceq} a system of coupled first order differential equations for the $N$  components of the spinor. To fully specify the solution we thus need to impose $N$ boundary conditions.

To calculate the retarded Green's functions for spinors we follow the prescription given by \cite{Iqbal:2009fd}. We first introduce the decomposition of the spinor in terms of the eigenvectors of the matrix $\grr$ defined by
\begin{align}
\label{eq:grrdecomp}
\psi = \psi_+ + \psi_- \,, \qquad  \grr \, \psi_{\pm} = \pm \psi_{\pm}\,,\qquad P_{\pm} \equiv  \frac{1}{2}\left( 1 \pm \grr\right)\,,
\end{align}
where $\psi_\pm$ each contain $N/2$ degrees of freedom.
Assuming that the metric components only depend on the $r$ coordinate, we make the plane wave ansatz $\psi = \psi(r) e^{-i \omega t + i \vec{k} \cdot \vec{x}}$\,  and solve the Dirac equation in Fourier space.
If we want to calculate the retarded Green's function, we need to choose the solution  that is  ingoing at the horizon.  This boundary condition usually reduces the number of free parameters in the solution to $N/2$.
We then evolve the solution to the AdS boundary ($r \rightarrow \infty$), where we find that in general it takes the following form\footnote{Note that the number $d$  in ref. \cite{Iqbal:2009fd} is equal to $d+1$ in our notation.}
\begin{align}
\label{eq:separation}
\psi_+  = A(k) r^{-\frac{d+1}{2}+ m} + B(k) r^{- \frac{d+1}{2}-m-1}\,, \qquad \psi_- = C(k) r^{- \frac{d+1}{2}+ m - 1} + D(k) r^{- \frac{d+1}{2}-m }\,,
\end{align}
with the Dirac equation imposing relations between the pairs $B(k)$, $D(k)$ and $A(k)$, $C(k)$.
For $m \geq 0$ the dominant contribution comes from the term multiplied by $A(k)$, which thus is identified with the source. The response is given by $D(k)$ as it is related to the finite term in the conjugate momentum to the field $\psi_+$ in the appropriate limit.
With this identification the mass $m$ of the  spinor in the bulk  and the conformal dimension $\Delta$ of its corresponding response in the boundary spinor are related via%
\footnote{There are some subtleties involved when the mass is in the ranges $ 0 \leq m \leq \frac12$ or  $m \leq 0$, but conceptually the prescription does not change. In these cases the relation between the mass of the fermion field and the conformal dimension of its dual boundary operator can be different. For more details see \cite{Iqbal:2009fd}.}
\begin{align}
\label{eq:adscftmatch}
\Delta = \frac{d+1}{2}+ m\,.
\end{align}
The prefactors $A(k)$ and $D(k)$ are spinors, and after imposing the ingoing condition one can find that they are related by a matrix $\cR(k)$ as
\begin{align}
\label{eq:dsa}
D(k) = \cR(k) \, A(k)\,.
\end{align}
The retarded Green's function in the boundary theory is given by
\begin{align}
G_R(k) \propto  i \cR(k)
\end{align}

It might be worth stressing how choosing the ingoing solution at the horizon renders the retarded Green's function unique for both scalar and fermion fields.
A scalar field has only one component, but since its dynamics is governed by a second order differential equation, we need two boundary conditions to fully determine the solution. The ingoing condition at the horizon imposes one constraint and thus the solution is effectively determined up to an overall normalization. As the correlator is a ratio between the two leading terms in the asymptotic expansion \eqref{eq:scalarasy}, this overall normalization cancels out and the Green's function is thus uniquely defined.

For a spinor field the procedure is conceptually the same as one can see the matrix $\cR(k)$ as a generalized ratio between two terms in the asymptotic expansion. However, the calculations are more involved.
The ingoing solution at the horizon fixes half of the degrees of freedom. This is usually achieved by transforming the Dirac equations into a second order equation for half of the components, say $\psi_+$, and then taking the ingoing solution. Putting the ingoing solution into the first order Dirac equation fixes the other half of the components, in this case $\psi_-$, in terms of the free parameters left  in $\psi_+$. Therefore the solution is completely determined up to an overall spinor with $N/2$ free parameters that multiplies both $\psi_\pm$.
When the solution is then evolved and expanded near the boundary, both $D(k)$ and $A(k)$ are proportional to this overall spinor, albeit the factor of proportionality can be a matrix in spinor space. This means that $\cR(k)$ does not depend on any free parameters and therefore the retarded Green's function is uniquely defined.

\section{Pole-skipping in asymptotically AdS$_3$ spaces}
\label{sec:AdS3fermionic}
We start with the simplest low-dimensional example, where the bulk theory is three-dimensional and the boundary theory has two spacetime dimensions. In this case both bulk and boundary spinors have two components. We will observe pole-skipping and develop a systematic approach to extract the location of the points in Fourier space  for any three-dimensional background.

\noindent
Let the background metric be given by
\begin{align}\label{eq:3dbackground}
ds^2 = - r^2 f(r) dv^2 + 2dv\,dr +  h(r) dx^2
\end{align}
where for now we leave $f(r)$ and $h(r)$ unspecified, apart from the  properties described in section~\ref{sec:Review}.
Let us choose the following frame
\begin{align}
\label{eq:3dframe}
E^{\uuv} =  \frac{1+ f(r)}{2}\,r dv - \frac{dr}{r}\,,\qquad E^{\uur} =  \frac{1- f(r)}{2}\, r dv + \frac{dr}{r}\,, \qquad E^{\uux} = \sqrt{h(r)} \, dx\,,
\end{align}
for which
\begin{align}
ds^2 = \eta_{ab} E^a\, E^b\,, \hspace{10ex} \eta_{ab} = \text{diag}(-1, 1,1)
\end{align}
We choose this frame firstly because neither the vielbein components nor any of their derivatives diverge at the horizon (assuming $\sqrt{h(r)}$ is regular at $r=r_0$). Secondly, we avoid any square roots of the emblackening factor $f(r)$ in the equations.  Furthermore, this vielbein reduces to a frame for AdS$_{3}$ at the leading order in the near-boundary limit $r \rightarrow \infty$.  In this frame, the spin connections are given by
\begin{align}
\omega_{\uuv \uur} = \frac{dr}{r}- \frac{2 r f(r) + r^2 f'(r)}{2} \,dv,\quad \omega_{\uuv \uux} =  \frac{r\,h'(r)\,\left(1 - f(r)\right)  }{4 \sqrt{h(r)}}\, dx, \quad \omega_{\uur \uux} = - \frac{r \,  h'(r) \, (1+ f(r))}{4 \sqrt{h(r)}}\, dx
\end{align}
with all other components, which are not related by symmetry to the ones above, vanishing. In this frame the Dirac equation is given by
\begin{align}
\label{eq:direqnf0}
\Biggr[ &\left( - \frac{r(1- f(r))}{2}\, \gvv + \frac{r(1+ f(r)) }{2}\, \grr\right) \pd_r + \frac{\grr + \gvv}{r}\pd_v+ \frac{\gxx}{\sqrt{h(r)}}\, \pd_x + \frac{1+ f(r) + r f'(r)}{4}\, \grr \nonumber\\
&- \frac{1- f(r) - r f'(r)}{4}\, \gvv - \frac{ r\, (1-f(r)) h'(r)}{8 h(r)}\, \gvv +  \frac{ r\, (1+f(r)) h'(r)}{8 h(r)}\, \grr- m \Biggr] \psi(r, v, x) = 0
\end{align}
Since the metric is independent of the coordinates $v$ and $x^i$, we introduce the plane wave ansatz $ \psi(r, v, x)  = \psi(r) e^{- i \omega v + i \vec{k} . \vec{x}}$.  Furthermore, we  separate the spinors according to their eigenvalues of the $\grr$ matrix. We define the two independent spinor components  associated with these eigenvalues as
\begin{align}
\label{eq:2dpsipmdef}
\psi = \psi_+ + \psi_- \,, \qquad  \grr \, \psi_{\pm} = \pm \psi_{\pm}\,,\qquad P_{\pm} \equiv  \frac{1}{2}\left( 1 \pm \grr\right)\,.
\end{align}
The spinors $\psi_\pm$ are two component objects, but contain only one independent degree of freedom each.  We insert this decomposition into \eqref{eq:direqnf0} and act on the equation with the projection operators defined in \eqref{eq:2dpsipmdef}.  After some algebra one can write the two resulting  equations as
\begin{subequations}
\label{eq:direqnf3}
\begin{align}
&r^2 f(r)\, \pd_r \psi_+ + \gvv\, \biggr[- i \omega +\frac{ r^2 f'(r)}{4} + \frac{m\, r(1-f(r))}{2}+ \frac{i kr(1+f(r))}{2 \sqrt{h(r)}} \biggr] \psi_-\nonumber\\*
&\quad + \biggr[ - i \omega + \frac{r^2 f'(r)}4 + \frac{r f(r)}{4}\left(2 + \frac{ r\, h'(r)}{h(r)}\right) - \frac{m\, r (1+ f(r))}{2}- \frac{i kr(1- f(r))}{2 \sqrt{h(r)}}\biggr] \psi_+ =0\,,\label{eq:direqnf3a} \\
&r^2 f(r)\, \pd_r \psi_- -  \gvv\, \biggr[- i \omega +\frac{ r^2 f'(r)}{4} - \frac{m\, r(1-f(r))}{2}- \frac{ i kr(1+f(r))}{2 \sqrt{h(r)}} \biggr] \psi_+\nonumber\\*
&\quad + \biggr[ - i \omega + \frac{r^2 f'(r)}4 + \frac{r f(r)}{4}\left(2 + \frac{\, r\, h'(r)}{h(r)}\right) + \frac{m\, r (1+ f(r))}{2}+ \frac{i k r(1- f(r))}{2 \sqrt{h(r)}}\biggr] \psi_- =0\,. \label{eq:direqnf3b}
\end{align}
\end{subequations}
Above  we have used the fact that the set of matrices  $(\mathbb{1}, \gvv, \gxx, \grr)$ forms a complete basis for all  $2\times2$ matrices, hence $\gvx$ can be rewritten as a linear combination of the matrices from the set. In fact, $\gvx = \pm \grr$ and we choose  a representation such that $\gvx = \grr$. For more details on gamma matrices and explicit representations, see appendix \ref{app:gammas}.

It is straightforward to transform \eqref{eq:direqnf3} into two decoupled second order ordinary differential equations for the spinors $\psi_\pm$.
Using these second order differential equations one can look for the leading behavior of the spinors at the horizon. In practice this is achieved by introducing an ansatz
\begin{align}
\psi_{+} \sim (r-r_0)^\alpha \xi_+\,,
\end{align}
where $\xi_+$ is a constant spinor satisfying $\grr\, \xi_+ = \xi_+$, and expanding the second order differential equations around the horizon $r = r_0$. One then finds that the equations are solved at first order for the exponents%
\footnote{In general $\psi_+$ is a two component spinor. However, the second order equations are diagonal, meaning that the linear differential operator acting on the spinor $\psi_+$ is proportional to the identity matrix.}
\begin{align}
\label{eq:nhexp}
\alpha_1 = 0\,,\qquad \alpha_2  =  - \frac12	+ \frac{i \omega}{2 \pi T}\,.
\end{align}
One can repeat the procedure for the $\psi_-$ spinor and obtain the same exponents as in the case of $\psi_+$. Recall that in order to obtain the retarded Green's function, we are supposed to select the ingoing solution at the horizon and evolve the solution towards the boundary. In \ief, this translates to taking the solution with  $\alpha_1$. However, naively both solutions are ingoing if $\omega$ is such  that $\alpha_2$ is a positive integer
which happens at
\begin{align}
\label{eq:highmatsfreq}
\omega = \omega_n \equiv - 2\pi i  T\left(n +\frac12\right)\,,\qquad n = 1, 2, 3, \ldots.
\end{align}
These are precisely the fermionic Matsubara frequencies \eqref{eq:fermmats}, with the exception of the lowest frequency $\omega = \omega_0 \equiv - i \pi T$, which appears to be missing.
Choosing such frequencies is not enough to produce two independent ingoing solutions. Similar to the scalar field,  a more thorough analysis shows that  logarithmic divergences appear in the expansions, making one of the solutions irregular.
If, in addition, we also tune the momentum $k$  to values such that these logarithmic divergences vanish, then there will be two independent ingoing solutions at the horizon. In this case the corresponding Green's function will show pole-skipping, as the ingoing solution and therefore the Green's function is not unique.

\subsection{Pole-skipping at the lowest Matsubara frequency}

In the case of the minimally coupled scalar field, the lowest Matsubara frequency is given by $\omega = 0$. No pole-skipping has been observed at this frequency \cite{Blake:2019otz}.
For the fermionic field, the lowest Matsubara frequency is given by $\omega_0 = - i \pi T$. The exponents \eqref{eq:nhexp} suggest that there is no pole-skipping at this frequency. However, this is not the case as we will soon see.

Pole-skipping at the lowest frequency occurs if there exist two independent ingoing solutions that behave as $(r-r_0)^0$ at the horizon. For the scalar field this actually implies that the two independent solutions are of the form
\begin{align}
\phi = (C + D \log r)\sum_{i=0}^{\infty}\phi_i(k) (r-r_0)^i, \,
\end{align}
with $C$ and $D$ being the free parameters associated with the two independent solutions.  $\phi_i(k)$  are coefficients fixed by the  equation of motion.  Unlike for any other bosonic Matsubara frequency $\omega_n = - 2\pi i T n\,, n \in \mathbb{Z}^+$, we cannot choose any value for $k$  that would give a vanishing prefactor multiplying the logarithmic term. The upshot of this is that for $\alpha = 0$, there is only one solution that is regular at the horizon, and thus no pole-skipping can be observed at this frequency.

The  spinor in  $d \geq 1$  is a multicomponent object which allows for pole-skipping to occur at the lowest Matsubara frequency. Let us introduce a series expansion for both spinor components
\begin{align}
\label{eq:serexp1}
\psi_+ = \sum_{j=0}^\infty \psi_+^{(j)} (r-r_0)^j\,,\qquad \psi_- = \sum_{j=0}^\infty \psi_-^{(j)} (r-r_0)^j, \,
\end{align}
where $\psi_{\pm}^{(j)}$ are constant spinors with definite $\grr$ eigenvalues. We put these expansions into \eqref{eq:direqnf3} and  expand the equations in a series around the horizon as
\begin{align}
\label{eq:serexp2}
\cS_+ = \sum_{j=0}^\infty \cS_+^{(j)} (\omega, k) \,(r-r_0)^j = 0\,,\qquad \cS_- = \sum_{j=0}^\infty \cS_-^{(j)}(\omega, k)\, (r-r_0)^j = 0\,.
\end{align}
In the above definitions, $\cS_+$ and $\cS_-$ are the horizon expansions of the equations \eqref{eq:direqnf3a} and \eqref{eq:direqnf3b} respectively and $\cS^{(j)}_{\pm}$ are series coefficients that can in principle depend on both $\omega$ and $k$. This dependence will be suppressed in the following.
%

We solve the equations \eqref{eq:serexp2} order by order. For the first instance of pole-skipping we only need to look at zeroth order coefficients. These are
\begin{subequations}
\label{eq:direqnc0}
\begin{align}
\cS_+^{(0)} &= \gvv\, \biggr[- i \omega +\frac{ r_0^2 f'(r_0)}{4} + \frac{m\, r_0}{2}+ \frac{i kr_0}{2 \sqrt{h(r_0)}} \biggr] \psi_-^{(0)}+ \biggr[ - i \omega + \frac{r^2_0 f'(r_0)}4  - \frac{m\, r_0 }{2}- \frac{i kr_0}{2 \sqrt{h(r_0)}}\biggr] \psi_+^{(0)} =0\,,\label{eq:direqnc01} \\
\cS_-^{(0)} &=-  \gvv\, \biggr[- i \omega +\frac{ r^2_0 f'(r_0)}{4} - \frac{m\, r_0}{2}- \frac{ i kr_0}{2 \sqrt{h(r_0)}} \biggr] \psi_+^{(0)}+ \biggr[ - i \omega + \frac{r^2_0 f'(r_0)}4  + \frac{m\, r_0 }{2}+ \frac{i k r_0}{2 \sqrt{h(r_0)}}\biggr] \psi_-^{(0)} =0\,.\label{eq:direqnc02}
\end{align}
\end{subequations}
We can immediately notice that
\begin{align*}
\cS_+^{(0)} = \gvv\, \cS_-^{(0)}
\end{align*}
and thus equations \eqref{eq:direqnc0} actually represent only a single constraint. This is not surprising, as the zeroth order should fix  one of the components in terms of the other so that we get a unique ingoing solution, up to an overall constant.
 If \eqref{eq:direqnc01} and \eqref{eq:direqnc02} were two independent equations they would completely fix $\psi_{\pm}^{(0)}$ leaving it with no free parameters.

To locate the pole-skipping points, we need the scalar coefficients multiplying $\psi_{\pm}^{(0)}$ to  vanish.  This happens precisely at
\begin{align}
\label{eq:zeropole}
\omega  = -  \pi i T\,, \qquad k = i m \sqrt{h(r_0)}\, ,
\end{align}
which is precisely the zeroth fermionic Matsubara frequency and the associated momentum. Here we have used the definition of the Hawking temperature \eqref{eq:hawk}.  At such points, equations \eqref{eq:direqnc0} are automatically satisfied and thus  $\psi_{\pm}^{(0)}$ both remain free and independent coefficients.

One can then take a look at the equations at higher orders in \eqref{eq:serexp2}. These relate the expansion coefficients $\psi^{(n)}_\pm$ to $\psi^{(0)}_\pm$, for $n >0$. Using these equations one can iteratively express all of the higher order coefficients as a linear combination of $\psi^{(0)}_\pm$ only.
In this way one can explicitly construct two independent solutions that are regular at the horizon with the leading behavior $(r-r_0)^0$. One of the solutions is parametrized by $\psi^{(0)}_+$ and the other by $\psi^{(0)}_-$. Therefore, at \eqref{eq:zeropole}, the retarded Green's function is not uniquely defined\footnote{Note that the solution parametrized by, for example, $\psi^{(0)}_+$, is \emph{not} a solution with a well defined eigenvalue under $\grr$ everywhere in the bulk. Setting $\psi_-^{(0)} = 0$ does mean that the coefficient multiplying $(r-r_0)^0$ has a positive eigenvalue under $\grr$. However, $\psi_\pm^{(1)}$ are already both non-vanishing. So, the leading component in the expansion has a well defined eigenvalue under $\grr$, but, as soon as we move away from the horizon the two components will start to mix.  The same is true for $\psi_-^{(0)}$.}.

\subsubsection*{Dealing with logarithmic divergences}

Finally, one may ask what happens to the logarithmic terms that one observes in the scalar field expansion at $\omega = 0$. As can be shown, such divergences also appear in the fermion field expansion and are the reason why for generic values of the momentum we do not find two independent ingoing solutions. This highlights the fact that one needs to tune both the frequency and momentum to obtain two non-divergent ingoing solutions, even in the fermionic case.

To see this explicitly, we are interested in the near-horizon solutions to Dirac equations at the frequency $\omega = \omega_0 = - \pi i T$. To leading order, the solutions to the equations take the following form
\begin{subequations}
\begin{align}
\psi_+ &= \psi_+^{(0)}  + \chi_+^{(0)} \log(r-r_0)+ \ldots\,,\\
\psi_- &= \psi_-^{(0)}  + \chi_-^{(0)} \log(r-r_0)+ \ldots\,,
\end{align}
\end{subequations}
with $\psi_\pm^{(0)}$ and $\chi_\pm^{(0)}$ being constant spinors of definite chirality.
We insert the expansion into \eqref{eq:direqnf3} and expand the equations in a series around the horizon. The equations now take the form
\begin{subequations}
\begin{align}
\widehat{\cS}_+ &= \widehat{\cS}_+^{(0)} + \widehat{\cS}_+^{(0l)}\log(r-r_0)+ \ldots = 0\,,\\
\widehat{\cS}_- &= \widehat{\cS}_-^{(0)} + \widehat{\cS}_-^{(0l)}\log(r-r_0)+ \ldots = 0\,,
\end{align}
\end{subequations}
and we solve them iteratively. At leading  order we get the following 4 equations
\begin{subequations}
\label{eq:aa}
\begin{align}
\widehat\cS_+^{(0)} &= - \frac{r_0 }{2 \sqrt{h(r_0)}}\, \left( i k + m \sqrt{h(r_0)}\right) \left( \psi_+^{(0)} - \gvv\, \psi_-^{(0)} \right)+ r_0^2 f'(r_0) \chi_+^{(0)}  = 0 \,,\label{eq:aaa} \\
\widehat  \cS_-^{(0)} &=   \frac{r_0 }{2 \sqrt{h(r_0)}}\, \left( i k + m \sqrt{h(r_0)}\right) \left( \psi_-^{(0)} + \gvv\, \psi_+^{(0)} \right)+ r_0^2 f'(r_0) \chi_-^{(0)}   = 0\,,\label{eq:aab} \\
\widehat\cS_+^{(0l)} &= - \frac{r_0 }{2 \sqrt{h(r_0)}}\, \left( i k + m \sqrt{h(r_0)}\right) \left( \chi_+^{(0)} - \gvv\, \chi_-^{(0)} \right) = 0 \,, \label{eq:aac} \\
\widehat  \cS_-^{(0l)} &=   \frac{r_0 }{2 \sqrt{h(r_0)}}\, \left( i k + m \sqrt{h(r_0)}\right) \left( \chi_-^{(0)} + \gvv\, \chi_+^{(0)} \right) = 0\,.\label{eq:aad}
\end{align}
\end{subequations}
The last two are not independent and are related via
\begin{align}
\widehat\cS_+^{(0l)} = \gvv \, \widehat\cS_-^{(0l)}\,.
\end{align}
In addition to that, inserting $\chi_+^{(0)} = \gvv\, \chi_-^{(0)}$, which is the solution of \eqref{eq:aac},  into  \eqref{eq:aaa} also gives
\begin{align}
\widehat\cS_+^{(0)} = \gvv \, \widehat\cS_-^{(0)}\,.
\end{align}
This means that for a generic value of $k$ there are only two independent equations in \eqref{eq:aa} and  there exist solutions with  $\chi_\pm^{(0)} \neq 0$. We have to set these coefficients to zero if we want a regular solution at the horizon. Thus for a general value of $k$ there is still a unique ingoing solution, as the other solution  contains logarithmic divergences.

If we set $k$ to \eqref{eq:zeropole}, then \eqref{eq:aac} and \eqref{eq:aad} are automatically satisfied. Furthermore, the remaining two equations   \eqref{eq:aaa} and \eqref{eq:aab} are now independent and in fact the first terms in both equations vanish. The solution to  \eqref{eq:aa} is then given  by $\chi_\pm^{(0)} = 0$ with $\psi_\pm^{(0)}$ being undetermined.
Thus we see explicitly that at the location of the pole-skipping point \eqref{eq:zeropole}, the logarithmic terms vanish and the two independent solutions are both regular at the horizon.

\subsection{Pole-skipping at higher Matsubara frequencies}

There are two equivalent ways to locate the pole-skipping points associated with higher Matsubara frequencies.
The first method is similar to the procedure used for the scalar field, as one uses the second order differential equations for half of the components.  This method is useful to determine the positions of the so-called \emph{anomalous} points, which are the locations of coinciding pole-skipping points. See appendix~\ref{app:form} for more details.

The second method is inspired by the lowest frequency pole-skipping point and uses only the first order Dirac equation.
This method completely bypasses the computational difficulties of obtaining a decoupled second order equation, however at the expense of working with higher dimensional systems of algebraic equations.

One can show that both methods yield  the same results and we will show in section~\ref{sec:Examples} that they exactly locate the points of intersection between the lines of poles and the lines of zeros for Green's function in a BTZ black hole background. There we will also illustrate the use of the  procedure using the first order Dirac equation.

Here we present both methods in turn. In both cases, we initially look at the lowest frequency pole-skipping location before generalising the procedure for arbitrary frequencies.

\subsubsection{Using the second-order differential equations}

The first method mimics the procedure of the scalar field reviewed in section~\ref{sec:Review}.  As mentioned above one can use \eqref{eq:direqnf3} to obtain decoupled second order differential equations for the components of one of the spinors. Without loss of generality, we work with $\psi_+$. The first order Dirac equations then completely determine the components of $\psi_-$ in terms of $\psi_+$.

We begin by expanding the second order differential equation of $\psi_+$ around the horizon. This can be schematically  written as
\begin{align}
\label{eq:serexpd0}
\cD_+ =  \sum_{j=0}^\infty \cD_+^{(j)} \,(r-r_0)^j=0\,,
\end{align}
where $\cD_+^{(j)}$ can in principle depend on both $\omega$ and $k$. These terms also depend on the expansion coefficients of $\psi_+$ defined in \eqref{eq:serexp1}.
We solve \eqref{eq:serexpd0} perturbatively by solving $\cD_+^{(j)} = 0$ for all $j$. The leading order equation reads
\begin{align}
\label{eq:serexpd1}
\cD_+^{(0)} = \left(3 \pi T - i \omega\right) \psi_+^{(1)} + \cM_+^{(00)}(\omega, k)\,  \psi_+^{(0)} = 0\,,
\end{align}
where  $\cM_+^{(00)}(\omega, k)$ is a scalar function of $\omega$ and $k$%
\footnote{To be precise $\cM_+^{(00)}$ is proportional to the two-dimensional identity matrix, as is the term multiplying $\psi_+^{(1)}$. However, in what follows, all the coefficients are proportional to the identity matrix. So when  we refer to a coefficient as a scalar function, it should be understood that it is multiplied by an identity matrix.}.
For generic values of $\omega$ and $k$, this equation determines $\psi_+^{(1)}$ in terms of $\psi_+^{(0)}$. And using the higher order equations  one can repeat the procedure and express all $\psi_+^{(j)}$ in terms of  $\psi_+^{(0)}$. In this manner one explicitly constructs an  ingoing solution which is unique up to an overall spinor and whose leading behavior at the horizon is $(r-r_0)^0$.

The above procedure fails if the frequency matches the first fermionic Matsubara frequency given by
\begin{align}
\label{eq:firstmats}
\omega_1 = - 3 \pi i T\,,
\end{align}
as in this case the coefficient of $\psi_+^{(1)}$  in \eqref{eq:serexpd1} vanishes.  For a generic value of $k$ this  sets $\psi_+^{(0)}$ to zero and $\psi_+^{(1)}$ is left undetermined. One can use the latter as the free parameter and again explicitly construct a regular solution that is determined up to an overall factor, $\psi_+^{(1)}$. The leading behavior at the horizon of such a solution is $(r-r_0)$, as the $(r-r_0)^0$ solution includes logarithmic divergences, as discussed in \cite{Blake:2019otz}.

However, if,  in addition to $\omega = \omega_1$, the momentum  $k$ is such that
\begin{align}
\label{eq:firstmatsk}
\cM_+^{(00)}(\omega_1, k) =0\,,
\end{align}
then the equation \eqref{eq:serexpd1} is automatically satisfied and both $\psi_+^{(0)}$ and $\psi_+^{(1)}$ remain unconstrained. Higher order equations are then used to determine all other series coefficients of $\psi_+$ in terms of both $\psi_+^{(0)}$ and $\psi_+^{(1)}$.  We hence construct two distinct regular solutions at the horizon, one with leading behavior $(r-r_0)^0$ and one with  $(r-r_0)$. Consequently, the retarded Green's function is not unique. Note that in general \eqref{eq:firstmatsk} is a third order polynomial in $k$ and  thus one expects three complex solutions for $k$.

One can go further in the series \eqref{eq:serexpd0}. At the linear order in the expansion coefficient one gets
\begin{align}
\label{eq:serexpd2}
\cD_+^{(1)} = \left(5 \pi T - i \omega\right) \psi_+^{(2)} + \cM_+^{(11)}(\omega, k)\,  \psi_+^{(1)}+  \cM_+^{(10)}(\omega, k)\,  \psi_+^{(0)}= 0\,.
\end{align}
At generic values of $\omega$ and $k$, \eqref{eq:serexpd2} combined with \eqref{eq:serexpd1} fix $\psi_+^{(1)}$ and $\psi^{(2)}$ in terms of $\psi_+^{(0)}$ and one can repeat the general procedure of obtaining a unique ingoing solution, as discussed above.

At the second fermionic Matsubara frequency
\begin{align}
\label{eq:secondmats}
\omega_2 = - 5 \pi i T\,,
\end{align}
the coefficient in front of $\psi_+^{(2)}$ vanishes. Then, for generic values of $k$, equations  \eqref{eq:serexpd1} and \eqref{eq:serexpd2} evaluated at $\omega= \omega_2$ set $\psi_+^{(0)}$ and $\psi_+^{(1)}$ to 0. $\psi_+^{(2)}$ is then used as the free parameter in the ingoing solution and the leading behavior at the horizon is $(r-r_0)^2$.
The exception are the values of $k$ for which the determinant of the matrix
\begin{align}
\cM_+^{(2)}(\omega_2, k)\equiv  \begin{pmatrix}
\cM_+^{(00)}(\omega_2, k) & - 2 \pi T\\
\cM_+^{(10)}(\omega_2, k) & \cM_+^{(11)}(\omega_2, k)
\end{pmatrix}
\end{align}
vanishes. Namely, at such points, equations \eqref{eq:serexpd1} and \eqref{eq:serexpd2} are not independent and allow us to express $\psi_+^{(1)}$ in terms of $\psi_+^{(0)}$. Higher order equations from \eqref{eq:serexpd0} then allow us to express all other coefficients of $\psi_+^{(n)}$ in terms of $\psi_+^{(0)}$ and $\psi_+^{(2)}$, meaning that we  again have two independent regular solutions at the horizon.

The procedure for finding the locations of pole-skipping points associated to higher Matsubara frequencies is easily generalized. The equation  \eqref{eq:serexpd0} at order $(n-1)$ is
\begin{align}
\label{eq:serexpdn}
\cD_+^{(n-1)} = \left(  (2n +1) \pi T - i \omega\right) \psi_+^{(n)} + \cM_+^{(n-1, n-1)} \psi_+^{(n-1)} + \ldots + \cM_+^{(n-1, 0)} \psi_+^{(0)} = 0\,.
\end{align}

The pole-skipping point is obtained when the coefficient multiplying $\psi^{(n)}_+$ vanishes and when not all of the equations $\cD^{(j)}_+$ with $j = 0, 1, \ldots n-1$ are independent. This is the case when  $\omega$ and $k$ are such that
\begin{align}
\label{eq:nthmats}
\omega = \omega_n = - 2 \pi i T\left( n + \frac12\right)\,,\qquad \det\cM_+^{(n)}(\omega_n, k) = 0\,,
\end{align}
where
\begin{align}
\label{eq:bigm1}
\hskip -0.5cm
\cM_+^{(n)}(\omega, k) \equiv \begin{pmatrix}
\cM_+^{(00)}(\omega, k) &3 \pi T - i \omega & 0 &\cdots & \cdots  &\cdots & 0 \\
\cM_+^{(10)}(\omega, k) & \cM_+^{(11)}(\omega, k)  & 5 \pi T - i \omega & 0 &\cdots  &\cdots& 0\\
\cM_+^{(20)}(\omega, k) & \cM_+^{(21)}(\omega, k)  & \cM_+^{(20)}(\omega, k)  &7 \pi T - i \omega & 0 &\cdots& 0\\
\vdots & \vdots & \vdots & \vdots & \ddots & \ddots&\vdots\\
& & & & & &  0 \\
\cM^{(n-1, 0)}(\omega, k) & \cdots & \cdots & \cdots & \cdots & \cdots & \cM^{(n-1, n-1)}(\omega, k)
\end{pmatrix}
\end{align}

Then $\psi_+^{(0)}$ and $\psi_+^{(n)}$ are the two independent free parameters that can be used to explicitly construct the two regular solutions at the horizon. As $\det\cM_+^{(n)}(\omega_n, k)$ is in general an $(2n +1)$-degree polynomial, we can expect the same number of complex roots and thus $(2n +1)$ pole-skipping locations associated to the frequency $\omega = \omega_n$.

So far we have not specified the representation for the gamma matrices. Thus  $\psi_+$ and all $\psi_+^{(k)}$ are two-component objects. Therefore, all entries in \eqref{eq:bigm1} are $2\times2$ matrices. However, the second order differential equations for $\psi_+$ are diagonal and consequently  the entries of \eqref{eq:bigm1} are proportional to two-dimensional identity matrices. The determinant \eqref{eq:nthmats} can then be calculated as if the coefficients were scalars.
This is not surprising. If we choose a gamma matrix representation in which the $\grr$ matrix is diagonal, the equations \eqref{eq:direqnf3} reduce to  scalar equations and all the entries in \eqref{eq:bigm1} become scalar functions as well.

\subsubsection{Using the first-order equations}

One can obtain  pole-skipping points at higher frequencies directly from the first order equations \eqref{eq:direqnf3} without having to transform them into second order equations.
Not only does this method provide an alternative to the previously mentioned one, but it is also the direct generalization of the method used to find the pole-skipping point at the lowest Matsubara frequency. Using this method one can thus find all pole-skipping points for the fermionic field.

We previously looked at the series expansion \eqref{eq:serexp2} at zeroth order where we found the first pole-skipping point \eqref{eq:zeropole}.
To obtain the locations with higher frequencies we look at the higher order terms in the expansions of the Dirac equations around the horizon.
We begin by looking at the linear terms. The two equations at this order can be written in a matrix form as
\begin{align}
\label{eq:direqnc1}
\begin{pmatrix}
\cS_+^{(1)} \\ \cS_-^{(1)}
\end{pmatrix} =
\cmt^{(11)}(\omega, k)\,
\begin{pmatrix}
\psi_+^{(1)} \\ \psi_-^{(1)}
\end{pmatrix}
 + \cmt^{(10)}( k)\,
 \begin{pmatrix}
\psi_+^{(0)} \\ \psi_-^{(0)}
\end{pmatrix}
= 0\,,
\end{align}
where $\cmt$ are $2\times2$ matrices whose elements are commuting $2\times2$ matrices%
\footnote{The elements are either proportional to the  identity matrix or  $\gvv$. These two form a set of commuting matrices.}.
 For example
\begin{align}
\cmt^{(11)}(\omega, k) = \begin{pmatrix}
- i\omega- \frac{m r_0}{2} - \frac{i k \, r_0}{\sqrt{h(r_0)}}+ 5 \pi T, & \left(- i\omega+  \frac{m r_0}{2} + \frac{i k \, r_0}{\sqrt{h(r_0)}}+  \pi T\right) \gvv \\
-\left( -i\omega -  \frac{m r_0}{2} -  \frac{i k \, r_0}{\sqrt{h(r_0)}}+  \pi T\right) \gvv \,, & - i\omega+ \frac{m r_0}{2} + \frac{i k \, r_0}{\sqrt{h(r_0)}}+ 5 \pi T
\end{pmatrix}\,,
\end{align}
while  $\cmt^{(10)}$  depends on $k$ but is independent of $\omega$. Its explicit form is not very illuminating, so we do not present it here.
As  $\cS_+^{(1)}$ is not proportional to $\cS_-^{(1)} $, there are   two independent equations at linear order in the series expansion of \eqref{eq:serexp2}.  This is expected, as for generic values of $\omega$ and $k$ these equations fully determine  $\psi_\pm^{(1)}$ in terms of the coefficient  left undetermined in \eqref{eq:direqnc0}. By repeating the procedure at higher orders we explicitly construct a solution that is regular at the horizon and determined up to an overall factor that contains half a spinor's worth of free parameters.

The above procedure fails if the equations \eqref{eq:direqnc1} do not provide two independent constraints on $\psi_\pm^{(1)}$. This is the case if one cannot rearrange \eqref{eq:direqnc1} to express $(\psi_+^{(1)}, \psi_-^{(1)})^T$ in terms of $(\psi_+^{(0)}, \psi_-^{(0)})^T$, in other words, when $\cmt^{(11)}$ is not invertible. Thus we are looking for values of the frequency at which the   determinant of the matrix  multiplying $(\psi_+^{(1)}, \psi_-^{(1)})^T$ vanishes.   One finds that
\begin{align}
\det \cmt^{(11)} = 8 \pi T (3 \pi T - i \omega) \, ,
\end{align}
which vanishes precisely at
\begin{align}
\omega = \omega_1 = - 3 \pi i T,
\end{align}
which is the same as \eqref{eq:firstmats}. At this frequency, there is only one independent equation relating $\psi_\pm^{(1)}$ to $\psi_\pm^{(0)}$. In other words, only a particular linear combination  of $\psi_+^{(1)}$ and $\psi_-^{(1)}$ will be constrained by the values of $\psi_\pm^{(0)}$. In this case the combination $\psi_c^{(1)}$ constrained by the equations is given by
\begin{align}
\psi_c^{(1)} = \psi_+^{(1)} - \gvv\, \psi_-^{(1)}\,.
\end{align}
Combining  \eqref{eq:direqnc0} and \eqref{eq:direqnc1} evaluated at $\omega = \omega_1$ thus yields a system of three independent equations for three variables, which can be schematically written as
\begin{align}
\label{eq:det22}
\begin{pmatrix}
\cS_+^{(0)} \\\cS_+^{(1)} \\ \cS_-^{(1)}
\end{pmatrix}
=
\cmt_{1}(\omega_2, k)\begin{pmatrix}
\psi_+^{(0)} \\
\psi_-^{(0)} \\
\psi_c^{(1)}
\end{pmatrix}\equiv
\begin{pmatrix}
\cmt_{++}^{(00)} & \cmt_{+-}^{(00)} & 0 \\
\cmt_{++}^{(10)} & \cmt_{+-}^{(10)} & \cmt^{(11)}_+\\
\cmt_{-+}^{(10)} & \cmt_{--}^{(10)} & \cmt^{(11)}_-\\
\end{pmatrix}
\begin{pmatrix}
\psi_+^{(0)} \\
\psi_-^{(0)} \\
\psi_c^{(1)}
\end{pmatrix}  =0\,.
\end{align}
The elements of the matrix  $\cmt_1$ are the appropriate coefficients from the equations  \eqref{eq:direqnc0} and \eqref{eq:direqnc1}  evaluated at the first Matsubara frequency. As such they are still commuting matrices, and their $k$ dependence has been suppressed.  Elements $\cmt^{(11)}_\pm$ are given by
\begin{align}
\cmt^{(11)}_+  = - \frac{m r_0}{2} - \frac{i k r_0}{2 \sqrt{h(r_0)}} + \frac{r_0^2 \, f'(r_0)}{2}\,, \quad  \cmt^{(11)}_-  = \left( \frac{m r_0}{2} + \frac{i k r_0}{2 \sqrt{h(r_0)}} + \frac{r_0^2 \, f'(r_0)}{2}\right)\, \gvv \,.
\end{align}

At generic values of $k$, the matrix $\cmt_{1}(\omega_2, k)$ is invertible and thus \eqref{eq:det22} sets all the series coefficients appearing in the equation to zero.
With that we see that at \eqref{eq:firstmats} the leading behavior of the solution at the horizon is $(r-r_0)$. Furthermore $\psi_c^{(1)} = 0$ implies that for such a solution
\begin{align}
\psi_+^{(1)} = \gvv \, \psi_-^{(1)}\,,
\end{align}
and thus we again obtain a unique ingoing solution that has half a spinor's worth of free parameters. With that, one can take, for example $\psi_+^{(1)}$ as the free parameter and then use the higher order equations to determine other coefficients and thus perturbatively construct a regular solution.

One obtains two independent regular solutions if the matrix $\cmt_1$ is not invertible. In that case not all three equations in \eqref{eq:det22} are independent. One obtains another free parameter, for example $\psi_+^{(0)}$, in addition to $\psi_+^{(1)}$.
The values at which we get two independent ingoing solutions are the values of $k$ for which
\begin{align}
\label{eq:det1lin}
\det \cmt_1(\omega_1,  k) = 0\,.
\end{align}
As a consistency check, the above equation yields the same roots as \eqref{eq:firstmatsk}. The determinant is a cubic function of $k$ so we expect three complex roots. These are the pole-skipping points associated with the frequency \eqref{eq:firstmats}.

As a side note, to find the locations of the pole-skipping points, in practice it is easier to simply set one of the $\psi_{\pm}^{(1)}$ to zero and treat the other variable as $ \psi_c^{(1)}$.  One finds that the roots of the equation \eqref{eq:det1lin} are independent of the choice of which variable we set to 0.

Pole-skipping points associated to higher Matsubara frequencies are located in a similar manner. We take the equations at order $n$ in the expansion \eqref{eq:serexp2} and write them schematically  as
\begin{align}
\label{eq:direqncn}
\begin{pmatrix}
\cS_+^{(n)} \\ \cS_-^{(n)}
\end{pmatrix} =
\cmt^{(nn)}(\omega, k)\,
\begin{pmatrix}
\psi_+^{(n)} \\ \psi_-^{(n)}
\end{pmatrix}
+ \ldots
 + \cmt^{(n0)}(k)\,
 \begin{pmatrix}
\psi_+^{(0)} \\ \psi_-^{(0)}
\end{pmatrix}
= 0\,,
\end{align}
with all $\cmt^{(jk)}$ being matrices whose elements are commuting matrices. Only the leading coefficient $\cmt^{(nn)}$ depends on both the frequency and momentum, while the remaining coefficients depend only on the momentum.

To get pole-skipping at $\omega = \omega_n$ we require that the equations \eqref{eq:direqncn} provide only one independent constraint for $\psi_\pm^{(n)}$, which translates to demanding that
\begin{align}
\det \cmt^{(nn)}(\omega, k) = 0\,.
\end{align}
One finds that for any $n$, the matrix $\cmt^{(nn)}$ has the form
\begin{align}
\label{eq:mnn}
\cmt^{(nn)} = \begin{pmatrix}
- i\omega- \frac{m r_0}{2} - \frac{i k \, r_0}{\sqrt{h{r_0}}}+ (4n +1) \pi T, & \left(- i\omega+  \frac{m r_0}{2} + \frac{i k \, r_0}{\sqrt{h{r_0}}}+  \pi T\right) \gvv \\
-\left( -i\omega -  \frac{m r_0}{2} -  \frac{i k \, r_0}{\sqrt{h{r_0}}}+  \pi T\right) \gvv \,, & - i\omega+ \frac{m r_0}{2} + \frac{i k \, r_0}{\sqrt{h{r_0}}}+ (4n +1) \pi T
\end{pmatrix}\,,
\end{align}
whose determinant is given by
\begin{align}
\label{eq:detmnn}
\det \cmt^{(nn)} = 8 \pi n T \left(\pi T(2n + 1) - i \omega\right)\,.
\end{align}
This vanishes at the fermionic Matsubara frequencies given by
\begin{align}
\label{eq:nmats}
\omega = \omega_n = - 2 \pi i T \left(n + \frac12\right)\,.
\end{align}
The corresponding momenta at which pole-skipping occurs are then found by constructing the analogue of the equation \eqref{eq:det22}.
We start by evaluating \eqref{eq:direqncn} at \eqref{eq:nmats}. Again, only a particular linear combination of the $n$-th order coefficients is constrained by the equations and is given by
\begin{align}
\label{eq:lincomb}
\psi_c^{(n)} = \psi_+^{(n)} - \gvv \, \psi_-^{(n)}\,.
\end{align}
One then combines all the equations at lower orders and evaluates them at the Matsubara frequency. These can be written in a schematic form as
\begin{align}
\label{eq:mateq1}
\begin{pmatrix}
\cS_+^{(0)} \\\cS_+^{(1)}  \\ \vdots \\  \cS_-^{(n)}
\end{pmatrix} = \cmt_n \begin{pmatrix}
\psi_+^{(0)} \\
\psi_-^{(0)} \\
\vdots\\
\psi_c^{(n)}
\end{pmatrix}
\equiv \begin{pmatrix}
\cmt_{++}^{(00)} & \cmt_{+-}^{(00)} &  0 & \cdots & \cdots & \cdots &0 \\
\cmt_{++}^{(10)} & \cmt_{+-}^{(10)} & \cmt^{(11)}_{++} &\cmt^{(11)}_{+-}& 0& \cdots & 0  \\
\vdots & \vdots & \vdots & \vdots & \vdots & \vdots & \vdots &\\
\cmt_{-+}^{(n0)} & \cmt_{--}^{(n0)} & \cdots & \cdots &\cdots & \cdots &  \cmt^{(nn)}_- \\
\end{pmatrix}
\begin{pmatrix}
\psi_+^{(0)} \\
\psi_-^{(0)} \\
\vdots\\
\psi_c^{(n)}
\end{pmatrix} =0\,,
\end{align}
with
\begin{align}
\cmt^{(nn)}_+  = - \frac{m r_0}{2} - \frac{i k r_0}{2 \sqrt{h(r_0)}} + \frac{n r_0^2 \, f'(r_0)}{2}\,, \quad  \cmt^{(11)}_-  = \left( \frac{m r_0}{2} + \frac{i k r_0}{2 \sqrt{h(r_0)}} + \frac{n r_0^2 \, f'(r_0)}{2}\right)\, \gvv\,.
\end{align}
Pole-skipping occurs when we have two independent regular solutions. This happens precisely when the matrix $\cmt_n$ is not invertible. Hence the locations of such points are given by the values of $k$ for which
 \begin{align}
 \det \cmt_n(\omega_n, k) = 0\,.
\end{align}
 We note that each $\cmt$ entry is a linear function in $k$ and
\eqref{eq:mateq1} is a system of $2n + 1$ equations, meaning that the determinant is an order $2n+1$ polynomial in $k$ and  has that many complex roots and thus for each frequency $\omega_n$, we find $2n + 1$ pole-skipping points.

\section{Pole-skipping in higher dimensions}
\label{sec:generalpoleskipping}

We now  generalize the procedure presented in the previous section  to higher dimensional spacetimes. We  work in $d+2$ bulk spacetime dimensions, which means that the boundary theory is formulated in $d+1$ dimensions.
Thus, the bulk spinor has $N = 2^{\lfloor \frac{d+2}{2}\rfloor}$ degrees of freedom and the boundary spinor has half as many.

We find that the equations split up into two decoupled subsystems both of which are related to the lower-dimensional case presented in the previous section. We also find that for generic values of $k$ the number of pole-skipping points at $\omega = \omega_n$ is doubled to $2(2n+1)$ and that the locations are in general different for the two different subsystems.

We work with the background metric in ingoing Eddington-Finkelstein coordinates \eqref{eq:backgroundmetric}. The orthonormal frame is taken to be
\begin{align}
\label{eq:general_frame}
E^{\uuv} =  \frac{1+ f(r)}{2}\,r dv - \frac{dr}{r}\,,\qquad E^{\uur} =  \frac{1- f(r)}{2}\, r dv + \frac{dr}{r}\,, \qquad E^{\uui} = \sqrt{h(r)} \, dx^{i}\,,
\end{align}
so that
\begin{align}
ds^2 = \eta_{ab} E^a\, E^b\,, \hspace{10ex} \eta_{ab} = \text{diag}(-1, 1,1, \ldots,1)\,.
\end{align}
This frame is the direct generalization of the frame \eqref{eq:3dframe} and shares all of its special properties.
The spin connections are given by
\begin{align}
\omega_{\uuv \uur} = \frac{dr}{r}- \frac{2 r f(r) + r^2 f'(r)}{2} \,dv,\quad \omega_{\uuv \uui} =  \frac{r\,h'(r)\,\left(1 - f(r)\right)  }{4 \sqrt{h(r)}}\, dx^i, \quad \omega_{\uur \uui} = - \frac{r \,  h'(r) \, (1+ f(r))}{4 \sqrt{h(r)}}\, dx^i \,,
\end{align}
with all other components not related by symmetry to the ones above being 0.

The calculation of the Dirac equation is conceptually the same as in section \ref{sec:AdS3fermionic}, so we don't repeat it in full.
We exploit again the fact that the metric does not depend on $v$ and $x^i$ and solve the equation in Fourier space by introducing
$ \psi(r, v, x^j)  = \psi(r) e^{- i \omega v + i k_i  x^i}$.
The Dirac equation then reads
\begin{align}
\label{eq:direqgf1}
\Biggr\{ &\Gamma^{\uuv}  \biggr[ - \frac{r(1-f(r))}{2}\, \pd_r - \frac{i \omega }{r}- \frac{1- f(r) - r f'(r)}{4} - \frac{d\, r\, (1- f(r)) h'(r)}{8 h(r)}\biggr]  \nonumber \\*
& \grr \biggr[ \frac{r (1+ f(r))}{2}\, \pd_r - \frac{i \omega}{r}+ \frac{1 + f(r) + r f'(r)}{4}+ \frac{d\, r\, (1+ f(r))h'(r)}{8 h(r)}\biggr] + \frac{i k_i \gii}{\sqrt{h(r)}}-m\Biggr\} \psi(r) = 0.
\end{align}
In general, this is a system of $N$ first order coupled ordinary differential equations for the components of the spinor. We  want to decouple them in a way that makes the pole-skipping mechanism manifest.
We begin by introducing the decomposition
\begin{align}
\label{eq:psipmdef}
\psi = \psi_+ + \psi_- \,, \qquad  \grr \, \psi_{\pm} = \pm \psi_{\pm}\,,\qquad P_{\pm} \equiv  \frac{1}{2}\left( 1 \pm \grr\right)\,,
\end{align}
where each component $\psi_{\pm}$ contains $N/2$ free parameters.
Furthermore, notice that for $ d \geq 2$ the two matrices $\grr$ and $k_i \gvi$ are independent and commuting\footnote{For $ d= 1$ which is the asymptotically AdS$_3$ case, we have $\gvi = \pm \grr$, as $(\mathbb{1}, \gvv, \gii, \grr)$ provide a complete basis for any $2\times2$ matrix.}.  Therefore, we can introduce an additional decomposition
\begin{align}
\label{eq:psipm2def}
\psi_{a} = \psi_a^{(+)} + \psi_{a}^{(-)}\,, \qquad \kh \gvi\, \psi_{a}^{(\pm)} = \pm \psi_{a}^{(\pm)}\,, \qquad P^{(\pm)} \equiv \frac12\left( 1 \pm \kh \gvi\right)\,,
\end{align}
where $a = \pm$, and we have used
\begin{align*}
\kh \gvi \equiv \frac{k_i}{k} \, \gvi\,, \qquad k = \sqrt{\sum_{i=1}^d k_i k_i}\,.
\end{align*}

 We now have divided the initial spinor $\psi$ that with $N$ degrees of freedom into four independent spinors $\psi_{\pm}^{(\pm)}$ which each contain $N/4$ independent degrees of freedom. Each $\psi_{\pm}^{(\pm)}$  has a set of definite eigenvalues under the action of  $\grr$ and $k_i \gvi$.
The $\grr$ matrix projects the spinor components along the radial direction and can be considered as the chirality projection, especially with respect to the boundary theory. We thus refer to $\psi_\pm$ as positive or negative chirality spinors.
Similarly $k_i \gvi$ can be considered as a projection of the components of the spinor along the direction of the momentum. In that way, it has a similar effect as a helicity projection. We  thus refer to spinors $\psi^{(\pm)}$ as positive (negative) helicity spinors.
As an example $\psi_+^{(-)}$ is a spinor with positive chirality but negative helicity.

Using this  decomposition into 4 independent components in the Dirac equation, one notices that the equations separate into two decoupled subsystems. One for $(\psi_+^{(+)}, \psi_-^{(-)})$ and
the other for $(\psi_+^{(-)}, \psi_-^{(+)})$. The equations for the first system can be written as
\begin{subequations}
\label{eq:direqgf4}
\begin{align}
\label{eq:direqgf41}
&r^2 f(r)\, \pd_r \psi_+^{(+)} + \gvv\, \biggr[- i \omega +\frac{ r^2 f'(r)}{4} + \frac{m\, r(1-f(r))}{2}+ \frac{ ik r(1+f(r))}{2 \sqrt{h(r)}}\biggr] \psi_-^{(-)}\nonumber\\*
&\quad + \biggr[ - i \omega + \frac{r^2 f'(r)}4 + \frac{r f(r)}{4}\left(2 + \frac{d\, r\, h'(r)}{h(r)}\right) - \frac{m\, r (1+ f(r))}{2}- \frac{ik r(1- f(r))}{2 \sqrt{h(r)}}\, \biggr] \psi_+^{(+)} =0\,,\\*
\label{eq:direqgf42}
&r^2 f(r)\, \pd_r \psi_-^{(-)} - \gvv\, \biggr[- i \omega +\frac{ r^2 f'(r)}{4} - \frac{m\, r(1-f(r))}{2}- \frac{ ik r(1+f(r))}{2 \sqrt{h(r)}}\biggr] \psi_+^{(+)}\nonumber\\*
&\quad + \biggr[ - i \omega + \frac{r^2 f'(r)}4 + \frac{r f(r)}{4}\left(2 + \frac{d\, r\, h'(r)}{h(r)}\right) + \frac{m\, r (1+ f(r))}{2}+ \frac{ik r(1- f(r))}{2 \sqrt{h(r)}}\, \biggr] \psi_-^{(-)} =0\,.
\end{align}
\end{subequations}
The equations for $(\psi_+^{(-)}, \psi_-^{(+)})$ are equivalent%
\footnote{The detailed derivation of these equations together with the explicit equations for $(\psi_+^{(+)}, \psi_-^{(-)})$ can be found in appendix~\ref{app:details}.},
but with $k \rightarrow -k$ and therefore we  focus only on the pair $(\psi_+^{(+)}, \psi_-^{(-)})$.

The equations \eqref{eq:direqgf4} are essentially the same  as \eqref{eq:direqnf3}. The only differences are the additional factor of $d$ in one of the terms of the equations, and that in higher dimensions the spinors $\psi_{\pm}^{(\pm)}$ are $N$-dimensional objects rather than 2-dimensional.
With that observation, most of what follows repeats itself from section  \ref{sec:AdS3fermionic}.

First, one can eliminate one of the spinors from \eqref{eq:direqgf4} to obtain a diagonal second order differential equation for the other.
One can expand the second order equations around the horizon region. Using the ansatz
\begin{align}
\psi_{+}^{(+)} \sim  (r-r_0)^\alpha\, \xi_+^{(+)}\,,
\end{align}
where $\xi_+^{(+)}$ is a constant spinor with definite chirality and helicity,
one finds that the second order equations are solved at leading order by the following exponents
\begin{align}
\label{eq:nhexpgen }
\alpha_1 = 0\,,\qquad \alpha_2  =  - \frac12	+ \frac{i \omega}{2 \pi T}\,,
\end{align}
with the same behavior being observed for $\psi_-^{(-)}$. We recall that the same exponents have been found in lower dimensional case \eqref{eq:nhexp}.
The exponent $\alpha_1$ is for generic values of the frequency associated with the ingoing solution. The exception are the cases where the frequency is such that the second exponent is equal to a positive integer. This is where we expect pole-skipping.
These special values for the frequency are again the fermionic Matsubara frequencies
\begin{align}
\omega_n = - 2\pi i  T\left(n +\frac12\right)\,,\qquad n = 1, 2, 3, \ldots\,,
\end{align}
which again does not include the zeroth Matsubara frequency $\omega_0 = - i \pi T$.

For pole-skipping to  occur  the momentum also needs to be set to special values. In the following, we briefly discuss how to locate the pole-skipping points in general dimensions. Since the equations governing the spinors are essentially the same as in the three-dimensional case, the procedure of finding the locations is also the same. In order  not to repeat too much of section \ref{sec:AdS3fermionic}, we  only outline the procedure and focus mainly on how to obtain the locations of the pole-skipping points. 
\subsection{Pole-skipping at the lowest Matsubara frequency}

We begin by expanding the spinors in a series around the horizon as
\begin{align}
\label{eq:serexp3}
\psi_+^{(+)}  = \sum_{j=0}^\infty \ppp{j} (r-r_0)^j\,,\qquad \psi_-^{(-)} = \sum_{j=0}^\infty \pmm{j} (r-r_0)^j\,.
\end{align}
We insert these expressions into the Dirac equations \eqref{eq:direqgf4} and  expand them around the horizon. Schematically these expansions are written as
\begin{align}
\label{eq:serexp4}
\cS_+^{(+)} = \sum_{j=0}^\infty \left(\cS_+^{(+)}\right)^{(j)}\, (r-r_0)^j \,, \qquad \cS_-^{(-)} = \sum_{j=0}^\infty \left(\cS_-^{(-)}\right)^{(j)}\, (r-r_0)^j\,.
\end{align}

In order to see pole-skipping  at the lowest frequency, we look at the zeroth order equations in \eqref{eq:serexp4} and find that there is only one independent equation at this order. It can be written as
\begin{align}
\label{eq:direqgc0}
\left(\cS_+^{(+)}\right)^{(0)} &= \gvv\, \biggr[- i \omega +\frac{ r_0^2 f'(r_0)}{4} + \frac{m\, r_0}{2}+ \frac{i kr_0}{2 \sqrt{h(r_0)}} \biggr] \pmm{0}\nonumber\\*
& \hspace{20ex}+ \biggr[ - i \omega + \frac{r^2_0 f'(r_0)}4  - \frac{m\, r_0 }{2}- \frac{i kr_0}{2 \sqrt{h(r_0)}}\biggr] \ppp{0} =0\,.
\end{align}
Notice that this equation is  equivalent to \eqref{eq:direqnc01}. Pole-skipping occurs if the coefficients multiplying $\ppp{0}$ and $\pmm{0}$ both vanish in which case the equation \eqref{eq:direqgc0} is automatically satisfied. This happens at
\begin{align}
\label{eq:zeromatspp}
\omega = \omega_0 = - \pi i T\,, \qquad k = i m \sqrt{h(r_0)}\,.
\end{align}
After repeating the same procedure for the components $\psi_+^{(-)}$ and $\psi_-^{(+)}$, one finds that pole-skipping occurs at
\begin{align}
\label{eq:zeromatspm}
\omega = \omega_0 = - \pi i T\,, \qquad k = -i m \sqrt{h(r_0)}\,.
\end{align}
Hence, combining these two results, one sees that for $d \geq 2$, the first occurrence of pole-skipping is at
\begin{align}
\label{eq:zeromatsg}
\omega = \omega_0 = - \pi i T\,, \qquad k = \pm i  m \sqrt{h(r_0)}\,.
\end{align}

Comparing the above results to those in asymptotically AdS$_3$ spaces we see that in higher dimensions there exists an additional pole-skipping point with negative imaginary momentum.
This feature  repeats itself with all other pole-skipping points. Fermions in higher dimensions have twice as many pole-skipping points than fermions in 3-dimensional spacetimes.
These additional locations appear due to the interaction between fermions whose chiralities are opposite to their helicities. In two dimensions such fermions are absent which explains why we observe only half as many pole-skipping points as in the general case.

Finally, each of $\left(\psi_a^{(b)}\right)^{(0)}$, with $a, b  =  \pm$ contains $N/4$ degrees of freedom. At a generic point in the $(\omega, k)$ space, equation \eqref{eq:direqgc0} (or its $\left(\cS_+^{(-)}\right)^{(0)}$ counterpart)  provides $N/2$ constraint equations, thus reducing the number of free parameters to $N/2$, which is enough to uniquely determine the boundary correlation function.
At a pole-skipping point, say, \eqref{eq:zeromatspp}, the equation \eqref{eq:direqgc0} is automatically satisfied, hence imposing no relation between $\ppp{0}$ and $\pmm{0}$. On the other hand, the  equation relating the  coefficients  $\ppm{0}$ and $\pmp{0}$ does not hold automatically at this pole-skipping point. The pole-skipping point associated to this subsystem has the opposite value of $k$, meaning that at zeroth order we will get a constraint equation for these two coefficients.
This means that although we are at a pole-skipping point, the Dirac equations still provide some constraints on the spinor, and the ingoing solution will thus have only $\tfrac{3N}{4}$ free parameters.
There is, however, a notable exception to this rule -- the case of the massless fermion. Taking $m =0$ in \eqref{eq:zeromatsg}, we notice that the two pole-skipping points merge into one, located at $\omega = \omega_0$ and $k =0$.  At this point in momentum space, the ingoing condition does not impose any constraints on the spinors.
This is unlike the scalar case,  where at any pole-skipping point the ingoing condition does not impose any constraints on the field regardless of the mass of the field.

\subsection{Higher order pole-skipping}

To get the higher frequency pole-skipping points  one can  either use the first order  or the diagonal second order differential equations, as they give the same locations.
The equations in higher dimensions \eqref{eq:direqgf4} have the same form as the ones in three spacetime dimensions \eqref{eq:direqnf3}.  Therefore both methods readily generalize to higher dimensions. 

Due to this similarity, we do not repeat the methods here. We only mention some of the differences.
The first difference is that in higher dimensions, spinors have $N$ components and separate into spinors containing $N/4$ degrees of freedom $\psi_\pm^{(\pm)}$. With that in mind, all the factors multiplying $\psi_\pm^{(\pm)}$ in the expansions around the horizon are $N\times N$ dimensional matrices. Thus the analogues of \eqref{eq:bigm1}, \eqref{eq:mnn} and \eqref{eq:mateq1} will be matrices whose elements are (commuting) $N\times N$ dimensional matrices.

The second difference is that the equations split into two independent subsystems which in general yield two independent sets of pole-skipping points. However, it is enough to find the locations for one of the subsystems as the pole-skipping points of the other are obtained by $k\leftrightarrow -k$, meaning that at the same frequency the two subsystems have opposite pole-skipping points.

Doing the explicit calculations, we find that all pole-skipping points occur at the fermionic Matsubara frequencies \eqref{eq:fermmats} regardless of the dimension of spacetime. At each frequency $\omega = \omega_n$ we get, for generic values of the mass, $2(2n+1)$ pole-skipping points.

As in the case with the lowest Matsubara frequency, at a generic pole-skipping point, only $N/4$ components of the spinor are constrained by the equations. Again this is related to the two subsystems experiencing  pole-skipping at different locations in momentum space.
However, explicitly working out the locations for the first few pole-skipping points, one again notices that the massless fermion is an exception.
In that case, one finds that the pole-skipping points associated to the $n$-th Matsubara frequency for one of the subsystems are given schematically at $\omega = \omega_n$ and $k =\lbrace 0, \pm k_1, \pm k_2\, \ldots, \pm k_n\rbrace$.
As we can see, this set of pole-skipping points is invariant under the reversal of the momentum and therefore the other subsystem will experience pole-skipping at the exact same locations in momentum space.
Thus, the number of pole-skipping points is halved to $(2n+1)$, yet, at each pole-skipping point we are left with an entire spinor's worth of free parameters.

While we are currently lacking a proof that this pattern continues for arbitrary $n$, we find no reason why this feature would cease to hold after the first few pole-skipping points, for which this was checked explicitly.
%

\section{Examples}
\label{sec:Examples}

The methods presented in the previous sections might be a bit abstract and an alert reader will realize that we have restrained from calculating any second order differential equations or determinants of matrices like \eqref{eq:det1lin}. While the methods are straightforward, the expressions quickly  become rather long.
To illuminate the procedure and show that our analysis matches the known results, we will consider some concrete examples.

First, we consider the case of a minimally coupled fermion on a non-rotating BTZ black hole background \cite{Banados:1992gq,Banados:1992wn}. In this case the fermionic Green's function is known explicitly  and we use it to verify our results.
We then consider the special case of a fermion with half-integer mass (or equivalently a dual operator with a half-integer conformal dimension). In that case, the correlation function takes a special form and we show that the near-horizon analysis still agrees with the exact result.
Finally, we briefly present the case of a massless fermion propagating in the planar Schwarzschild black hole in anti-de Sitter spacetime in general dimension and show that the locations of the pole-skipping points pair up, as discussed in the previous section.
\subsection{BTZ black hole}

\subsubsection{Pole-skipping points}

For the non-spinning BTZ black hole,  two functions that appear in the metric \eqref{eq:3dbackground} are given by
\begin{align}
f(r) =1 - \left(\frac{r_0 }{r}\right)^2\,,\qquad \text{and} \qquad  h(r) = r^2\,,
\end{align}
which implies that the Hawking temperature is given by $T = r_0/2\pi$. We  use the following set of gamma matrices
\begin{align}
(\Gamma^{\uuv}, \Gamma^{\uur}, \Gamma^{\uux})= (i \sigma^2, \sigma^3, \sigma^1)\,,
\end{align}
where $\sigma^i$ are the  Pauli matrices. In this case $\grr$ is diagonal and thus the two Weyl fermions are given by
\begin{align}
 \psi_+ =  \psi_+(r)\begin{pmatrix}
 1\\0
 \end{pmatrix}\,, \qquad
 \psi_- (r) =  \psi_-(r)\begin{pmatrix}
 0 \\ 1
 \end{pmatrix}\,,
\end{align}
where $\psi_\pm(r)$ are scalar functions. Because of the favorable choice of gamma matrices, the Dirac equation can be reduced to two coupled scalar differential equations that read
\begin{subequations}
\label{eq:direqBTZ}
\begin{align}
&\hspace*{-5ex}(r^2 - r_0^2) \pd_r\psi_+ + \left[ - i \omega - (m -1) r + \frac{\left((m-1)r- i k \right) r_0^2}{2r^2}\right]\, \psi_+  + \left[ik - i \omega + \frac{(- i k + (m+1)r) r_0^2}{2r^2}\right]\, \psi_- = 0\\
&\hspace*{-5ex}(r^2 - r_0^2) \pd_r\psi_- + \left[ - i \omega + (m +1) r - \frac{\left((m+1)r- i k \right) r_0^2}{2r^2}\right]\, \psi_- + \left[- i k - i \omega + \frac{(ik - (m-1)r)r_0^2}{2r^2}\right]\, \psi_+ = 0\,.
\end{align}
\end{subequations}

To get the first pole-skipping point,  we then follow the  procedure given in section~\ref{sec:AdS3fermionic}.
The point in momentum space where the first pole-skipping occurs is
\begin{align}
\label{eq:firstmatsBTZ}
\omega = - \frac{i r_0}{2}= - i \pi T\,,\qquad k = i m r_0 = 2 \pi i m T\,,
\end{align}
which agrees with the general result \eqref{eq:firstmats} and \eqref{eq:firstmatsk}.
Since $\grr$ is diagonal, we note that at \eqref{eq:firstmatsBTZ}, the two independent solutions that are regular at the horizon can be written as
\begin{align}
\psi(r) = \begin{pmatrix}
\chi_+ \\ \chi_-
\end{pmatrix}\, \left( 1 + \chib_1 (r-r_0) + \chib_2 (r- r_0)^2+ \ldots\right)\,,
\end{align}
where $\chi_\pm$ are the two free parameters and $\chib_n$ are two dimensional spinors whose components are fully determined in terms of $\chi_\pm$. In general, $\chib_n$ are not eigenstates of the chirality matrix.

We then use either of the two procedures presented in section~\ref{sec:AdS3fermionic} to find the locations of other pole-skipping points.
Here, we will explicitly calculate the locations of the pole-skipping points associated with the next lowest frequency using the first order differential equations, while merely stating the locations of the higher frequency pole-skiping points.

The equations \eqref{eq:direqnc1} for this example read
\begin{align}
\label{eq:expBTZ2}
M^{(11)}
\begin{pmatrix}
\psi_+^{(1)}  \\ \psi_-^{(1)}
\end{pmatrix}
+
M^{(10)}
\begin{pmatrix}
\psi_+^{(0)}  \\ \psi_-^{(0)}
\end{pmatrix}
=0\,,
\end{align}
with the two matrices being
\begin{align*}
M^{(11)} &= \begin{pmatrix}
- i k - (m-5)r_0 - 2i \omega, & i k +(m+1)r_0 - 2 i \omega \\
- i k - (m-1) r_0 - 2 i \omega, & i k + (m+5)r_0 - 2 i \omega
\end{pmatrix}\,,\\
M^{(10)} &= \frac{1}{r_0}\begin{pmatrix}
-3(m-1) r_0 + 2 i k, & -(m+1)r_0 + 2 i k \\
(m-1) r_0 - 2 i k, & 3(m+1) r_0 - 2 i k
\end{pmatrix}\,.
\end{align*}
All of the elements of the matrices are in this case scalar functions.
The frequency of the next pole-skipping point is given by the value at which the determinant of the coefficient in front of $(\psi_+^{(1)}, \psi_-^{(1)})^T$ vanishes. One finds that
\begin{align}
\det M^{(11)} = 8 r_0 ( 3 r_0 - 2i \omega)\,,
\end{align}
which vanishes at
\begin{align}
\omega = \omega_1 \equiv  -  \frac{3 i r_0}{2} = - 3 \pi i T\,.
\end{align}
The easiest way to obtain the corresponding momenta is to set one of $\psi_\pm^{(1)}$ to 0 and combine  \eqref{eq:expBTZ2} with the zeroth order equation%
\footnote{The zeroth order equation reads $(-i k + r_0 - m r_0 - 2 i \omega) \psi_+^{(0)} + (i k + r_0 + m r_0 - 2 i \omega) \psi_-^{(0)} = 0$}
(both evaluated at $\omega = \omega_1$) to obtain a system of three equations for three variables. For example, setting $\psi_-^{(1)} = 0$  gives
\begin{align*}
\label{eq:blah1}
\begin{pmatrix}
-(m+2) r_0-i k & (m-2) {r_0}+i k & 0 \\
 \frac{2 i k}{{r_0}}-3 m+3 & \frac{2 i k}{{r_0}}-m-1 & -(m-2) {r_0}-i k \\
 -\frac{2 i k}{{r_0}}+m-1 & -\frac{2 i k}{{r_0}}+3 m+3 & -(m+2) {r_0}-i k \\
\end{pmatrix}\,
\begin{pmatrix}
\psi_+^{(0)} \\ \psi_-^{(0)} \\ \psi_+^{(1)}
\end{pmatrix}
=0\,.
\end{align*}
The momentum values for the pole-skipping points are obtained by looking for the values at which the determinant of the matrix vanishes.  One finds that
\begin{align}
&\det\begin{pmatrix}
-(m+2) r_0-i k & (m-2) {r_0}+i k & 0 \\
 \frac{2 i k}{{r_0}}-3 m+3 & \frac{2 i k}{{r_0}}-m-1 & -(m-2) {r_0}-i k \\
 -\frac{2 i k}{{r_0}}+m-1 & -\frac{2 i k}{{r_0}}+3 m+3 & -(m+2) {r_0}-i k \\
\end{pmatrix}\nonumber \\
&\hspace{20 ex}= -\frac{8i}{r_0}( k - i (m-1)r_0) (k + i m r_0) ( k - (m+1)i r_0)\,,
\end{align}
which vanishes at
\begin{align}
k = - i m r_0 = - 2 \pi i T\,,\qquad k = i (m\pm 1) r_0 = 2 \pi i T (m \pm 1)\,.
\end{align}
If we set $\psi_+^{(1)}$ to 0  and include $\psi_-^{(1)}$ in the matrix \eqref{eq:blah1}, the determinant switches sign. This is because the coefficients multiplying $\psi_\pm^{(1)}$ in \eqref{eq:expBTZ2}, evaluated at $\omega = \omega_1$, only differ by a sign.  This obviously does not change values of the momenta.
The fact that we can simply set one of $\psi_\pm^{(1)}$ to 0 and calculate the pole-skipping points in the above way  is a consequence of the fact that at any fermionic Matsubara frequency, only the combination \eqref{eq:lincomb} is constrained, which in our case is simply $\psi_+^{(1)} - \psi_-^{(1)}$.

Finding the locations of other pole-skipping points  follows the same pattern. We find that first few pole-skipping points are then located at
\begin{subequations}
\label{eq:pspointsBTZ}
\begin{align}
&\omega =  \omega_0 = -\pi  i T\,, \qquad\quad  k  =  2 \pi i m T\, \\
&\omega = \omega_1 = - 3 \pi i T\,, \qquad  k = \begin{cases}
2 \pi i (m \pm 1) T\\
- 2 \pi i m T
\end{cases}\\
& \omega = \omega_2 = - 5 \pi i T\,, \qquad  k = \begin{cases}
2 \pi i m T \\
2 \pi i (m \pm 2) T\\
- 2 \pi i (m \pm 1) T
\end{cases}\\
& \omega = \omega_3 = - 7 \pi i T\,, \qquad  k = \begin{cases}
2\pi i (m \pm 3) T\\
2 \pi i (m \pm 1) T\\
-2 \pi i(m \pm 2) T\\
- 2 \pi i m T\\
\end{cases}\\*
& \qquad\hspace{5ex} \vdots \hspace{25ex} \vdots\nonumber
\end{align}
\end{subequations}

As a final remark, one can notice that unlike in the case of a minimally coupled scalar in the BTZ black hole background (see eq (4.6) of \cite{Blake:2019otz}), the pole-skipping points do not occur in pairs of positive and negative imaginary momenta for a general mass.
The exception is when $m = 0$, where one can see from \eqref{eq:pspointsBTZ} that one of the momenta vanishes and the others form pairs of the form $k = \pm i k_n$, just as in the scalar case.
This is the same phenomenon observed in higher dimensions, where it is also associated with a decreased number of pole-skipping points and an increase of undetermined parameters from the near-horizon analysis.

\subsubsection{Comparison with the exact Green's function}

 \begin{figure}[htb!]
\includegraphics[width= \textwidth]{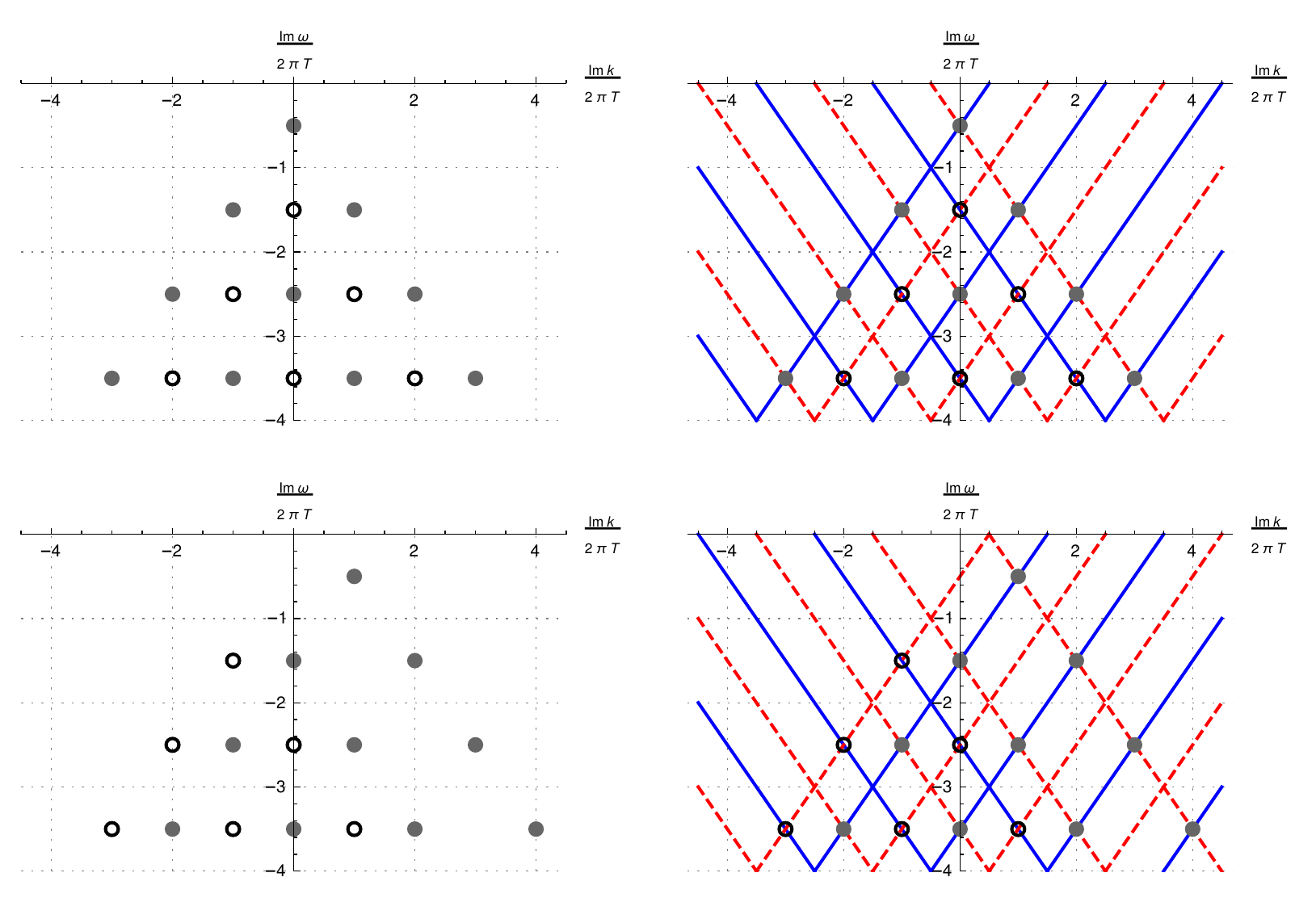}
\caption{Plots of the locations of pole-skipping points for the fermionic Green's function in the BTZ black hole background. The top row is for $m=0$ and the bottom row shows the locations for $m=1$. The left column shows only the locations of the pole-skipping points as predicted from the near horizon analysis.  The gray points correspond to the momentum written with a positive sign in  and the hollow points correspond to the momenta with a negative sign as written in \eqref{eq:pspointsBTZ}. Comparing the top left and bottom left panel we notice that, by increasing the mass, the gray points get rigidly translated to the right by the value of $m$ and the hollow points get translated by an equal amount to the left. The right column has superimposed the lines of zeros (red, dashed) from \eqref{eq:btzgzeros} and lines of poles (blue) from \eqref{eq:btzgpoles}. For both values of the mass the near-horizon analysis predicts the location of the intersections of lines of zeros and lines of poles. }
\label{fig:btz1}
\end{figure}
 
The exact retarded Green's function for the BTZ black hole was derived in \cite{Iqbal:2009fd}. For non-half-integer mass fermions, it is given by
\begin{align}
\label{eq:btzretard}
G_R(\om, k) = -i\,
{
 \Gamma\left(\frac{1}{2} - m\right)\Gamma\le({m\ov 2}+{1\ov 4}+{i(k-\om)\ov 4\pi T}\ri)
 \Gamma\le({m\ov 2}+{3\ov 4}-{i(k+\om)\ov 4\pi T}\ri) \ov
\Gamma\left(\frac{1}{2} + m\right) \Gamma\le(-{m\ov 2}+{3\ov 4}+{i(k-\om)\ov 4\pi T}\ri)
 \Gamma\le(-{m\ov 2}+{1\ov 4}-{i(k+\om)\ov 4\pi T}\ri) }\,.
\end{align}
It has a pole whenever  the argument of any of the gamma functions in the numerator hits a non-positive integer. Similarly, it has a zero whenever an argument of any of the gamma functions in the denominator is equal to a non-positive integer.

Assuming that the mass $m$ is fixed and is not half-integer valued,  we get two infinite families of lines of poles and   two infinite families lines of zeros in the $(\omega, k)$ plane.
The poles are located at
\label{eq:btzgpoles}
\begin{align}
\omega^P_1 = k - \pi i  T( 4n + 2m +1)\,, \qquad \omega^P_2 = -k - \pi  i T (4n + 2m + 3)\,,
\end{align}
and the zeros  can be found at
\label{eq:btzgzeros}
\begin{align}
\omega^Z_1 = k - \pi i T(4n - 2m + 3)\,, \qquad \omega^Z_2 = -k -\pi i T(4n -2m +1)\,,
\end{align}
where in all cases $n = 0, 1, 2, \ldots$. Pole-skipping is observed whenever a line of poles and a line of zeros intersect and thus the Green's function skips a pole. This can be shown to happen precisely at
\begin{eqnarray}
\label{locationsferm}
  \om_n = -i\pi T (2n+1), \qquad k_{n,q_1} &=&  2\pi i T(m+n-2q_1), \nonumber \\  k_{n,q_2} &=&  - 2 \pi i T(m +n + 1 - 2q_2),
\end{eqnarray}
for any $n \in \{0,1,\ldots\}$ and with $q_1 \in \{ 0, \ldots, n \}$, $q_2 \in \{ 1, \ldots, n \}$,\footnote{For $n=0$, there are no solutions in the $k_{n,q_2}$ branch of \eqref{locationsferm}.} which precisely matches our near-horizon analysis from \eqref{eq:pspointsBTZ}.

\begin{figure}[t]
\includegraphics[width= \textwidth]{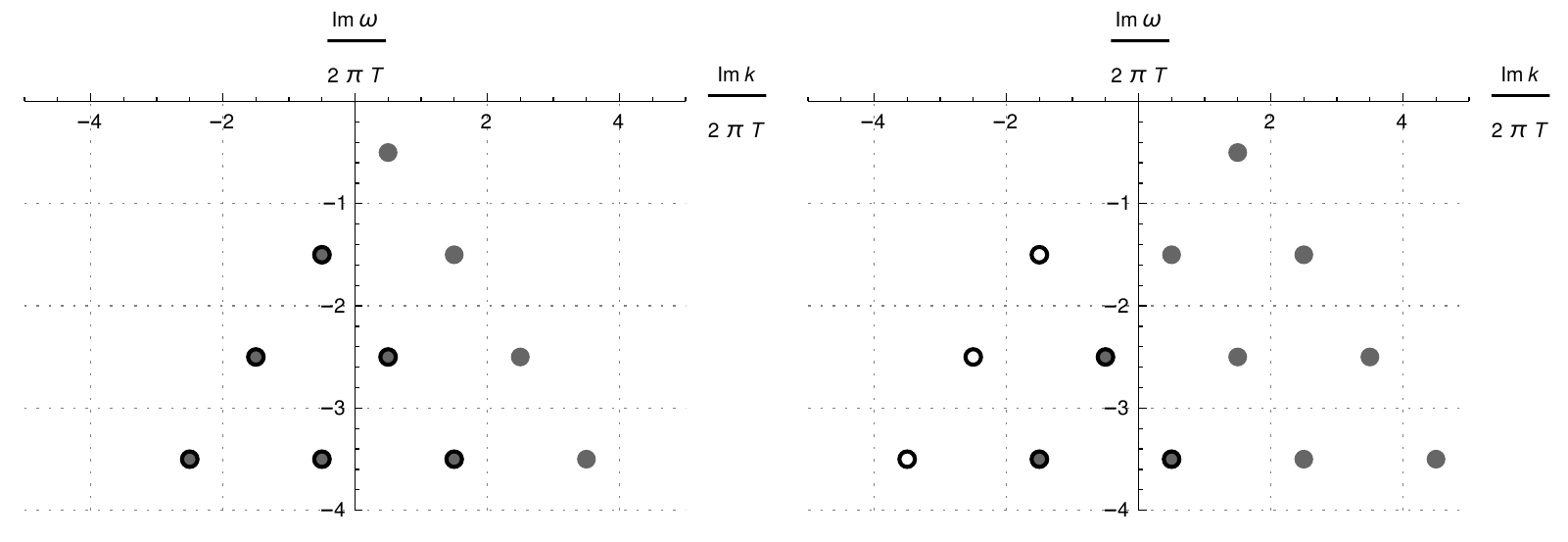}
\caption{Pole-skipping points as predicted from the near-horizon analysis for half-integer mass values. The left plot are the locations for $m = \tfrac12$ and the right plot contains the locations for $m = \tfrac32$.  The gray points correspond to the momentum written with a positive sign in  and the hollow points correspond to the momenta with a negative sign as written in \eqref{eq:pspointsBTZ}. We see that at half-integer values of the mass, some of the locations overlap (black circles with gray filling). These cases correspond to so-called anomalous points (see appendix~\ref{app:form} for details) and signal that a more thorough analysis of the boundary Green's function is needed.}
\label{fig:btz2}
\end{figure}

\subsubsection{Green's function at half-integer conformal dimensions}

When the mass $m$ (or equivalently the scaling dimension $\Delta$ of the dual operators) is half-integer valued, the near-boundary expansion contains logarithmic terms and therefore the boundary retarded Green's function takes a different form. We focus on the case of $m >0$ or equivalently  $\Delta >1$. The boundary retarded Green's function is then given by
\begin{align} \label{eq:retardedgreenhalfinteger}
G_{R} (\omega, k) &\propto  \,
{
 \Gamma\left(\frac{\Delta}{2} - {1 \ov 4} + i{(k - \omega) \ov 4\pi T}\right)\Gamma\left(\frac{\Delta}{2} + {1 \ov 4} - i{(k + \omega) \ov 4\pi T}\right) \ov
\Gamma\left(-\frac{\Delta}{2} + {5 \ov 4} + i{(k - \omega) \ov 4\pi T}\right)\Gamma\left(-\frac{\Delta}{2} + {3 \ov 4} - i{(k + \omega) \ov 4\pi T}\right)} \times \nonumber\\
&  \hspace{20ex} \,
\left[\psi \left(\frac{\Delta}{2} - {1 \ov 4} + i{(k - \omega) \ov 4\pi T}\right)+  \psi \left(\frac{\Delta}{2} + {1 \ov 4} - i{(k + \omega) \ov 4\pi T}\right)\right],
\end{align}
where $\psi(z)$ is the digamma function and we have written the mass $m$ in terms of the scaling dimension $\Delta$. For a more explicit derivation of the Green's functions for half-integer mass fermions, see Appendix \ref{app:ExactBTZ}.

Because $\Delta$ is half-integer valued, the arguments of the gamma functions in the denominator and numerator of \eqref{eq:retardedgreenhalfinteger} differ pairwise by an integer.
Thus, we can expand the ratio of the gamma functions into a product of finitely many terms as
\begin{align}
{\Gamma \left( {\Delta \ov 2} - {1 \ov 4} + {i(k - \omega) \ov 4\pi T}\right) \ov \Gamma \left(-{\Delta \ov 2} + {5 \ov 4} + {i(k - \omega) \ov 4\pi T}\right)
 } = \prod_{n = 1}^{\Delta- \tfrac32} \left( \frac{\Delta}{2}- \frac14 - n + \frac{i(k-\omega)}{4 \pi T}\right)\,.
\end{align}
The ratio of the other two gamma functions can be found in a similar way to be
\begin{align}
{\Gamma \left( {\Delta \ov 2} + {1 \ov 4} - {i(k + \omega) \ov 4\pi T}\right) \ov \Gamma \left(-{\Delta \ov 2} + {3 \ov 4} - {i(k + \omega) \ov 4\pi T}\right)
 } = \prod_{n = 0}^{\Delta - \tfrac32} \left( \frac{\Delta}{2}- \frac34 - n - \frac{i(k+\omega)}{4 \pi T}\right)\,.
\end{align}
This means that the retarded Green's function will have a family of $2 \Delta - 2$ lines of zeros, given by the equations
\begin{align}
\label{eq:btzgzerosHALFINTEGER}
\omega^Z_1 = k - 2\pi i T\left(2n - \Delta + \frac12\right)\,, \qquad \omega^Z_2 = -k -2\pi i T\left( 2n - \Delta + \frac32\right)\,,
\end{align}
where $n \in \left\{0, 1, \ldots, \Delta -{3 \ov 2} \right \}$ and there is no solution for $n=0$ in $\omega_1^Z$.

As all gamma functions cancel out, poles arise only when the argument of any of the two digamma functions is a non-positive integer.
Thus there are two infinite families of  lines of poles located at
\begin{align}
\label{eq:btzgpolesHALFINTEGER}
\omega^P_1 = k -2 \pi i  T\left(2n + \Delta - \frac12\right)\,, \qquad \omega^P_2 = -k - 2\pi  i T \left(2n + \Delta + \frac12\right)\,,
\end{align}
for $n = 0, 1, 2, \ldots$.
One can look for intersections between the lines of zeros and the lines of poles. These occur at the following values for the frequency and momentum
\begin{eqnarray}
\label{locationsfermHALFINTEGER}
 \om_n = -i\pi T (2n+1), \qquad k_{n,q_1} &=&  2\pi i T(n + \Delta - 2 q_1 - 1), \nonumber \\  k_{n,q_2} &=&  -2\pi i T(n + \Delta - 2 q_2),
\end{eqnarray}
where $n \in \{0,1,\ldots\}$,  $q_1 \in \{ 0, \ldots,  \text{min}\left(n, \Delta - {3\ov 2}\right) \}$, $q_2 \in \{ 1, \ldots, \text{min}\left(n, \Delta - {3\ov 2}\right) \}$  and again there is no pole-skipping point at $n =0$ for the momenta given by $k_{0,q_2}$.

To see that even these special cases are predicted by the near-horizon behavior, we must mention the occurrence of so-called \emph{anomalous} pole-skipping points. Namely, at points in the momentum space that are infinitesimally close to a pole-skipping point, the boundary Green's function takes on a certain form, which was dubbed the pole-skipping form and is given as
\begin{align}
G_R(\omega_n + \delta\omega, k_n + \delta k) \propto \frac{\delta\omega - \left(\frac{\delta \omega}{\delta k}\right)_z \delta k }{\delta\omega - \left(\frac{\delta \omega}{\delta k}\right)_p \delta k}\,,
\end{align}
where $\delta\omega$ and $\delta k$ are the directions in momentum space in which we move away from a pole-skipping point and $(\delta\omega/\delta k)_{p,z}$ correspond to the slope of the lines of poles and lines of zeros going through the pole-skipping point.

The locations where the near-horizon analysis predicts pole-skipping, but the correlator does not take on the pole-skipping point form are called anomalous. For the fermionic field, these occur when two pole-skipping points overlap (see figure~\ref{fig:btz2}) and can only occur for $n \geq 1$. The detailed analysis of the pole-skipping form for the fermionic field is given in appendix~\ref{app:form}.

\begin{figure}[tb]
\includegraphics[width= \textwidth]{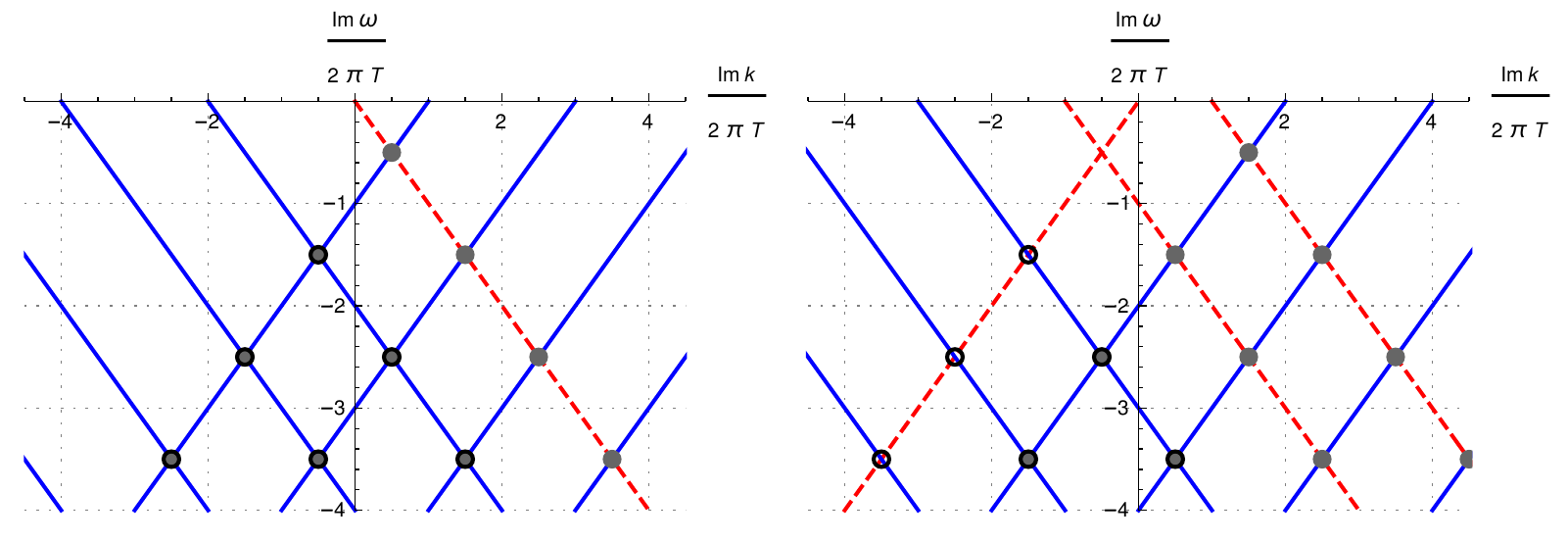}
\caption{Comparison of the locations of the pole-skipping points predicted by the near-horizon analysis (gray and hollow points) and the locations of the intersections of the lines of poles (blue) and lines of zeros (red, dashed) of the exact boundary retarded Green's function for half-integer values of the conformal dimension. We see that the non-anomalous pole-skipping points (either hollow or gray, but not gray with black circle) perfectly match the locations of the intersections. The anomalous pole-skipping points (gray with black boundary) correspond to the locations where two lines of poles intersect. The physical interpretation of these anomalous points is still unclear.}
\label{fig:newpoles}
\end{figure}

Let us assume that the mass of the bulk fermionic field is a half-integer number and focus on $m >0$. The analysis of anomalous points shows that for $n<m + 1/2 $, there are only non-anomalous pole-skipping points. For $n \geq m + 1/2 $, the non-anomalous pole-skipping points are given by
\begin{subequations}
\begin{align}
k_{n,q_1} &=  2\pi i T(m+n-2q_1)\,,&q_1 \in \lbrace 0, 1, \ldots m-1/2\rbrace\,,\\
 k_{n,q_2} &=  - 2 \pi i T(m +n + 1 - 2q_2)\,,  & q_2 \in \lbrace  1, 2,\ldots m-1/2 \rbrace\,,
\end{align}
\end{subequations}
where for $m  = 1/2$, there are no solutions in the second branch.
This implies that the anomalous points are given by
\begin{align}
k_{n,q_1} &=  2\pi i T(m+n-2q_1)\,,\qquad q_1 \in \lbrace m+1/2,m+ 3/2, \ldots n\rbrace\,.
\end{align}
Therefore, all in all, the near-horizon analysis predicts that the non-anomalous pole-skipping points are located at
\begin{eqnarray}
\label{locationsferm}
  \om_n = -i\pi T (2n+1), \qquad k_{n,q_1} &=&  2\pi i T(m+n-2q_1), \nonumber \\  k_{n,q_2} &=&  - 2 \pi i T(m +n + 1 - 2q_2),
\end{eqnarray}
with $n \in \{0,1, \ldots\}$ and $q_1 \in \{ 0, \ldots,  \text{min}\left(n, m - {1\ov 2}\right) \}$, $q_2 \in \{ 1, \ldots, \text{min}\left(n, m - {1\ov 2}\right) \}$. Since $m = \Delta - 1$, we see that these results completely match the positions of intersections of the lines of poles and lines of zeros of the exact boundary retarded Green's function (see figure~\ref{fig:newpoles}).

\subsection{Schwarzschild black hole in AdS$_{d+2}$}

Let the background metric be the Schwarzschild-AdS black hole in $d+2$  dimensions. In this case, the functions determining the metric are given by
\begin{align}
f(r) = 1 - \left(r_0 \over r\right)^{d+1}\,, \qquad h(r) = r^2\,,
\end{align}
with the Hawking temperature defined by $(d+1)r_0 = 4 \pi T$. For a convenient choice of gamma matrices in any dimension and the discussion of how to calculate the pole-skipping points in practice, see appendix~\ref{app:gammas}.

Following the procedure outlined in section~\ref{sec:generalpoleskipping}, the pole-skipping points at the lowest frequency are located at
\begin{align}
\omega = \omega_0 =  - \pi i T\,, \qquad k = \pm \frac{4 \pi i }{d + 1} m T\,.
\end{align}
These include the locations for both subsystems that we discussed.
Notice that if we set $m =0$, these two points merge into a single point with momentum given by $k =0$.

Now, let us focus on the case of a massless fermion ($m =0$). The next pole-skipping points are located at
\begin{subequations}
\label{eq:psmaslessssads}
\begin{align}
&\omega = \omega_1 = - 3 \pi i T\,, \qquad k = \begin{cases} 0\,,\\
k = \pm 2\sqrt{2} i \, \sqrt{\frac{d}{d+1}} \, \pi T\,,
\end{cases}\\
&\omega = \omega_2  = - 5\pi i T\,, \qquad  k  = \begin{cases} 0\,,\\
\pm 2 i \sqrt{\frac{5 d+\sqrt{d (d+8)}}{d+1}} \, \pi T\,,\\
\pm 2 i \sqrt{\frac{5 d-\sqrt{d (d+8)}}{d+1}} \, \pi T\,,
\end{cases}\\
&\hspace{10 ex}\vdots \hspace{25ex}\vdots\nonumber
\end{align}
\end{subequations}

An interesting observation is that the non-zero momenta at $\omega_1$ for the massless fermion field coincide with the  pole-skipping momenta associated with first bosonic Matsubara frequency $\omega = \omega_1^B = - 2 \pi i T$  for the massless bosonic field in the Schwarzschild-AdS background \cite{Blake:2019otz}. This is not the case if the fields are massive. Furthermore, this ceases to hold when one compares higher frequencies.
\section{Discussion}
\label{sec:Discussion}

In this paper we have investigated the near-horizon behavior of a minimally coupled fermion in  asymptotically anti-de-Sitter spacetimes.
The thermal Green's function of the dual fermionic operator exhibits an ambiguity: at certain values of the frequency and the momentum, there exist multiple independent solutions to the Dirac equations that are ingoing at the horizon. As a consequence, the Green's function is not uniquely defined at these points. A pole and a zero of the Green's function collides which results in the pole not appearing. Hence, this phenomenon was termed `pole-skipping' in the literature.
The special frequencies where this happens are precisely the negative fermionic Matsubara frequencies
\be
  \omega_n = - 2 \pi i T \left(n + \frac12\right)\,,
\ee
where $n$ is a non-negative integer.
At each of these frequencies, there are in general $2(2n+1)$ associated values of the momentum, at which pole-skipping takes place.
Generically, the ingoing boundary condition at the horizon fixes half of the components of a spinor, whereas at pole-skipping points, the ingoing condition only fixes a quarter.

Interesting exceptional cases include that of a spinor in three-dimensional spacetime and the case of a massless spinor field in any dimension where there are only $2n+1$ pole-skipping points for each $n$. These scenarios are analyzed in section \ref{sec:AdS3fermionic} and \ref{sec:generalpoleskipping}, respectively.

The fermionic case is conceptually similar to the bosonic case \cite{Blake:2019otz}.
In both cases, the near-horizon behavior of the fields determines the behavior of the boundary field theory correlators {\it away from the origin in Fourier space}.
Furthermore, there is a similarity in that the higher the frequency of the pole-skipping point, the farther we probe into the spacetime, away from the horizon. This is manifested in the fact that the special momenta depend on higher and higher derivatives of metric functions evaluated at the horizon.

In addition, we see that the pole-skipping points in general have a similar structure in both cases. The frequency is determined purely by the temperature of the black hole and thus the surface gravity at the horizon. The momentum has two general contributions, one coming from the mass term and the other which is independent of the mass.

Despite all the similarities, there are also some differences between the bosonic and the fermionic cases. The first one is that in the scalar case, we have fewer pole-skipping points:
at any strictly negative imaginary bosonic Matsubara frequency, i.e. $\omega = - 2 \pi i \tilde n T$, with $\tilde n = 1, 2,3 \ldots$, there are $2\tilde n$ values for the momentum where the Green's function exhibits pole-skipping.

Another interesting difference is the existence of the pole-skipping point at the zeroth Matsubara frequency for fermions. This is due to the fact that the spinors are multi-component objects, and thus there can exist two linearly-independent solutions which have the same behavior near the horizon. Higher (bosonic or fermionic) pole-skipping points depend in some way on the derivatives of metric functions. Since the zeroth-order fermionic pole-skipping point is independent of these derivatives, it is the most localized probe at the horizon.
Furthermore, as we discuss in appendix~\ref{app:form}, this pole-skipping point can never be anomalous and thus it is a robust feature of holographic Green's functions of fermionic operators for any value of the conformal dimension.

There are a few potential pathways in which one could generalize the above results.
Pole-skipping has now been observed and analyzed for both bosonic and fermionic fields. It should be possible to extend the analysis to the case of a gravitino field. Furthermore, using a 2-dimensional CFT, it has been shown \cite{Das:2019tga} that pole-skipping is also seen for frequencies which are non-integer multiples of $\pi i T$. These are neither bosonic nor fermionic Matsubara frequencies and could be associated with non-half integer spin particles: anyons.  It would be interesting to see whether there is a corresponding bulk object, whose near-horizon behavior would explain pole-skipping at such frequencies.

Our hope is that one can get a better understanding of pole-skipping by considering more complicated, yet soluble models, such as the axion model \cite{Andrade:2013gsa, Davison:2014lua}. This model contains an additional parameter which regulates the strength of the energy dissipation in the boundary theory. Ref. \cite{Blake:2018leo} discusses pole-skipping in this model for the energy density function and finds that the pole-skipping point does not change as the dissipation is increased and correctly predicts the dispersion relation of the collective excitations in the boundary for both the weakly and strongly dissipating regime. It would be interesting to see whether such statements could be translated to the scalar or spinor field case.

Another point of interest might be the interpretation of the anomalous points. Anomalous points occur whenever two pole-skipping points overlap. From the example of the BTZ black hole we see that such points correspond to the locations in momentum space where two lines of poles overlap. It would be interesting to see if there is some additional physics that happens at such points.

The detailed analysis of the Green's function revealed that at (bosonic or fermionic) Matsubara frequencies, the retarded and advanced Green's function are equal. Another interesting aspect worth looking into is to see how this is manifested in the boundary theory.

Finally, we have added to the literature of properties of the boundary theories that are encoded in the near-horizon region. One may wonder if there are other universal properties of holographic theories that can be seen from simple near-horizon analysis of bulk fields.

\section*{Acknowledgements}
We would like to thank Mike Blake, Richard Davison, Sa\v{s}o Grozdanov, Costis Papageorgakis, Rodolfo Russo for helpful discussions and comments on the manuscript.
DV is supported by the STFC Ernest Rutherford grant ST/P004334/1.

\clearpage
\appendix
\section{Gamma matrices in various dimensions}
\label{app:gammas}
The gamma matrices used in the calculations satisfy the Clifford algebra relations
\begin{align}
\left\{ \Gamma^a, \Gamma^b\right\} = 2 \eta^{ab}\,,
\end{align}
where  $\eta_{ab} = \text{diag}(-1, +1, +1, \ldots, +1)$. In particular this means that the gamma matrix associated with the time direction\footnote{In our case this is the $\gvv$ matrix.} squares to $-1$ while the gamma matrices associated to the spatial directions square to $1$.

Most of the calculations in the main text were done without referring to a particular representation of the gamma matrices.
However, in practice it might be useful to choose a nice representation  to easily extract the locations of the pole-skipping points. Here we present a choice we found particularly useful.

Recall that in AdS/CFT, a bulk spinor and a boundary spinor can have a different number of components, depending on the number of dimensions of spacetime.
If the boundary theory is even-dimensional ($d+1$ is even) and thus the bulk theory is odd dimensional, then the boundary and bulk spinors have an equal number of components.
If the boundary theory is odd-dimensional ($d+1$ is odd) and the dual bulk theory is even dimensional, then the bulk spinor has twice as many components as the boundary theory.

Our choice of gamma matrices should reflect this counting. Following \cite{Iqbal:2009fd}, we choose the representations of the bulk gamma matrices $\Gamma^a$ in terms of the boundary matrices $\gamma^a$ in the case of even $(d+1)$ as
\begin{align}
  \grr = \gamma_{d+2}\,,\qquad\Gamma^a = \gamma^a\,, \quad  a = \uuv, 1, 2, \ldots, d\,,
\end{align}
where $\gamma_{d+2}$ is the analogue of the usual $\gamma_5$ matrix in flat space quantum field theories\footnote{One can take for example $\gamma_{d+2} = i^{-\frac{d-1}{2}}\gamma^{0}\gamma^{1}\ldots\gamma^{d}$}. In the case of odd $d+1$, the bulk theory has spinors with twice as many components. We can make the choice
\begin{align}
\label{eq:AoddevenI}
\grr = \begin{pmatrix}
1 & 0\\
0 & -1
\end{pmatrix}\,, \qquad \Gamma^a = \begin{pmatrix}
0 & \gamma^a \\ \gamma^a & 0
\end{pmatrix}\,, \quad  a = \uuv, 1, 2, \ldots, d\,.
\end{align}
In the latter case, the $\grr$ matrix is explicitly diagonal, which is not necessarily the case in the former.
Here, we go a step further. We construct a representation in which the $\grr$ matrix is diagonal in any dimension and in which the matrix $\kh\gvi$ is also diagonal. In this way, we can show that any fermionic system can be effectively reduced not only to 2 subsystems, each involving  $N/2$ degrees of freedom, but to $N/2$ subsystems each containing only 2 degrees of freedom. In that way every system can be reduced to solving equations similar to the BTZ example in section \ref{sec:Examples}.

We start with a 3-dimensional bulk spacetime. The spinors are two dimensional and so we can use the gamma matrices
\begin{align}
\label{eq:Arep3}
\Gamma^{\uuv} = i \sigma^2\,,\qquad  \Gamma^{\uur}= \sigma^3\,,\qquad  \Gamma^{\uux} = \sigma^1\,,
\end{align}
where $\sigma^i$ are the usual Pauli matrices given by
\begin{align}
\sigma^1 = \begin{pmatrix}
0 & 1\\
1 & 0
\end{pmatrix}\,, \quad
  \sigma^2 = \begin{pmatrix}
 0 & -i\\
 i & 0
 \end{pmatrix}\,, \quad
 \sigma^3 = \begin{pmatrix}
 1 & 0 \\
 0 & -1
 \end{pmatrix}\,.
\end{align}
We see that $\grr$ is a diagonal matrix, despite the bulk theory being odd-dimensional. Furthermore, we see that in this case, we have $\gvv\, \gxx = \grr$, which is something we have demanded in the main text. One could also change the sign of any of the three matrices and still get a possible representation%
\footnote{However, the relation $\gvv\, \gxx = \grr$ would hold only if we simultaneously change the sign of two of the matrices in \eqref{eq:Arep3}. If we change the sign of only one or all three, then $\gvv\, \gxx = -\grr$.}.

Let us now look at a 4-dimensional bulk theory, where the spinors have four components and the associated boundary theory has spinors with only two components. In order to make the $\grr$ matrix diagonal, we choose the following representation
\begin{align}
&\gvv = \sigma^1 \otimes i \sigma^2 = \begin{pmatrix}
0 & i\sigma^2 \\ i\sigma^2 & 0
\end{pmatrix}
\,,\quad \gxx = \sigma^1 \otimes \sigma^1 =
\begin{pmatrix}
0 & \sigma^1 \\ \sigma^1 & 0
\end{pmatrix}\,,\nonumber \\
&\Gamma^{\underline{y}} = \sigma^1 \otimes  \sigma^3 = \begin{pmatrix}
0 & \sigma^3\\ \sigma^3 & 0
\end{pmatrix}\,, \qquad
 \grr = \sigma^3 \otimes \mathbb{1} = \begin{pmatrix}
\mathbb 1 & 0 \\ 0 & -\mathbb 1
\end{pmatrix}\,. \label{eq:Arep4}
\end{align}
In this case, $\grr$ is diagonal, while all other matrices are obtained by tensor multiplying (from the left) the gamma matrices in the 3-dimensional theory $\eqref{eq:Arep3}$ with $\sigma^1$. In fact this is the representation used by \cite{Iqbal:2009fd}.

In a bulk theory with 5-dimensions, both the boundary and the bulk spinors have four components. We get a representation for the 5-dimensional case by adding the $\pm \Gamma_5$ matrix to the 4-dimensional set of matrices \eqref{eq:Arep4}. In this case, we can choose to add
\begin{align}
\Gamma^{\underline{z}} &= i \gvv \, \gxx\,\Gamma^{\underline{y}}\,   \grr =  \sigma^2 \otimes  \mathbb{1}\,.
\end{align}
Using the above definition, we can easily see that
\begin{align}
\grr = - i \gvv \, \gxx\,\Gamma^{\underline{y}}\,   \Gamma^{\underline{z}}\,,
\end{align}
which means that we have constructed a representation where $\grr$ is again diagonal and again the analogue of the $\Gamma_5$ matrix.

The generalization to higher dimensional cases is straightforward. When constructing the gamma matrices for a bulk theory in even dimensions, we pick
\begin{align}
\label{eq:Aoddeven}
\grr = \begin{pmatrix}
1 & 0\\
0 & -1
\end{pmatrix}\,,\qquad \Gamma^a = \sigma^1 \otimes \tg^a = \begin{pmatrix}
0 & \tg^a \\ \tg^a & 0
\end{pmatrix}\,, \quad a = \uuv, 1, 2, \ldots, d\,,
\end{align}
where $\tg^a$ are the gamma matrices from the bulk theory in one dimension lower. %
In particular, notice that in this case the $\gvv$ and $\gxx$ matrices have the following forms
\begin{align}
\label{eq:Agvvgx}
\gvv= \sigma^1 \otimes \sigma^1 \otimes \ldots \otimes i \sigma^2 = \begin{pmatrix}
 & & & & 1\\
 & & & -1 & \\
 & & $\reflectbox{$\ddots$}$  & & \\
 & 1 & & & &\\
 -1 & & & & &
\end{pmatrix}\,, \quad \gxx = \sigma^1\otimes \ldots \otimes \sigma^1 = \begin{pmatrix}
 & & 1\\
& $\reflectbox{$\ddots$}$  & \\
1 & & \\
\end{pmatrix}\,,
\end{align}
and consequently
\begin{align}
\label{eq:Agvx}
\gvx = \mathbb{1} \otimes \ldots \otimes \mathbb{1} \otimes \sigma^3  = \begin{pmatrix}
1 & & &  & \\
& -1 & & & \\
& & \ddots & & \\
& &  &  1 & \\
& &  &  & -1
\end{pmatrix}\,.
\end{align}
When constructing gamma matrices for a bulk theory in odd dimensions, we pick
\begin{align}
\Gamma^a = \tg^a\,, a = \uuv, \uur, 1, 2,\ldots d-1\,, \qquad \Gamma^{\underline{d}} = - i^{\frac{d-1}{2}} \gvv \, \Gamma^{\underline{1}} \, \ldots \Gamma^{\underline{d-1}}\,\grr\,,
\end{align}
so that
\begin{align}
\Gamma^{\underline{r}} = i^{\frac{d-1}{2}} \gvv \, \Gamma^{\underline{1}} \, \ldots \Gamma^{\underline{d-1}}\,\Gamma^{\underline{d}}\,.
\end{align}
We see that this way, in both even and odd dimensions, $\grr$, $\gvv$ and $\gxx$  have the same form.

As we saw in section~\ref{sec:generalpoleskipping}, in higher dimensional cases we needed to split the spinor according to two projections, $\grr$ and $\kh \,\gvi$ which are independent for $ d\geq2$.
In practice, we can simplify these conditions by using the symmetry under the rotations in the $d$-dimensional subspace. This means that we can always rotate the system in such a way that the momentum points along the $x$-direction, or in other words, $k_x = k$ and $k_i = 0$, for $i \neq x$.

In such a case, $\kh \gvi \Rightarrow \gvx$, with $\gvx$ being given in \eqref{eq:Agvx}. Thus, using symmetry and a clever choice of gamma matrices, both projection matrices become diagonal.
Furthermore, the only four matrices that are of importance are the projection matrices $\grr$  and $\gvx$  (given in \eqref{eq:Aoddeven} and \eqref{eq:Agvx} respectively) and $\gvv$ and $\gxx$ (given in \eqref{eq:Agvvgx})  that mix up the components.

Using these matrices, the Dirac equations do not only separate into 2 subsystems, each containing half of the degrees of freedom, as was the generic case presented in section~\ref{sec:generalpoleskipping}, but rather, the equations separate into $N/2$ subsystems\footnote{Where $N$ is the total number of degrees of freedom in the spinor.}, each containing 2 degrees of freedom, similar to the BTZ case discussed in \ref{sec:Examples}.
The decoupled subsystems of two degrees of freedom are the first and the last component, the second and second-to-last, the third and third-to-last and so on. Effectively, one "peels" the matrices off layer by layer.

This introduces  two-dimensional  "effective" gamma matrices for the subsystems. One notices that the odd numbered subsystems  have the same effective matrices, which  differ from the even numbered subsystems, whose matrices in turn are also all the same.
For the odd numbered layers (e.g. first and last, third and third-to-last) the effective gamma matrices are given by
\begin{align}
\grr_o = \gvr_o=  \sigma^3\,,\qquad \gvv_o = i \sigma^2\,, \qquad \gxx_o = \sigma^1\,,
\end{align}
while for the even numbered layers, the two gamma matrices are given by
\begin{align}
\grr_e = -  \gvr_e=  \sigma^3\,,\qquad \gvv_e =  - i \sigma^2\,, \qquad \gxx_e = \sigma^1\,,
\end{align}
and the subscript denotes either even or odd.

Thus, in practice, solving the equations of motion always reduces to solving a system of two coupled first order ordinary differential equations for scalar functions.
This does not mean that the number of pole-skipping points  or the number of free parameters at a pole-skipping point changes as, although we have $N/2$ independent subsystems of equations, $N/4$ of those  produce the same pole-skipping points, while the other $N/4$  have pole-skipping points at the same frequency, but the opposite momenta.

As an example, we can look at the case of a 5-dimensional bulk theory with a 4-dimensional spinor. According to the above procedure, one can use the following representation
\begin{align}
\gvv = \sigma^1 \otimes i \sigma^2 = \begin{pmatrix}
0 & i\sigma^2 \\ i\sigma^2 & 0
\end{pmatrix}
\,,\quad \gxx = \sigma^1 \otimes \sigma^1 =
\begin{pmatrix}
0 & \sigma^1 \\ \sigma^1 & 0
\end{pmatrix}\,, \quad \grr = \sigma^3 \otimes \mathbb{1} = \begin{pmatrix}
\mathbb 1 & 0 \\ 0 & -\mathbb 1
\end{pmatrix}\,,
\end{align}
and $\Gamma^{\underline{y}} = \sigma^1 \otimes  \sigma^3$ and $\Gamma^{\underline{z}} = \sigma^2 \otimes  \mathbb{1}$.
We choose the momentum to be along the $x$-direction ($k_x = k$ and $k_y = k_z = 0$) so that the helicity matrix becomes
\begin{align}
\kh \gvi \Rightarrow \gvx = \mathbb{1} \otimes \sigma^3 = \begin{pmatrix}
\sigma^3 & 0 \\ 0 & \sigma^3
\end{pmatrix}\,.
\end{align}
This means that the four-component spinor can be written as
$\psi(r) = (\psi_+^{(+)} ,\psi_+^{(-)} ,\psi_-^{(+)} ,\psi_-^{(-)})^T$,
with each $\psi_a^{(b)}$ denoting an independent degree of freedom with well-defined eigenvalues under $\grr$ and $\gvx$. The Dirac equations are then split into two subsystems, one involving $\psi_+^{(+)}$ and $\psi_-^{(-)}$, and the other containing $\psi_+^{(-)}$ and $\psi_-^{(+)}$. Each is governed by two coupled, first order differential equations for scalars.

\section{Green's function near pole-skipping points and anomalous points}
\label{app:form}

In the main text we have described how to obtain the location of the pole-skipping points and claimed that at such points a line of poles and a line of zeros of the boundary Green's function intersect. Here we  explicitly show that this is the case by moving infinitesimally away from the pole-skipping point in momentum space. Furthermore, we show that the solution depends on the direction of the move and thus near the special locations in Fourier space, the correlator takes the \emph{pole-skipping form}
\begin{align}
\label{eq:Cpsform}
G_R(\omega_n + \delta\omega, k_n + \delta k) \propto \frac{\delta\omega - \left(\frac{\delta \omega}{\delta k}\right)_z \delta k }{\delta\omega - \left(\frac{\delta \omega}{\delta k}\right)_p \delta k}\,,
\end{align}
where $\omega_n$ and $k_n$ are the frequency and momentum of a pole-skipping point respectively and $(\delta\omega/\delta k)_{p,z}$ correspond to the directions in which we need to move away from the pole-skipping point in order to obtain a normalizable or a non-normalizable solution at the boundary.
As normalizable solutions correspond to poles in the Green's function, the associated direction is the slope of the line of poles, passing through the pole-skipping point. Non-normalizable solutions are related to zeros in the correlator and thus the associated direction is the slope of the line of zeros passing through the pole-skipping point.

Originally, all these calculations were performed for the energy-density component of the stress-energy tensor \cite{Blake:2018leo} and the minimally coupled scalar field \cite{Blake:2019otz}. Here we will show that analogous calculations can be done for the minimally coupled fermionic field as well. 

\subsection{Near the lowest Matsubara frequency}
We have seen that the pole-skipping point at $\omega = \omega_0 = - \pi i T$ is different from  other points and we thus consider it separately. We also saw that this pole-skipping point comes from the interaction of the zeroth order coefficients in the spinor expansion \eqref{eq:serexp1}%
\footnote{Here we limit ourselves to the case of a 3-dimensional bulk spacetime with a 2-dimensional boundary. The generalization to higher dimensions is trivial.}.
At the pole-skipping point $\omega = \omega_0 = - \pi i T$ and $k = k_0 = i m \sqrt{h({r_0})}$, the system has two independent solutions that are regular at the horizon and thus the boundary retarded Green's function is  ill-defined. However, at any point infinitesimally close to the pole-skipping location, there exists only one independent ingoing solution. To see this, let us look at the leading order in the series expansion of the equations of motion \eqref{eq:serexp2} at
\begin{align}
\label{eq:smallmove}
\omega = \omega_0 + \epsilon\, \delta \omega\,, \qquad k = k_0 + \epsilon\, \delta k\,,
\end{align}
where $\epsilon$ is a small dimensionless parameter. 
If $\epsilon = 0$, the equations \eqref{eq:direqnc0} are automatically satisfied. But at linear order in $\epsilon$, we get a constraint relating $\psi_\pm^{(0)}$ as
\begin{align}
\left( \delta \omega + \frac{r_0}{2 \sqrt{h(r_0)}}\, \delta k \right) \psi_+^{(0)} + \left(\delta \omega -  \frac{r_0}{2 \sqrt{h(r_0)}}\, \delta k \right) \gvv \, \psi_-^{(0)}=0\,.
\end{align}
This allows us to express one of $\psi_\pm^{(0)}$ in terms of the other. The relation can be written as
\begin{align}
\label{eq:Czeroexp}
\psi_-^{(0)} = \frac{\left( \frac{\delta \omega}{\delta k}\right) + \frac{r_0}{2 \sqrt{h(r_0)}}}{\left(\frac{\delta \omega}{\delta k}\right) - \frac{r_0}{2 \sqrt{h(r_0)}}}\, \gvv \, \psi_+^{(0)} \,,\quad  \text{or equivalently}\quad \psi_+^{(0)} = -\frac{\left( \frac{\delta \omega}{\delta k}\right) - \frac{r_0}{2 \sqrt{h(r_0)}}}{\left(\frac{\delta \omega}{\delta k}\right) + \frac{r_0}{2 \sqrt{h(r_0)}}}\, \gvv \, \psi_-^{(0)}\,.
\end{align}
The relation between $\psi_\pm^{(0)}$ explicitly depends on the direction $(\delta\omega/\delta k)$ in which we move away from the pole-skipping point.
One can  interpret this relation in a different way, as one can think of the aforementioned slope as the additional undetermined parameter of the regular solution at the pole-skipping point. The solution then has two free parameters -- the overall normalization, and the direction in which we move away from the pole-skipping point in Fourier space.

One can use the relations \eqref{eq:Czeroexp} in the reverse way.
Let us assume we found a particular solution to the bulk equations of motion, specified by certain boundary conditions, and let's expand it around the horizon. The equations \eqref{eq:Czeroexp}  allow us to determine the slope at which the solution will approach the pole-skipping point in Fourier space.
This is important because from the near-boundary analysis we know that the spinors separate into a normalizable part, which is related to the poles of the Green's function and a non-normalizable part, related to the zeros of the Green's function (see \eqref{eq:separation}). By a choice of appropriate boundary conditions, we can therefore find bulk solutions which are either fully normalizable or non-normalizable at the boundary. Both can be expanded near the horizon as
\begin{subequations}
\label{eq:Cnnnexp}
\begin{align}
\psi^{(n)}& = \begin{pmatrix}
\psi_+^{(n)} \\ \psi_-^{(n)}
\end{pmatrix}  = \begin{pmatrix}
\psi_+^{(n)} \\ \psi_-^{(n)}
\end{pmatrix}^{(0)}
 + \begin{pmatrix}
\psi_+^{(n)} \\ \psi_-^{(n)}
\end{pmatrix}^{(1)}(r-r_0) + \ldots\\
\psi^{(nn)} &= \begin{pmatrix}
\psi_+^{(nn)} \\ \psi_-^{(nn)}
\end{pmatrix}  = \begin{pmatrix}
\psi_+^{(nn)} \\ \psi_-^{(nn)}
\end{pmatrix}^{(0)}
 + \begin{pmatrix}
\psi_+^{(nn)} \\ \psi_-^{(nn)}
\end{pmatrix}^{(1)}(r-r_0) + \ldots\,,
\end{align}
\end{subequations}
where $\psi^{(n)}$ denotes the normalizable and $\psi^{(nn)}$ the non-normalizable solution.
Near the pole-skipping point, the components of the zeroth order coefficient are then related by
\begin{subequations}
\label{eq:Czeroexp1}
\begin{align}
\left(\psi_-^{(n)}\right)^{(0)} &= \frac{\left( \frac{\delta \omega}{\delta k}\right)_p + \frac{r_0}{2 \sqrt{h(r_0)}}}{\left(\frac{\delta \omega}{\delta k}\right)_p - \frac{r_0}{2 \sqrt{h(r_0)}}}\, \gvv \,\left(\psi_+^{(n)}\right)^{(0)} \,,\\
\left(\psi_+^{(nn)}\right)^{(0)} &= -\frac{\left( \frac{\delta \omega}{\delta k}\right)_z - \frac{r_0}{2 \sqrt{h(r_0)}}}{\left(\frac{\delta \omega}{\delta k}\right)_z + \frac{r_0}{2 \sqrt{h(r_0)}}}\, \gvv \, \left(\psi_-^{(nn)}\right)^{(0)}\,,
\end{align}
\end{subequations}
and we take $\left(\psi_+^{(n)}\right)^{(0)}$ and $\left(\psi_-^{(nn)}\right)^{(0)}$ as the two free parameters associated with the normalizable and the non-normalizable solution. As mentioned above, $(\delta\omega/\delta k)_{p,z}$ correspond to the directions in which we need to move away from the pole-skipping point in order to obtain a normalizable or a non-normalizable solution at the boundary. The meaning of the subscripts will become apparent momentarily. 

Let us assume that we are near the location of the first pole-skipping point at \eqref{eq:smallmove}. At linear order in $\epsilon$, the ingoing solution can be written as a linear combination of the normalizable and non-normalizable component
\begin{align}
\label{eq:Clincomb}
\psi = \psi^{(n)} + \psi^{(nn)}\,,
\end{align}
where neither of the components is normalized. Following the above argument, both $\psi^{(n)}$ and $\psi^{(nn)}$ contain one free parameter,  $\left(\psi_+^{(n)}\right)^{(0)}$, and $\left(\psi_-^{(nn)}\right)^{(0)}$, respectively. If both were left undetermined, the solution would have too many free parameters. However, the direction in which we move away from the pole-skipping point determines one free parameter in terms of the other. One can write the relation between the two as 
\begin{align}
\label{eq:Crel1}
\psi^{(nn)} = \cR\left(\frac{\delta \omega}{\delta k}\right)\,\cdot \psi^{(n)}\,,
\end{align}
where $\cR(\delta \omega / \delta k)$ is a matrix that depends on the slope. 
Using the prescription of \cite{Iqbal:2009fd}, the boundary Green's function is then proportional to this matrix 
\begin{align}
\label{eq:CpresIqbal}
G_R \propto \cR\left(\frac{\delta \omega}{\delta k}\right)\,,
\end{align}
meaning that the Green's function depends on the slope as well.

To obtain the explicit form of the Green's function, we insert \eqref{eq:Clincomb} into the Dirac equations, and expand them around the horizon. At linear order in $\epsilon$, the two undetermined sets of parameters $\left(\psi_+^{(n)}\right)^{(0)}$ and $\left(\psi_-^{(nn)}\right)^{(0)}$ are related by
\begin{align}
\label{eq:Crelation1}
\left(\psi_-^{(nn)}\right)^{(0)} = - \frac{\left(\frac{\delta \omega}{\delta k}\right)_p + \frac{r_0}{2 \sqrt{h(r_0)}}}{\left(\frac{\delta \omega}{\delta k}\right)_z - \frac{r_0}{2 \sqrt{h(r_0)}}} \, \frac{\delta \omega - \left(\frac{\delta \omega}{\delta k}\right)_z \, \delta k}{\delta \omega - \left(\frac{\delta \omega}{\delta k}\right)_p \, \delta k }\,\gvv\, \left(\psi_+^{(n)}\right)^{(0)} \,.
\end{align}
Ignoring all the unimportant factors, one can see that  the Green's function is proportional to
\begin{align}
\label{eq:Cgreensform}
G_R(\omega_n + \delta\omega, k_n + \delta k)  \propto  \frac{\delta \omega - \left(\frac{\delta \omega}{\delta k}\right)_z \, \delta k}{\delta \omega - \left(\frac{\delta \omega}{\delta k}\right)_p \, \delta k }\,,
\end{align}
which is precisely the pole-skipping form. In addition to this, the details of the location of the pole-skipping point do not enter  the calculation at any point. Thus, this pole-skipping point will never be \emph{anomalous}. Here, we follow the definition from \cite{Blake:2019otz}, where a pole-skipping point was called anomalous if it appeared as a possible location from the near-horizon analysis, but the Green's function near such a point did not take the pole-skipping form \eqref{eq:Cpsform}.
For a scalar field such anomalous points usually appeared when two pole-skipping points collided. We will shortly see that this is the case for the fermionic field as well. With that, one can understand that the pole-skipping point at $\omega = \omega_0$ can never be anomalous as it is the only pole skipping point with such frequency, hence there is no other pole-skipping point that it can collide with.
Thus there will always be a line of zeros and a line of poles that will intersect at this pole-skipping point.

Finally, one might wonder if the prefactors in \eqref{eq:Crelation1} cause some trouble. They are related to the prefactors in \eqref{eq:Czeroexp1} and one notices that if they vanish, $\left(\psi_-^{(nn)}\right)^{(0)} $ and/or $\left(\psi_+^{(n)}\right)^{(0)}$ vanish as well. In that case, these particular spinor components cannot be taken as undetermined free parameters. One must rather use the other half of the spinor as the free parameter.
In fact, the slopes $(\delta \omega/\delta k) = \pm r_0/(2 \sqrt{h(r_0)}) $ denote the two special cases where the leading orders of the normalizable and non-normalizable solutions have a well defined eigenvalue under $\grr$ at leading order expansion around the horizon.

\subsection{Near higher Matsubara frequencies}
To analyze the form of the Green's function at higher fermionic Matsubara frequencies, we use the method involving second order differential equations, as it allows us to draw close comparisons with the scalar field case. For higher frequencies we find that  pole-skipping points can be anomalous. For simplicity, we will again work only in the 3-dimensional bulk spacetime with two dimensional spinors. Without loss of generality, let us look at the variable $\psi_+$. The variable $\psi_-$ is fully determined by $\psi_+$ through the first order differential equations.

Let us analyze the form of the Green's function at
\begin{align}
\label{eq:smallmoven}
\omega = \omega_q + \epsilon\, \delta \omega\,, \qquad k = k_q + \epsilon\, \delta k\,,
\end{align}
where $\omega_q$, $k_q$ is the location of a pole-skipping point with $q >0$ and $\epsilon$ is a small parameter.
If $\omega$ and $k$ were generic points in momentum space, then we could use equations like \eqref{eq:serexpd1} and \eqref{eq:serexpd2} to iteratively express all $\psi_+^{(s)}$ with $s>0$ in terms of $\psi_+^{(0)}$. Solving the equations order by order, one finds that for $q>0$, the relation between $\psi_+^{(0)}$ and $\psi_+^{(q)}$ is given by
\begin{align}
\label{eq:Cexpn1}
\frac{1}{N^{(q)}(\omega)}\, \det \cM^{(q)} (\omega, k) \, \psi_+^{(0)} + \left( (2q+1)\pi T - i \omega\right) \, \psi^{(q)}_+ = 0 \,,
\end{align}
where $\cM^{(q)}$ is the matrix defined in \eqref{eq:bigm1} and
\begin{align}
N^{(q)} = (i \omega - 3\pi T) (i \omega - 5 \pi T) \ldots ( i \omega - (2q-1)\pi T)\,,
\end{align}
where we assumed that $\omega \neq \omega_s$ with $s < q$.
Now we want to evaluate the equation $\eqref{eq:Cexpn1}$ in the vicinity of a pole-skipping point with $\omega = \omega_q$. At the pole-skipping point ($\epsilon =0$), the equation is automatically satisfied, however, at linear order in $\epsilon$, the equation becomes
\begin{align}
\label{eq:Crel2}
 \frac1{N(\omega_q)} \left( \pd_k \det\cM(\omega_q, k_q) \, \delta k + \pd_\omega \det \cM(\omega_q, k_q) \, \delta \omega\right) \, \psi_+^{(0)} - i\delta \omega \psi^{(q)}_+= 0\,,
\end{align}
where $N(\omega_q) = (q-1)! (2 \pi T)^{q-1}$. It immediately follows that
\begin{align}
\label{eq:crel3}
\psi_+^{(q)} = - i\, \frac{\pd_k \det\cM(\omega_q, k_q)  + \pd_\omega \det \cM(\omega_q, k_q) \,\left( \dfrac{\delta \omega}{\delta k}\right)}{N(\omega_q)\left(\dfrac{\delta \omega}{\delta k}\right) }\, \psi_+^{(0)}\,.
\end{align}
Again, the direction in which we move away from the pole-skipping point determines the relation between the two coefficients.

In particular, there exist slopes associated to normalizable ($\psi_+^{(n)}$) and non-normalizable ($\psi_+^{(nn)}$) solutions at the boundary. In these cases the explicit relations are given by
\begin{subequations}
\label{eq:Crel4}
\begin{align}
\left(\psi^{(n)}_+\right)^{(q)} &= - i \,\frac{\pd_k \det\cM(\omega_q, k_q)  + \pd_\omega \det \cM(\omega_q, k_q) \,\left( \dfrac{\delta \omega}{\delta k}\right)_p}{N(\omega_q)\left(\dfrac{\delta \omega}{\delta k}\right)_p }\,\left(\psi^{(n)}_+\right)^{(0)}\,,\\
\left(\psi^{(nn)}_+\right)^{(q)} &= - i\, \frac{\pd_k \det\cM(\omega_q, k_q)  + \pd_\omega \det \cM(\omega_q, k_q) \,\left( \dfrac{\delta \omega}{\delta k}\right)_z}{N(\omega_q)\left(\dfrac{\delta \omega}{\delta k}\right)_z }\,\left(\psi^{(nn)}_+\right)^{(0)}\,.
\end{align}
\end{subequations}
At linear order in $\epsilon$, the normalizable and non-normalizable solution have one free parameter which we take to be $\psi_+^{(0)}$ for both solutions.

If we move away now from the location of the pole-skipping in a general direction, then at linear order in $\epsilon$, the solution will be a linear combination of the normalizable and non-normalizable solution
\begin{align}
\label{eq:Clincombq}
\psi = \psi^{(n)} + \psi^{(nn)}\,,
\end{align}
where again neither of the components are normalized and hence naively the above solution has two free parameters.  However, at linear order in $\epsilon$, we get a relation between the free parameters of the normalizable and non-normalizable solution, which depends on the direction in which we move away from the pole skipping point and is given by
\begin{align}
\left(\psi^{(nn)}_+\right)^{(0)} = - \left(\frac{\left(\dfrac{\delta \omega}{\delta k}\right)_p}{\left(\dfrac{\delta \omega}{\delta k}\right)_z}\right)\, \frac{\delta \omega - \left(\dfrac{\delta \omega}{\delta k}\right)_z\, \delta k }{\delta \omega - \left(\dfrac{\delta \omega}{\delta k}\right)_p \, \delta k }\, \left(\psi^{(n)}_+\right)^{(0)}\,.
\end{align}
Using the prescription \eqref{eq:CpresIqbal},  the retarded Green's function is proportional to the multiplicative factor relating the non-normalizable and normalizable component and thus has precisely the pole-skipping form \eqref{eq:Cpsform}.

However, in the case of higher Matsubara frequencies, we may have anomalous points. These occur whenever the determinant of the matrix \eqref{eq:bigm1} satisfies both
\begin{align}
\det \cM(\omega_n, k_n) = 0\,, \qquad \text{and} \qquad \pd_k \det \cM(\omega_n, k_n) = 0\,.
\end{align}
This occurs whenever we have a repeated root or in other words, when two pole-skipping points overlap. Notice that the above condition does not automatically include $k_n = 0$ roots. This is the consequence of the determinant being a function of $k$ and not $k^2$, which is the case for the scalar field.
An example of anomalous roots is given in section~\ref{sec:Examples}, where anomalous pole-skipping points occur in the case of a fermion with half-integer mass (in units of the AdS radius) propagating in the BTZ background.

\section{Details of the calculations}
\label{app:details}
Here we present the detailed calculations that lead towards pole-skipping in asymptotically AdS$_{d+2}$ spacetimes. In principle, the same equation also applies in the asymptotically  AdS$_3$ case and we  point out where the two cases differ. We  repeat some of the steps from the main text, in order to make this calculation more or less self-contained.

We work with the background metric in the ingoing Eddington-Finkelstein coordinates given by
\begin{align}\label{eq:Bbackgroundmetric}
ds^2 = - r^2 f(r) dv^2 + 2dv\,dr +  h(r) d\vec{x}^2\,.
\end{align}
We choose the orthonormal frame to be
\begin{align}
\label{eq:Bgeneral_frame}
E^{\uuv} =  \frac{1+ f(r)}{2}\,r dv - \frac{dr}{r}\,,\qquad E^{\uur} =  \frac{1- f(r)}{2}\, r dv + \frac{dr}{r}\,, \qquad E^{\uui} = \sqrt{h(r)} \, dx^{i}\,,
\end{align}
so that
\begin{align}
ds^2 = \eta_{ab} E^a\, E^b\,, \hspace{10ex} \eta_{ab} = \text{diag}(-1, 1,1, \ldots,1)\,.
\end{align}
The spin connections for this frame are given by
\begin{align}
\omega_{\uuv \uur} = \frac{dr}{r}- \frac{2 r f(r) + r^2 f'(r)}{2} \,dv,\quad \omega_{\uuv \uui} =  \frac{r\,h'(r)\,\left(1 - f(r)\right)  }{4 \sqrt{h(r)}}\, dx^i, \quad \omega_{\uur \uui} = - \frac{r \,  h'(r) \, (1+ f(r))}{4 \sqrt{h(r)}}\, dx^i\,,
\end{align}
with all other components, which are not related by symmetry to the ones above, being 0.
Using these spin connections, one can calculate the Dirac equation to be 
\begin{align}
\label{eq:Bdireqgf0}
\Biggr[ &\left( - \frac{r(1- f(r))}{2}\, \gvv + \frac{r(1+ f(r)) }{2}\, \grr\right) \pd_r + \frac{\grr + \gvv}{r}\pd_v+ \frac{\gii}{\sqrt{h(r)}}\, \pd_i + \frac{1+ f(r) + r f'(r)}{4}\, \grr \nonumber\\
&- \frac{1- f(r) - r f'(r)}{4}\, \gvv - \frac{d\, r\, (1-f(r)) h'(r)}{8 h(r)}\, \gvv +  \frac{d\, r\, (1+f(r)) h'(r)}{8 h(r)}\, \grr- m \Biggr] \psi(r, v, x^j) = 0\,.
\end{align}
Since the metric is independent of the coordinates $v$ and $x^i$, one can insert the plane wave ansatz $ \psi(r, v, x^j)  = \psi(r) e^{- i \omega v + i k_i  x^i}$. The Dirac equation in Fourier space then reads
\begin{align}
\label{eq:Bdireqgf1}
\Biggr\{ &\Gamma^{\uuv}  \biggr[ - \frac{r(1-f(r))}{2}\, \pd_r - \frac{i \omega }{r}- \frac{1- f(r) - r f'(r)}{4} - \frac{d\, r\, (1- f(r)) h'(r)}{8 h(r)}\biggr]  \nonumber \\*
& \grr \biggr[ \frac{r (1+ f(r))}{2}\, \pd_r - \frac{i \omega}{r}+ \frac{1 + f(r) + r f'(r)}{4}+ \frac{d\, r\, (1+ f(r))h'(r)}{8 h(r)}\biggr] + \frac{i k_i \gii}{\sqrt{h(r)}}-m\Biggr\} \psi(r) = 0\,.
\end{align}
In general, this is a system of $N$ first order coupled ordinary differential equations. In order to proceed, we want to decouple them in a way that  makes the pole-skipping mechanism manifest.

We start by separating the spinors according to the eigenvalues of the $\grr$ matrix.
Since $(\grr)^2 = 1$ and $\text{Tr}(\grr) =0$, this implies that exactly half of the eigenvalues are $+1$ while the other half are $-1$. Therefore, we introduce
\begin{align}
\label{eq:Bpsipmdef}
\psi = \psi_+ + \psi_- \,, \qquad  \grr \, \psi_{\pm} = \pm \psi_{\pm}\,,\qquad P_{\pm} \equiv  \frac{1}{2}\left( 1 \pm \grr\right)\,,
\end{align}
and insert this decomposition into  \eqref{eq:Bdireqgf1}. This allows us to split the Dirac equations into two independent equations according to the subspaces for $\psi_{\pm}$, which we obtain by acting on \eqref{eq:Bdireqgf1} with the two projection operators \eqref{eq:Bpsipmdef}.
Notice, however, that $\grr \Gamma^{\uua} \psi_\pm = \mp \Gamma^{\uua} \psi_{\pm}$ for $ a \neq r$, meaning that any action of a gamma matrix that is not $\grr$ changes the subspace in which the spinor lives.
The two independent  equations then read
\begin{subequations}
\label{eq:Bdireqgf2}
\begin{align}
&\Biggr[\left( - \frac{r(1-f(r))}{2}\, \pd_r - \frac{i \omega }{r}- \frac{1- f(r) - r f'(r)}{4} - \frac{d r (1- f(r)) h'(r)}{8 h(r)}\right)  \gvv + \frac{i k_i \gii }{\sqrt{h(r)}} \Biggr]\psi_-\nonumber\\*
&\quad + \Biggr[ \frac{r (1+ f(r))}{2}\, \pd_r - \frac{i \omega}{r}+ \frac{1 + f(r) + r f'(r)}{4}+ \frac{d r (1+ f(r))h'(r)}{8 h(r)}-m\Biggr]\psi_+  = 0\,,\\
&\Biggr[\left( - \frac{r(1-f(r))}{2}\, \pd_r - \frac{i \omega }{r}- \frac{1- f(r) - r f'(r)}{4} - \frac{d r (1- f(r)) h'(r)}{8 h(r)}\right)  \gvv + \frac{i k_i \gii }{\sqrt{h(r)}} \Biggr]\psi_+\nonumber\\*
&\quad - \Biggr[ \frac{r (1+ f(r))}{2}\, \pd_r - \frac{i \omega}{r}+ \frac{1 + f(r) + r f'(r)}{4}+ \frac{d r (1+ f(r))h'(r)}{8 h(r)}+m\Biggr]\psi_-  = 0\,.
\end{align}
\end{subequations}
We can see that these equations  in general have a linear combination of derivatives of different spinor components. However, we can transform them into a form where each equation only contains the derivative of a single component
\begin{subequations}
\label{eq:Bdireqgf3}
\begin{align}
&r^2 f(r)\, \pd_r \psi_+ + \gvv\, \biggr[- i \omega +\frac{ r^2 f'(r)}{4} + \frac{m\, r(1-f(r))}{2}- \frac{r(1+f(r))}{2 \sqrt{h(r)}}\, i k_i \gvi\biggr] \psi_-\nonumber\\*
&\quad + \biggr[ - i \omega + \frac{r^2 f'(r)}4 + \frac{r f(r)}{4}\left(2 + \frac{d\, r\, h'(r)}{h(r)}\right) - \frac{m\, r (1+ f(r))}{2}- \frac{r(1- f(r))}{2 \sqrt{h(r)}}\, i k_i \gvi\biggr] \psi_+ =0\,, \\
&r^2 f(r)\, \pd_r \psi_- -  \gvv\, \biggr[- i \omega +\frac{ r^2 f'(r)}{4} - \frac{m\, r(1-f(r))}{2}- \frac{r(1+f(r))}{2 \sqrt{h(r)}}\, i k_i \gvi\biggr] \psi_+\nonumber\\*
&\quad + \biggr[ - i \omega + \frac{r^2 f'(r)}4 + \frac{r f(r)}{4}\left(2 + \frac{d\, r\, h'(r)}{h(r)}\right) + \frac{m\, r (1+ f(r))}{2}- \frac{r(1- f(r))}{2 \sqrt{h(r)}}\, i k_i \gvi\biggr] \psi_- =0\,.
\end{align}
\end{subequations}

We also observe that for $ d \geq 2$, the two matrices $\grr$ and $\gvi$ are independent and commuting\footnote{For $ d= 1$, which is the asymptotically AdS$_3$ case, we have $\gvi = \pm \grr$, as $(\mathbb{1}, \gvv, \gii, \grr)$ provide a complete basis for any $2\times2$ matrix.}.
The matrix
\begin{align}
\kh \gvi \equiv \frac{k_i}{k} \, \gvi\,,
\end{align}
where
\begin{align*}
k = \sqrt{\sum_{i=1}^d k_i k_i}\,,
\end{align*}
squares to unity and is traceless. Since it is commuting with $\grr$, we can find common eigenvectors. Thus, we define
\begin{align}
\label{eq:Bpsipm2def}
\psi_{a} = \psi_a^{(+)} + \psi_{a}^{(-)}\,, \qquad \kh \gvi\, \psi_{a}^{(\pm)} = \pm \psi_{a}^{(\pm)}\,, \qquad P^{(\pm)} \equiv \frac12\left( 1 \pm \kh \gvi\right)\,,
\end{align}
where $a = \pm$.  We have now divided the initial spinor $\psi$ that with $N$ degrees of freedom into four independent parts $\psi_{\pm}^{(\pm)}$ that each contain $N/4$ independent components.  Inserting this decomposition into \eqref{eq:Bdireqgf3} and separating each of the equations with the projectors defined in \eqref{eq:Bpsipm2def} give four independent equations each containing only one derivative term
\begin{subequations}
\label{eq:Bdireqgf4}
\begin{align}
\label{eq:Bdireqgf41}
&r^2 f(r)\, \pd_r \psi_+^{(+)} + \gvv\, \biggr[- i \omega +\frac{ r^2 f'(r)}{4} + \frac{m\, r(1-f(r))}{2}+ \frac{ ik r(1+f(r))}{2 \sqrt{h(r)}}\biggr] \psi_-^{(-)}\nonumber\\*
&\quad + \biggr[ - i \omega + \frac{r^2 f'(r)}4 + \frac{r f(r)}{4}\left(2 + \frac{d\, r\, h'(r)}{h(r)}\right) - \frac{m\, r (1+ f(r))}{2}- \frac{ik r(1- f(r))}{2 \sqrt{h(r)}}\, \biggr] \psi_+^{(+)} =0\,,\\
\label{eq:Bdireqgf42}
&r^2 f(r)\, \pd_r \psi_-^{(-)} - \gvv\, \biggr[- i \omega +\frac{ r^2 f'(r)}{4} - \frac{m\, r(1-f(r))}{2}- \frac{ ik r(1+f(r))}{2 \sqrt{h(r)}}\biggr] \psi_+^{(+)}\nonumber\\*
&\quad + \biggr[ - i \omega + \frac{r^2 f'(r)}4 + \frac{r f(r)}{4}\left(2 + \frac{d\, r\, h'(r)}{h(r)}\right) + \frac{m\, r (1+ f(r))}{2}+ \frac{ik r(1- f(r))}{2 \sqrt{h(r)}}\, \biggr] \psi_-^{(-)} =0\,,\\
\label{eq:Bdireqgf43}
&r^2 f(r)\, \pd_r \psi_+^{(-)} + \gvv\, \biggr[- i \omega +\frac{ r^2 f'(r)}{4} + \frac{m\, r(1-f(r))}{2}- \frac{ ik r(1+f(r))}{2 \sqrt{h(r)}}\biggr] \psi_-^{(+)}\nonumber\\*
&\quad + \biggr[ - i \omega + \frac{r^2 f'(r)}4 + \frac{r f(r)}{4}\left(2 + \frac{d\, r\, h'(r)}{h(r)}\right) - \frac{m\, r (1+ f(r))}{2}+ \frac{ik r(1- f(r))}{2 \sqrt{h(r)}}\, \biggr] \psi_+^{(-)} =0\,,\\
\label{eq:Bdireqgf44}
&r^2 f(r)\, \pd_r \psi_-^{(+)} - \gvv\, \biggr[- i \omega +\frac{ r^2 f'(r)}{4} - \frac{m\, r(1-f(r))}{2}+ \frac{ ik r(1+f(r))}{2 \sqrt{h(r)}}\biggr] \psi_+^{(-)}\nonumber\\*
&\quad + \biggr[ - i \omega + \frac{r^2 f'(r)}4 + \frac{r f(r)}{4}\left(2 + \frac{d\, r\, h'(r)}{h(r)}\right) + \frac{m\, r (1+ f(r))}{2}- \frac{ik r(1- f(r))}{2 \sqrt{h(r)}}\, \biggr] \psi_-^{(+)} =0\,.
\end{align}
\end{subequations}
The equations split into two decoupled subsystems, one for the pair $(\psi_+^{(+)}, \psi_-^{(-)})$ and one for the pair $(\psi_+^{(-)}, \psi_-^{(+)})$. Furthermore, we observe that equations \eqref{eq:Bdireqgf41} and \eqref{eq:Bdireqgf43} are equivalent, except that $k \rightarrow -k$. The same can be said for the pair of equations \eqref{eq:Bdireqgf42} and \eqref{eq:Bdireqgf44}. This allows us to focus only on the subsystem \eqref{eq:Bdireqgf41} and \eqref{eq:Bdireqgf42}, while in order to obtain the solutions for the other subsystem (\eqref{eq:Bdireqgf43} and \eqref{eq:Bdireqgf44}), we only need to change $k$ into $-k$.

Using \eqref{eq:Bdireqgf41} and \eqref{eq:Bdireqgf42}, it is pretty straightforward to eliminate one of the spinors to obtain a decoupled and diagonal second order differential equation for a single spinor, either $\psi_{+}^{(+)}$ or $\psi_{-}^{(-)}$.
As a check, one can expand the spinor around the boundary $r \rightarrow \infty$, and look for the leading behavior of the spinors. One finds that
\begin{align}
\label{eq:Basyexp}
\psi_{+}^{(+)} \sim r^{- \frac{d+1}{2}+ m} + r^{- \frac{d+ 3}{2}- m}\,,\qquad \psi_{-}^{(-)} \sim r^{- \frac{d+ 3}{2}+ m} + r^{- \frac{d+1}{2}- m}\,,
\end{align}
which is in agreement with the results obtained  in \cite{Iqbal:2009fd}.
 Note that since the equations \eqref{eq:Bdireqgf43} and \eqref{eq:Bdireqgf44} are essentially the same except for $k \rightarrow - k$, the same asymptotic behavior is observed for $\psi_{\pm}^{(\mp)}$ as well.

\section{Exact fermionic Green's function for BTZ black hole}
\label{app:ExactBTZ}

The metric of a spinning BTZ black hole \cite{Banados:1992gq, Banados:1992wn} can be defined as follows
\bea
\label{eq:Cmetold}
ds^{2} = -\frac{(r^{2} - r^{2}_{+})(r^{2} - r^{2}_{-})}{r^{2}} \, dt^{2} + \frac{r^{2}dr^{2}}{(r^{2} - r^{2}_{+})(r^{2} - r^{2}_{-})} + r^{2}\left(d\phi - \frac{r_{+}r_{-}}{r^2} \, dt\right)^{2},
\eea
where we define $\phi$ as an angular coordinate having a period of $2\pi$. There are several parameters of the system that are given by
\bea
M =  \frac{r^{2}_{+} + r^{2}_{-} }{8G}\,, \qquad
J = \frac{r_{+}r_{-}}{4G}\,, \qquad
T_{L} =  \frac{r_{+} - r_{-}}{2\pi}\,, \qquad
T_{R} = \frac{r_{+} + r_{-}}{2\pi}\,,
\eea
where $M$ is the mass, $J$ is the angular momentum, $T_{L}$ and $T_{R}$ are the left and right moving temperature of the system respectively, and $G$ is the Newton constant in 3-dimensions.

It is convenient to change to a new coordinate system $(r, t, \phi) \rightarrow (\rho, T, X)$, in which the variables are defined as
\begin{subequations}
\begin{align}
r^{2} =  r^{2}_{+}\cosh^{2}\rho - r^{2}_{-}\sinh^{2}\rho\,, \qquad
T + X =  (r_{+} - r_{-})(t + \phi)\,,  \qquad
T - X =  (r_{+} + r_{-})(t - \phi).
\end{align}
\end{subequations}
In these new coordinates the metric is written as
\be
\label{eq:Cmetnew}
ds^2 = -\sinh^{2}{\rho}\, dT^2 + \cosh^{2}\rho \,dX^2 + d\rho^{2}.
\ee
We then choose the diagonal frame, such that
\begin{align}
E^T = - \sinh \rho\, dT\,, \qquad E^X = \cosh\rho\, dX\,, \qquad E^\rho = d \rho\,.
\end{align}
The spin connections in this frame are given by
\bea
\nonumber
\omega_{T{\rho}} = & -\cosh\rho\, dT  \\
\omega_{X{\rho}} = & -\sinh\rho\, dX.
\eea
In the new coordinates, the metric depends only on the coordinate $\rho$ (in the old coordinates the metric depends only on $r$), and hence we can expand the solutions in the basis of plane waves as
\bea
\psi(T, X, \rho) = e^{-ik_{T}T + ik_{X}X} \psi(\rho, k_{\mu}) = e^{-i\omega t + ik\phi} \psi(\rho, k_{\mu})\,.
\eea
The  momenta $(\omega, k)$ and  $(k_{T}, k_{X})$ are related via
\bea
\label{eq:Ckom}
k_{T} + k_{X} = \frac{\omega + k}{2 \pi T_{R}}, && k_{T} - k_{X} = \frac{\omega - k}{2 \pi T_{L}}.
\eea
The Dirac equations in Fourier space are then given by
\bea
\left[\Gamma^{\rho}\left(\partial_{\rho} + \frac{1}{2}\left(\frac{\cosh \rho}{\sinh \rho} + \frac{\sinh \rho}{\cosh \rho}\right) \right) + i \left(\frac{k_{X}\Gamma^{X}}{\cosh \rho} - \frac{k_{T}\Gamma^{T}}{\sinh \rho}\right) - m\right] \psi = 0.
\eea
We can choose a matrix representation such that $\Gamma^{\rho} = \sigma^{3}$, $\Gamma^{T} = i \sigma^{2}$, $\Gamma^{X} = \sigma^{1}$ and write $\psi = (\psi_{+}, \psi_{-})^T$. Note that the subscript of the components denotes the eigenvalue under the action of the $\Gamma^\rho$ matrix. We rescale the two degrees of freedom by introducing
\bea
\label{eq:Cchi12}
\psi_{\pm} \equiv \sqrt{\frac{\cosh\rho \pm \sinh \rho}{\cosh\rho \sinh\rho}} \left(\chi_{1} \pm \chi_{2}\right) , && z = \tanh ^{2} \rho\,.
\eea
In the coordinate $z$, the asymptotic boundary is located at $z =1$ and the horizon of the black hole is at $ z = 0$. In these coordinates, after some algebra, the Dirac equations can be written as
\bea
\nonumber
2(1 - z)\sqrt{z}\partial_{z}\chi_{1} - i\left(\frac{k_{T}}{\sqrt{z}} + k_{X}\sqrt{z}\right)\chi_{1} = \left(m -\frac{1}{2} + i(k_{T} + k_{X})\right)\chi_{2} \\
2(1 - z)\sqrt{z}\partial_{z}\chi_{2} + i\left(\frac{k_{T}}{\sqrt{z}} + k_{X}\sqrt{z}\right)\chi_{2} = \left(m -\frac{1}{2} - i(k_{T} + k_{X})\right)\chi_{1}.
\eea
It is straightforward to transform these two equations into second order differential equations for a single variable. One finds that the solutions to these equations are the hypergeometric functions ${}_2F_1(a, b,c;z)$.

We first wish to calculate the retarded correlator, so we choose the solutions that are ingoing at the horizon. Such solutions are of the form
\begin{subequations}
\label{eq:Csol1}
\begin{align}
\chi_{1}(z)& =  \left(\frac{a - c}{c} \right)z^{\frac{1}{2} + \alpha}(1 - z)^{\beta}F(a, b + 1; c + 1; z) \\
\nonumber
\chi_{2}(z) &=  z^{\alpha}(1 - z)^{\beta}F(a, b; c; z)\,,
\end{align}
\end{subequations}
where the constants are  given by
\bea
\alpha = -\frac{ik_{T}}{2}, && \beta = -\frac{1}{4} + \frac{m}{2} \,,
\eea
and
\begin{subequations}
\label{eq:Cabc}
\begin{align}
a =& \frac{1}{2}\left(m + \frac{1}{2}\right) - \frac{i}{2}\left(k_{T} - k_{X}\right), \\
b =& \frac{1}{2}\left(m - \frac{1}{2}\right) - \frac{i}{2}\left(k_{T} + k_{X}\right), \\
c =& \frac{1}{2} - ik_{T}.
\end{align}
\end{subequations}
Inserting the solutions \eqref{eq:Csol1} into \eqref{eq:Cchi12} and expanding them around the asymptotic boundary, one finds that the two spinor components behave as
\bea \label{eq:Cpsi+-}
	\psi_{+} \sim A(1 - z)^{\frac{1}{2} - \frac{m}{2}} + B(1 - z)^{1 + \frac{m}{2}} && \psi_{-} \sim C(1 - z)^{1 - \frac{m}{2}} + D(1 - z)^{\frac{1}{2} + \frac{m}{2}}.
\eea
It is important to note that when $0 \leq m < \frac{1}{2}$, every term in $\psi_{\pm}$ is normalizable. Therefore, either $A$ or $D$ can be chosen to be the source, and the other as the corresponding response.

Recall that in general bulk dimensions, the mass $m$ of the fermionic field and the scaling dimension $\Delta$ of the dual operator are related via
\be
\Delta = \frac{d+1}{2} + m.
\ee
In the case of the BTZ black hole, $d=1$, and we get the relation
\begin{align}
\Delta = 1 + m\,.
\end{align}
If $m > 0$, the source is taken to be $A$ and the expectation value is $D$ and the retarded Green's function in this case is given by their ratio
\bea
G_{R} = i\frac{D}{A} =  -i \frac{\Gamma\left(\frac{1}{2} - m\right) \Gamma\left(\frac{1 - 2i(k_{T} - k_{X}) + 2m}{4}\right) \Gamma\left(\frac{3 - 2i(k_{T} + k_{X}) + 2m}{4}\right)}{\Gamma\left(\frac{1}{2} + m\right) \Gamma\left(\frac{ 1 - 2i(k_{T} + k_{X}) - 2m}{4}\right) \Gamma\left(\frac{3 - 2i(k_{T} - k_{X}) - 2m}{4}\right)}\,.
\eea
Using the relations \eqref{eq:Ckom} one then finds that the retarded Green's function in terms of the frequency $\omega$ and momentum $k$ is given by
\begin{align}
G_{R} =-i\,
{
 \Gamma\left(\frac{1}{2} - m\right)\Gamma\le({m\ov 2}+{1\ov 4}+{i(k-\om)\ov 4\pi T}\ri)
 \Gamma\le({m\ov 2}+{3\ov 4}-{i(k+\om)\ov 4\pi T}\ri) \ov
\Gamma\left(\frac{1}{2} + m\right) \Gamma\le(-{m\ov 2}+{3\ov 4}+{i(k-\om)\ov 4\pi T}\ri)
 \Gamma\le(-{m\ov 2}+{1\ov 4}-{i(k+\om)\ov 4\pi T}\ri) }\,.
\end{align}

To calculate the advanced Green's function, we need the solutions that are outgoing at the horizon. In this case, the solutions take on the same form as \eqref{eq:Cchi12}, only that $\chi_1(z) \leftrightarrow \chi_2(z)$ and the parameters in the solutions are now
\bea
\alpha = \frac{ik_{T}}{2}, && \beta = -\frac{1}{4} + \frac{m}{2} \,,
\eea
and
\begin{subequations}
\label{q:CabcA}
\begin{align}
a =& \frac{1}{2}\left(m + \frac{1}{2}\right) + \frac{i}{2}\left(k_{T} - k_{X}\right),\\
b =& \frac{1}{2}\left(m - \frac{1}{2}\right) + \frac{i}{2}\left(k_{T} + k_{X}\right), \\
c =& \frac{1}{2} + ik_{T}.
\end{align}
\end{subequations}

Following the same steps as in the calculation of the retarded Green's function, the advanced Green's function works out to be
\bea
G_{A} = i \, \frac{\Gamma\left(\frac{1}{2} - m\right) \Gamma\left(\frac{ 1 + 2i(k_{T} - k_{X}) + 2m}{4}\right) \Gamma\left(\frac{ 3 + 2i(k_{T} + k_{X}) + 2m}{4}\right)}{\Gamma\left(\frac{1}{2} + m\right) \Gamma\left(\frac{ 1 + 2i(k_{T} + k_{X}) - 2m}{4}\right) \Gamma\left(\frac{ 3 + 2i(k_{T} - k_{X}) - 2m}{4}\right)}.
\eea

In terms of the frequency $\omega$ and momentum $k$, the advanced Green's function can be written as
\begin{align}
G_{A} = i \,
{
 \Gamma\left(\frac{1}{2} - m\right)\Gamma\le({m\ov 2}+{1\ov 4}-{i(k-\om)\ov 4\pi T}\ri)
 \Gamma\le({m\ov 2}+{3\ov 4}+{i(k+\om)\ov 4\pi T}\ri) \ov
\Gamma\left(\frac{1}{2} + m\right) \Gamma\le(-{m\ov 2}+{3\ov 4}-{i(k-\om)\ov 4\pi T}\ri)
 \Gamma\le(-{m\ov 2}+{1\ov 4}+{i(k+\om)\ov 4\pi T}\ri) }\,.
\end{align}

\subsection{At Matsubara frequencies}

The hypergeometric functions in \eqref{eq:Csol1} are generally well-defined at generic $k_X$ unless their third arguments are non-positive integers. By taking the limit $ik_T \rightarrow \frac{1}{2}+ n$, where $n \in \{0,1, \dots\}$, we can investigate what happens at these points, where the values correspond exactly to the Matsubara frequencies $\omega_n = -i \pi T(2n + 1)$. The ingoing solutions blow up as $ik_T \rightarrow \frac{1}{2} +n$, so we divide by another infinite factor to give a finite limit
\bea
\widetilde{\psi}_{in}(z) \equiv \lim_{ik_T \rightarrow \frac{1}{2} +n} \frac{\psi_{ in}(z)}{\Gamma\left(\frac{1}{2} - ik_{T}\right)}.
\eea

The ingoing and outgoing solutions are now degenerate
\begin{align}
\widetilde{\psi}_{in}(z) &= \frac{\Gamma\left(\frac{\Delta + n - ik_{X}}{2}\right) \Gamma\left(\frac{\Delta + 1 + n + ik_{X}}{2}\right)}{\Gamma (1 + n) \, \Gamma\left(\frac{\Delta - n - ik_{X}}{2}\right) \Gamma\left(\frac{\Delta - 1 - n + ik_{X}}{2}\right)}\,\psi_{out}(z)\,.
\end{align}
Since the ingoing and outgoing solutions are proportional to each other and their ratios are independent of $z$, we conclude the equivalence of the retarded and advanced Green's functions at these Matsubara frequencies.

\subsection{At half-integer conformal dimension}
We can rewrite \eqref{eq:Cpsi+-} in terms of $\Delta = 1 + m$ as
\bea \label{Eq:psidelta}
\psi_{+} \sim A(1 - z)^{1-\tfrac{\Delta}{2}} + B(1 - z)^{\tfrac{1}{2} + \tfrac{\Delta}{2}}&& \psi_{-} \sim C(1 - z)^{\frac{3}{2}- \tfrac{\Delta}{2}} + D(1 - z)^{\tfrac{\Delta}{2}}\,.
\eea
At half-integer $\Delta$ values, calculating the Green's function is not as straightforward because the exponents in \eqref{Eq:psidelta} differ by an integer and this results in logarithms appearing in the expansion near the boundary at $z \approx 1$.
The expansion of the hypergeometric function that is appropriate to use here is
\begin{eqnarray}
\label{Eq:HypergeoPosn}
\nonumber
&_{2}F_{1}(a, b; a + b - n; z) = \frac{(n - 1)!\Gamma(a + b - n)}{\Gamma(a)\Gamma(b)}(1-z)^{-n} \sum_{j=0}^{n-1}\frac{(a - n)_{j} (b - n)_{j}(1 - z)^{j}}{j!(1 - n)_{j}} + (-1)^n\frac{\Gamma(a + b - n)}{\Gamma(a - n)\Gamma(b - n)} \times \\*
&\hspace{-5ex} \sum_{j=0}^{\infty} \frac{(a)_{j}(b)_{j}}{j!(j + n)!}\biggr[-\log(1 - z) + \psi(j + 1) + \psi(j + n + 1) - \psi(a + j) - \psi(b + j)\biggr](1 - z)^{j}.
\end{eqnarray}
where the arguments of the hypergeometric function take the same form as the ones in \eqref{eq:Csol1}. $n$ is an integer and in this case,  $n= m - \frac{1}{2}$, $(x)_{j} \equiv \frac{\Gamma(x + j)}{\Gamma(x)}$ is the Pochhammer symbol and $\psi(x)$ is the digamma function.

To calculate the retarded Green's function, we consider terms up to order $(1 -z)^{-n}$ in the above expansion. The leading term is the source and the sum of non-logarithmic terms multiplying $(1 - z)^{-n}$ is the expectation value. Then, the Green's function is simply the ratio of the expectation value to the source up to a normalization factor.

\subsubsection*{Retarded Green's function}
To compute the retarded Green's function, we take $j = 0$. The terms in the expansion of the prefactors only contribute to the contact terms so we can ignore them%
\footnote{The prefactors $z^{\alpha}$ and $z^{\frac{1}{2} + \alpha}$ in \eqref{eq:Csol1} have contributions to the overall expansion of $\psi_{\pm}$ and can typically be expanded near the boundary in the following way
\bea
\nonumber
z^{\gamma} = \sum_{j=0}^{\infty} \frac{1}{j!} \frac{\Gamma(\gamma + 1)}{\Gamma(\gamma -j +1)} (z - 1)^{j}.
\eea
In our analysis, we do not take into account these additional terms.}.
The expectation value is
\begin{align}
\nonumber
D = (-1)^n \left({a - c} \ov c \right) \frac{\Gamma(a + b + 1 - n)}{\Gamma(a - n)\Gamma(b + 1 - n)} \frac{1}{n!}\left[\psi(1) + \psi(n + 1) - \psi(a) - \psi(b + 1)\right] & \\
- (-1)^n \frac{\Gamma(a + b - n)}{\Gamma(a - n)\Gamma(b - n)}\frac{1}{n!}\left[\psi(1) + \psi(n + 1) - \psi(a) - \psi(b)\right]
\end{align}
where the first term comes from $\chi_{1}(z)$ and the second from $\chi_{2}(z)$.
The source is
\begin{align}
& A =  \left(\frac{a-c}{c}\right) \frac{(n - 1)!\Gamma(a + b + 1 + n)}{\Gamma(a)\Gamma(b + 1)} + \frac{(n - 1)!\Gamma(a + b - n)}{\Gamma(a)\Gamma(b)},
\end{align}
where again the first term comes from $\chi_{1}(z)$ and the second from $\chi_{2}(z)$.
The retarded Green's function is then given by
\begin{align}
\label{eq:Cretardedgreenhalfinteger}
G_{R} (\omega, k) &\propto  \,
{
 \Gamma\left(\frac{\Delta}{2} - {1 \ov 4} + i{(k - \omega) \ov 4\pi T}\right)\Gamma\left(\frac{\Delta}{2} + {1 \ov 4} - i{(k + \omega) \ov 4\pi T}\right) \ov
\Gamma\left(-\frac{\Delta}{2} + {5 \ov 4} + i{(k - \omega) \ov 4\pi T}\right)\Gamma\left(-\frac{\Delta}{2} + {3 \ov 4} - i{(k + \omega) \ov 4\pi T}\right)} \times \nonumber\\
&  \hspace{20ex} \,
\left[\psi \left(\frac{\Delta}{2} - {1 \ov 4} + i{(k - \omega) \ov 4\pi T}\right)+  \psi \left(\frac{\Delta}{2} + {1 \ov 4} - i{(k + \omega) \ov 4\pi T}\right)\right],
\end{align}
up to contact terms.

\subsubsection*{Advanced Green's function}
Following the same steps as above but keeping in mind that $\chi_{1}(z) \leftrightarrow \chi_{2}(z)$ in the case of the outgoing solutions, the advanced Green's function turns out to be
\begin{align}
G_{A} (\omega, k)& \propto \,
{
 \Gamma\left(\frac{\Delta}{2} + {1 \ov 4} - i{(k - \omega) \ov 4\pi T}\right)\Gamma\left(\frac{\Delta}{2} - {1 \ov 4} + i{(k + \omega) \ov 4\pi T}\right) \ov
\Gamma\left(\frac34 -\frac{\Delta}{2}  - i{(k - \omega) \ov 4\pi T}\right)\Gamma\left(\frac54-\frac{\Delta}{2} + i{(k + \omega) \ov 4\pi T}\right)} \,\times \nonumber\\*
&  \hspace{20ex} \,
\left[\psi \left(\frac{\Delta}{2} + {1 \ov 4} - {i(k - \omega) \ov 4\pi T}\right) + \psi \left(\frac{\Delta}{2} - {1 \ov 4} + {i(k + \omega) \ov 4\pi T}\right)\right],
\end{align}
up to contact terms.


\begin{adjustwidth}{-1mm}{-1mm} 

\bibliographystyle{utphys}

\bibliography{adscft}

\providecommand{\href}[2]{#2}\begingroup\raggedright\begin{thebibliography}{10}

\bibitem{Maldacena:1997re}
J.~M. Maldacena, ``{The Large N limit of superconformal field theories and
  supergravity},'' \href{http://dx.doi.org/10.1023/A:1026654312961,
  10.4310/ATMP.1998.v2.n2.a1}{{\em Int. J. Theor. Phys.} {\bfseries 38} (1999)
  1113--1133}, \href{http://arxiv.org/abs/hep-th/9711200}{{\ttfamily
  arXiv:hep-th/9711200 [hep-th]}}.
[Adv. Theor. Math. Phys.2,231(1998)].

\bibitem{Gubser:1998bc}
S.~S. Gubser, I.~R. Klebanov, and A.~M. Polyakov, ``{Gauge theory correlators
  from noncritical string theory},''
  \href{http://dx.doi.org/10.1016/S0370-2693(98)00377-3}{{\em Phys. Lett.}
  {\bfseries B428} (1998) 105--114},
\href{http://arxiv.org/abs/hep-th/9802109}{{\ttfamily arXiv:hep-th/9802109
  [hep-th]}}.

\bibitem{Witten:1998qj}
E.~Witten, ``{Anti-de Sitter space and holography},''
  \href{http://dx.doi.org/10.4310/ATMP.1998.v2.n2.a2}{{\em Adv. Theor. Math.
  Phys.} {\bfseries 2} (1998) 253--291},
\href{http://arxiv.org/abs/hep-th/9802150}{{\ttfamily arXiv:hep-th/9802150
  [hep-th]}}.

\bibitem{Son:2002sd}
D.~T. Son and A.~O. Starinets, ``{Minkowski space correlators in AdS / CFT
  correspondence: Recipe and applications},''
  \href{http://dx.doi.org/10.1088/1126-6708/2002/09/042}{{\em JHEP} {\bfseries
  09} (2002) 042},
\href{http://arxiv.org/abs/hep-th/0205051}{{\ttfamily arXiv:hep-th/0205051
  [hep-th]}}.

\bibitem{Horowitz:1999jd}
G.~T. Horowitz and V.~E. Hubeny, ``{Quasinormal modes of AdS black holes and
  the approach to thermal equilibrium},''
  \href{http://dx.doi.org/10.1103/PhysRevD.62.024027}{{\em Phys. Rev.}
  {\bfseries D62} (2000) 024027},
\href{http://arxiv.org/abs/hep-th/9909056}{{\ttfamily arXiv:hep-th/9909056
  [hep-th]}}.

\bibitem{Herzog:2002pc}
C.~P. Herzog and D.~T. Son, ``{Schwinger-Keldysh propagators from AdS/CFT
  correspondence},''
  \href{http://dx.doi.org/10.1088/1126-6708/2003/03/046}{{\em JHEP} {\bfseries
  03} (2003) 046},
\href{http://arxiv.org/abs/hep-th/0212072}{{\ttfamily arXiv:hep-th/0212072
  [hep-th]}}.

\bibitem{Skenderis:2008dh}
K.~Skenderis and B.~C. van Rees, ``{Real-time gauge/gravity duality},''
  \href{http://dx.doi.org/10.1103/PhysRevLett.101.081601}{{\em Phys. Rev.
  Lett.} {\bfseries 101} (2008) 081601},
\href{http://arxiv.org/abs/0805.0150}{{\ttfamily arXiv:0805.0150 [hep-th]}}.

\bibitem{Skenderis:2008dg}
K.~Skenderis and B.~C. van Rees, ``{Real-time gauge/gravity duality:
  Prescription, Renormalization and Examples},''
  \href{http://dx.doi.org/10.1088/1126-6708/2009/05/085}{{\em JHEP} {\bfseries
  05} (2009) 085},
\href{http://arxiv.org/abs/0812.2909}{{\ttfamily arXiv:0812.2909 [hep-th]}}.

\bibitem{Son:2009vu}
D.~T. Son and D.~Teaney, ``{Thermal Noise and Stochastic Strings in AdS/CFT},''
  \href{http://dx.doi.org/10.1088/1126-6708/2009/07/021}{{\em JHEP} {\bfseries
  07} (2009) 021},
\href{http://arxiv.org/abs/0901.2338}{{\ttfamily arXiv:0901.2338 [hep-th]}}.

\bibitem{Iqbal:2009fd}
N.~Iqbal and H.~Liu, ``{Real-time response in AdS/CFT with application to
  spinors},'' \href{http://dx.doi.org/10.1002/prop.200900057}{{\em Fortsch.
  Phys.} {\bfseries 57} (2009) 367--384},
\href{http://arxiv.org/abs/0903.2596}{{\ttfamily arXiv:0903.2596 [hep-th]}}.

\bibitem{vanRees:2009rw}
B.~C. van Rees, ``{Real-time gauge/gravity duality and ingoing boundary
  conditions},''
  \href{http://dx.doi.org/10.1016/j.nuclphysbps.2009.07.078}{{\em Nucl. Phys.
  Proc. Suppl.} {\bfseries 192-193} (2009) 193--196},
\href{http://arxiv.org/abs/0902.4010}{{\ttfamily arXiv:0902.4010 [hep-th]}}.

\bibitem{Glorioso:2018mmw}
P.~Glorioso, M.~Crossley, and H.~Liu, ``{A prescription for holographic
  Schwinger-Keldysh contour in non-equilibrium systems},''
\href{http://arxiv.org/abs/1812.08785}{{\ttfamily arXiv:1812.08785 [hep-th]}}.

\bibitem{Liu:2018crr}
H.~Liu and J.~Sonner, ``{Holographic systems far from equilibrium: a review},''
\href{http://arxiv.org/abs/1810.02367}{{\ttfamily arXiv:1810.02367 [hep-th]}}.

\bibitem{deBoer:2018qqm}
J.~de~Boer, M.~P. Heller, and N.~Pinzani-Fokeeva, ``{Holographic
  Schwinger-Keldysh effective field theories},''
  \href{http://dx.doi.org/10.1007/JHEP05(2019)188}{{\em JHEP} {\bfseries 05}
  (2019) 188},
\href{http://arxiv.org/abs/1812.06093}{{\ttfamily arXiv:1812.06093 [hep-th]}}.

\bibitem{Banados:1992gq}
M.~Banados, M.~Henneaux, C.~Teitelboim, and J.~Zanelli, ``{Geometry of the
  (2+1) black hole},'' \href{http://dx.doi.org/10.1103/PhysRevD.48.1506,
  10.1103/PhysRevD.88.069902}{{\em Phys. Rev.} {\bfseries D48} (1993)
  1506--1525}, \href{http://arxiv.org/abs/gr-qc/9302012}{{\ttfamily
  arXiv:gr-qc/9302012 [gr-qc]}}.
[Erratum: Phys. Rev.D88,069902(2013)].

\bibitem{Banados:1992wn}
M.~Banados, C.~Teitelboim, and J.~Zanelli, ``{The Black hole in
  three-dimensional space-time},''
  \href{http://dx.doi.org/10.1103/PhysRevLett.69.1849}{{\em Phys. Rev. Lett.}
  {\bfseries 69} (1992) 1849--1851},
\href{http://arxiv.org/abs/hep-th/9204099}{{\ttfamily arXiv:hep-th/9204099
  [hep-th]}}.

\bibitem{Birmingham:2001hc}
D.~Birmingham, ``{Choptuik scaling and quasinormal modes in the AdS / CFT
  correspondence},'' \href{http://dx.doi.org/10.1103/PhysRevD.64.064024}{{\em
  Phys. Rev.} {\bfseries D64} (2001) 064024},
\href{http://arxiv.org/abs/hep-th/0101194}{{\ttfamily arXiv:hep-th/0101194
  [hep-th]}}.

\bibitem{Cardoso:2001hn}
V.~Cardoso and J.~P.~S. Lemos, ``{Scalar, electromagnetic and Weyl
  perturbations of BTZ black holes: Quasinormal modes},''
  \href{http://dx.doi.org/10.1103/PhysRevD.63.124015}{{\em Phys. Rev.}
  {\bfseries D63} (2001) 124015},
\href{http://arxiv.org/abs/gr-qc/0101052}{{\ttfamily arXiv:gr-qc/0101052
  [gr-qc]}}.

\bibitem{Birmingham:2001pj}
D.~Birmingham, I.~Sachs, and S.~N. Solodukhin, ``{Conformal field theory
  interpretation of black hole quasinormal modes},''
  \href{http://dx.doi.org/10.1103/PhysRevLett.88.151301}{{\em Phys. Rev. Lett.}
  {\bfseries 88} (2002) 151301},
\href{http://arxiv.org/abs/hep-th/0112055}{{\ttfamily arXiv:hep-th/0112055
  [hep-th]}}.

\bibitem{Blake:2019otz}
M.~Blake, R.~A. Davison, and D.~Vegh, ``{Horizon constraints on holographic
  Green's functions},''
\href{http://arxiv.org/abs/1904.12883}{{\ttfamily arXiv:1904.12883 [hep-th]}}.

\bibitem{Kovtun:2004de}
P.~Kovtun, D.~T. Son, and A.~O. Starinets, ``{Viscosity in strongly interacting
  quantum field theories from black hole physics},''
  \href{http://dx.doi.org/10.1103/PhysRevLett.94.111601}{{\em Phys. Rev. Lett.}
  {\bfseries 94} (2005) 111601},
\href{http://arxiv.org/abs/hep-th/0405231}{{\ttfamily arXiv:hep-th/0405231
  [hep-th]}}.

\bibitem{Grozdanov:2017ajz}
S.~Grozdanov, K.~Schalm, and V.~Scopelliti, ``{Black hole scrambling from
  hydrodynamics},''
  \href{http://dx.doi.org/10.1103/PhysRevLett.120.231601}{{\em Phys. Rev.
  Lett.} {\bfseries 120} no.~23, (2018) 231601},
\href{http://arxiv.org/abs/1710.00921}{{\ttfamily arXiv:1710.00921 [hep-th]}}.

\bibitem{Blake:2017ris}
M.~Blake, H.~Lee, and H.~Liu, ``{A quantum hydrodynamical description for
  scrambling and many-body chaos},''
  \href{http://dx.doi.org/10.1007/JHEP10(2018)127}{{\em JHEP} {\bfseries 10}
  (2018) 127},
\href{http://arxiv.org/abs/1801.00010}{{\ttfamily arXiv:1801.00010 [hep-th]}}.

\bibitem{Blake:2018leo}
M.~Blake, R.~A. Davison, S.~Grozdanov, and H.~Liu, ``{Many-body chaos and
  energy dynamics in holography},''
  \href{http://dx.doi.org/10.1007/JHEP10(2018)035}{{\em JHEP} {\bfseries 10}
  (2018) 035},
\href{http://arxiv.org/abs/1809.01169}{{\ttfamily arXiv:1809.01169 [hep-th]}}.

\bibitem{Grozdanov:2018kkt}
S.~Grozdanov, ``{On the connection between hydrodynamics and quantum chaos in
  holographic theories with stringy corrections},''
  \href{http://dx.doi.org/10.1007/JHEP01(2019)048}{{\em JHEP} {\bfseries 01}
  (2019) 048},
\href{http://arxiv.org/abs/1811.09641}{{\ttfamily arXiv:1811.09641 [hep-th]}}.

\bibitem{Shenker:2013pqa}
S.~H. Shenker and D.~Stanford, ``{Black holes and the butterfly effect},''
  \href{http://dx.doi.org/10.1007/JHEP03(2014)067}{{\em JHEP} {\bfseries 03}
  (2014) 067},
\href{http://arxiv.org/abs/1306.0622}{{\ttfamily arXiv:1306.0622 [hep-th]}}.

\bibitem{Shenker:2013yza}
S.~H. Shenker and D.~Stanford, ``{Multiple Shocks},''
  \href{http://dx.doi.org/10.1007/JHEP12(2014)046}{{\em JHEP} {\bfseries 12}
  (2014) 046},
\href{http://arxiv.org/abs/1312.3296}{{\ttfamily arXiv:1312.3296 [hep-th]}}.

\bibitem{Shenker:2014cwa}
S.~H. Shenker and D.~Stanford, ``{Stringy effects in scrambling},''
  \href{http://dx.doi.org/10.1007/JHEP05(2015)132}{{\em JHEP} {\bfseries 05}
  (2015) 132},
\href{http://arxiv.org/abs/1412.6087}{{\ttfamily arXiv:1412.6087 [hep-th]}}.

\bibitem{Roberts:2014isa}
D.~A. Roberts, D.~Stanford, and L.~Susskind, ``{Localized shocks},''
  \href{http://dx.doi.org/10.1007/JHEP03(2015)051}{{\em JHEP} {\bfseries 03}
  (2015) 051},
\href{http://arxiv.org/abs/1409.8180}{{\ttfamily arXiv:1409.8180 [hep-th]}}.

\bibitem{Maldacena:2015waa}
J.~Maldacena, S.~H. Shenker, and D.~Stanford, ``{A bound on chaos},''
  \href{http://dx.doi.org/10.1007/JHEP08(2016)106}{{\em JHEP} {\bfseries 08}
  (2016) 106},
\href{http://arxiv.org/abs/1503.01409}{{\ttfamily arXiv:1503.01409 [hep-th]}}.

\bibitem{Balm:2019dxk}
F.~Balm, A.~Krikun, A.~Romero-Bermúdez, K.~Schalm, and J.~Zaanen, ``{Isolated
  zeros destroy Fermi surface in holographic models with a lattice},''
\href{http://arxiv.org/abs/1909.09394}{{\ttfamily arXiv:1909.09394 [hep-th]}}.

\bibitem{Grozdanov:2019uhi}
S.~Grozdanov, P.~K. Kovtun, A.~O. Starinets, and P.~Tadić, ``{The complex life
  of hydrodynamic modes},''
\href{http://arxiv.org/abs/1904.12862}{{\ttfamily arXiv:1904.12862 [hep-th]}}.

\bibitem{Natsuume:2019sfp}
M.~Natsuume and T.~Okamura, ``{Holographic chaos, pole-skipping, and
  regularity},''
\href{http://arxiv.org/abs/1905.12014}{{\ttfamily arXiv:1905.12014 [hep-th]}}.

\bibitem{Natsuume:2019xcy}
M.~Natsuume and T.~Okamura, ``{Nonuniqueness of Green's functions at special
  points},''
\href{http://arxiv.org/abs/1905.12015}{{\ttfamily arXiv:1905.12015 [hep-th]}}.

\bibitem{Henningson:1998cd}
M.~Henningson and K.~Sfetsos, ``{Spinors and the AdS / CFT correspondence},''
  \href{http://dx.doi.org/10.1016/S0370-2693(98)00559-0}{{\em Phys. Lett.}
  {\bfseries B431} (1998) 63--68},
\href{http://arxiv.org/abs/hep-th/9803251}{{\ttfamily arXiv:hep-th/9803251
  [hep-th]}}.

\bibitem{Mueck:1998iz}
W.~Mueck and K.~S. Viswanathan, ``{Conformal field theory correlators from
  classical field theory on anti-de Sitter space. 2. Vector and spinor
  fields},'' \href{http://dx.doi.org/10.1103/PhysRevD.58.106006}{{\em Phys.
  Rev.} {\bfseries D58} (1998) 106006},
\href{http://arxiv.org/abs/hep-th/9805145}{{\ttfamily arXiv:hep-th/9805145
  [hep-th]}}.

\bibitem{Das:2019tga}
S.~Das, B.~Ezhuthachan, and A.~Kundu, ``{Real Time Dynamics in Low Point
  Correlators},''
\href{http://arxiv.org/abs/1907.08763}{{\ttfamily arXiv:1907.08763 [hep-th]}}.

\bibitem{Andrade:2013gsa}
T.~Andrade and B.~Withers, ``{A simple holographic model of momentum
  relaxation},'' \href{http://dx.doi.org/10.1007/JHEP05(2014)101}{{\em JHEP}
  {\bfseries 05} (2014) 101},
\href{http://arxiv.org/abs/1311.5157}{{\ttfamily arXiv:1311.5157 [hep-th]}}.

\bibitem{Davison:2014lua}
R.~A. Davison and B.~Gouteraux, ``{Momentum dissipation and effective theories
  of coherent and incoherent transport},''
  \href{http://dx.doi.org/10.1007/JHEP01(2015)039}{{\em JHEP} {\bfseries 01}
  (2015) 039},
\href{http://arxiv.org/abs/1411.1062}{{\ttfamily arXiv:1411.1062 [hep-th]}}.

\end{thebibliography}\endgroup

\end{adjustwidth}


\end{document}